\documentclass[12pt]{article}
\usepackage{amssymb}
\usepackage{mathtools}
\usepackage{amsmath}
\usepackage[colorlinks,linkcolor=red,anchorcolor=black,citecolor=green]{hyperref}
\usepackage{xcolor}
\usepackage{graphicx}
\usepackage{multirow}
\usepackage{array}
\usepackage{cite}
\usepackage{pifont}
\usepackage{mathrsfs}
\usepackage{soul}
\usepackage{hepnames}
\usepackage{subcaption}
\usepackage[T1]{fontenc}
\usepackage{lmodern}
\usepackage{pifont}
\allowdisplaybreaks
\usepackage[normalem]{ulem}
\usepackage{listings}
\newcommand{\ignore}[1]{}

\usepackage{tabularx}
\usepackage{physics}

\def\be{\begin{equation}}
\def\ee{\end{equation}}

\allowdisplaybreaks

\begin{document}

\title{
\begin{flushright}
\hfill\mbox{\small USTC-ICTS/PCFT-21-40 } \\[5mm]
\begin{minipage}{0.2\linewidth}
\normalsize
\end{minipage}
\end{flushright}

{\Large \bf
 Decomposition of $d=9$ short-range $0\nu\beta\beta$ decay operators at one-loop level \\[2mm]} }

\date{}

\author{
Ping-Tao Chen$^{1,2}$\footnote{E-mail: {\tt
chenpt@mail.ustc.edu.cn}}, \
Gui-Jun~Ding$^{1,2}$\footnote{E-mail: {\tt
dinggj@ustc.edu.cn}},  \
Chang-Yuan~Yao$^{3}$\footnote{E-mail: {\tt
yaocy@nankai.edu.cn}}
\\*[20pt]
\centerline{
\begin{minipage}{\linewidth}
\begin{center}
$^1${\it \small
Interdisciplinary Center for Theoretical Study and  Department of Modern Physics,\\
University of Science and Technology of China, Hefei, Anhui 230026, China}\\[2mm]
$^2${\it\small Peng Huanwu Center for Fundamental Theory, Hefei, Anhui 230026, China} \\[2mm]
$^3${\it \small
School of Physics, Nankai University, Tianjin 300071, China}\\
\end{center}
\end{minipage}}
\\[10mm]}

\maketitle

\begin{abstract}

We perform a systematical study of the dimension-9 short-range $0\nu\beta\beta$ decay operators at one-loop level. There are only six genuine topologies which generate eight diagrams, and the recipe to identify the possible one-loop realizations of the $0\nu\beta\beta$ decay operators is sketched. Certain hypercharge assignments are excluded by the absence of tree-level diagrams in a genuine one-loop model. The mediators of each decomposition can generate Majorana neutrino masses which are discussed up to two-loop level. We present an example of $0\nu\beta\beta$ decay model in which the neutrino masses are generated at two-loop level, and the short-range contribution can be comparable with the mass mechanism in some region of parameter space.

\end{abstract}
\thispagestyle{empty}
\vfill

\newpage
\section{Introduction}

The discovery of neutrino oscillations shows that neutrinos have tiny masses. In order to generate the tiny neutrino masses, one must extend the filed content or gauge symmetry of the standard model, or depart from renormalizability~\cite{ParticleDataGroup:2020ssz}. Depending on whether neutrinos are Dirac or Majorana particles, the neutrino mass terms can be constructed in different ways. For the case of Majorana neutrinos, neutrinos are identical with their antiparticles and the lepton number would be violated by two units $\Delta L=2$. As a consequence, the exchange of light Majorana neutrino can lead to the well-known neutrinoless double beta ($0\nu\beta\beta$) decay
\begin{equation}
(A, Z)\rightarrow (A, Z+2)+e^{-}+e^{-} \,,
\end{equation}
which violates the lepton number by two units as well. This is the standard
mass mechanism. Conversely, if the rare $0\nu\beta\beta$ decay is observed, neutrinos must be Majorana particles~\cite{Schechter:1981bd}. The new physics giving rise to $0\nu\beta\beta$ decay will always produce a non-zero mass of Majorana neutrinos. Under the assumption that the Majorana neutrino mass is the unique source of $0\nu\beta\beta$ decay, the decay rate is fixed by the phase space factor, the nuclear matrix element and the effective Majorana mass $m_{\beta\beta}$ with
\begin{equation}
\label{eq:mbb-eff}m_{\beta\beta}= \sum^{3}_{i}U_{ei}^{2}m_{i}\,,
\end{equation}
where $U$ is the lepton mixing matrix and $m_i (i=1,2,3)$ are the light neutrino masses. The effective mass $m_{\beta\beta}$ is the (11) entry of the light neutrino mass matrix in the basis where the charged lepton mass matrix is diagonal.
%Hence non-zero neutrino masses can results in the $0\nu\beta\beta$ mass %mechanism,
Hence $0\nu\beta\beta$ decay can provide valuable information on both neutrino mass spectrum and Majorana CP phases. The experimental search for the $0\nu\beta\beta$ decay is extremely challenging, and no positive signal has been observed so far. The most stringent bound is given by the KamLAND-Zen Collaboration with the half-life $T^{0\nu}_{1/2}> 1.07\times10^{26}$~years for $^{136}$Xe at $90\%$ confidence level, this bound implies a limit on the effective Majorana neutrino mass $|m_{\beta\beta}|<(61-165)$ meV~\cite{KamLAND-Zen:2016pfg}. The next generation tonne-scale $0\nu\beta\beta$ decay experiments such as LEGEND~\cite{LEGEND:2021bnm} and nEXO~\cite{nEXO:2021ujk} aim at the half-life sensitivity of approximately $10^{28}$ years. If all the future results from $0\nu\beta\beta$ decays, neutrino oscillation experiments and cosmology observation on the light neutrino mass sum are compatible with each other, it would be a strong support to the mass mechanism mentioned above.

However, it is notable that the $0\nu\beta\beta$ decay could possibly be dominantly induced by other new physics effects beyond that of Majorana neutrino masses. The contributions to the $0\nu\beta\beta$ decay rate can be generally divided into the
short-range part~\cite{Pas:2000vn} and long-range part~\cite{Pas:1999fc}. See Refs.~\cite{Elliott:2002xe,Rodejohann:2011mu,Deppisch:2012nb,DellOro:2016tmg,Dolinski:2019nrj} for review.  The short-range contributions arise from the lepton number violating couplings or the Majorana mass terms of new particles in possible extensions of the Standard Model (SM).
%, and the exchanged particles are all much heavier
The short-range dynamics of the $0\nu\beta\beta$ decay is expected to be at a scale $\Lambda$ higher than the weak scale,
and it can be described by a set of $\Delta L=2$ and baryon conserving operators which are invariant under the SM gauge symmetry $SU(3)_C\times SU(2)_L \times U(1)_Y$ .
At leading order, there are 11 dimension-9 operators which give rise to $0\nu\beta\beta$ decay~\cite{Graesser:2016bpz,Liao:2020jmn,Li:2020xlh}:
\begin{subequations}
\begin{eqnarray}
\mathcal{O}^{0\nu}_{1} &=&  \epsilon_{ij} (\overline{Q}_{i}  \gamma^\mu Q_{m})
( \overline{u}_R  \gamma_\mu d_R)( \overline{\ell}_j  \ell^c _m ) \,, \\
\mathcal{O}^{0\nu}_{2}&=&  \epsilon_{i j} (\overline{Q}_{i}  \gamma^\mu \lambda^A Q_{m})(\overline{u}_R  \gamma_\mu \lambda^A d_R)(\overline{\ell}_j  \ell^c _m ) \,,  \\
\mathcal{O}^{0\nu}_{3}&=&(\overline{u}_R Q_{i} ) (\overline{u}_R Q_{j})
(\overline{\ell}_{i} \ell^c_{j}) \,,   \\
\mathcal{O}^{0\nu}_{4}&=&(\overline{u}_R \lambda^A Q_{i} ) (\overline{u}_R  \lambda^A Q_{j})(\overline{\ell}_{i}\ell^c_{j}) \,,  \\
\mathcal{O}^{0\nu}_{5}&=&\epsilon_{ij} \epsilon_{mn} (\overline{Q}_i d_R) (\overline{Q}_{m} d_R)(\overline{\ell} _{j}  {\ell}^c_n) \,,   \\
\mathcal{O}^{0\nu}_{6}&=&\epsilon_{ij} \epsilon_{mn} (\overline{Q}_i \lambda^A d_R) (\overline{Q}_{m}\lambda^A d_R)
(\overline{\ell} _{j}  {\ell}^c_n) \,,     \\
\mathcal{O}^{0\nu}_{7}&=&(\overline{u}_R \gamma^\mu d_R)(\overline{u}_R \gamma_\mu d_R)(\overline{e}_R  e^c_R) \,,  \\
\mathcal{O}^{0\nu}_{8}&=&\epsilon_{ij} (\overline{u}_R \gamma^\mu d_R) (\overline{Q}_{i} d_R)(\overline{\ell}_{j} \gamma_\mu  e^c_R)\,,      \\
\mathcal{O}^{0\nu}_{9}&= &\epsilon_{ij} (\overline{u}_R \gamma^\mu \lambda^A d_R) (\overline{Q}_{i}  \lambda^A d_R)
(\overline{\ell}_{j} \gamma_\mu  e^c_R)\,,   \\
\mathcal{O}^{0\nu}_{10}&=& (\overline{u}_R \gamma^\mu d_R)(\overline{u}_R Q_i)(\overline{\ell}_i \gamma_\mu  e^c_R) \,,   \\
\mathcal{O}^{0\nu}_{11}&= & (\overline{u}_R \gamma^\mu \lambda^A d_R)(\overline{u}_R \lambda^A Q_i)(\overline{\ell}_i \gamma_\mu  e^c_R)\,,
\end{eqnarray}
\label{eqn-basis-summary-dim9}
\end{subequations}
where $Q=(u_L, ~d_L)^{T}$ and $\ell=(\nu_L,~l_L)^{T}$ stand for the quark and lepton doublets of $SU(2)_L$ respectively, and we denote the charge conjugation fields $e^c_R=(e_R)^c$ and $\ell^c =((\nu_L)^{c},~(l_L)^c)^T$. The Roman letters $i, j, m, n=1,2$ refer to $SU(2)_L$ indices, and $\epsilon$ is the 2-component antisymmetric tensor with $\epsilon_{12}=-\epsilon_{21}=1$, and $\lambda^A$ are the Gell-Mann matrices. Here we only consider operators involving first generation fields, since we are concerned with the operators which directly contribute to the $0\nu\beta\beta$ decay. In addition to $0\nu\beta\beta$ decays, these operators in Eq.~\eqref{eqn-basis-summary-dim9} can produce the signals of same sign dilepton, single lepton $+$ MET or 2jets+MET at hadron colliders~\cite{Graesser:2016bpz}. The different possible ways to generate the above dimension nine ($d=9$) $0\nu\beta\beta$ decay operators at tree level have been studied in Ref.~\cite{Bonnet:2012kh}. All the tree-level decomposition of the long-range $0\nu\beta\beta$ decay operators has been discussed in Ref.~\cite{Helo:2016vsi}. The one-loop contribution to the
$0\nu\beta\beta$ amplitude was studied in the radiative neutrino mass models~\cite{Gustafsson:2014vpa,Liu:2016mpf,Alcaide:2017xoe,Rodejohann:2019quz}. In this paper, we shall study the systematical decomposition of the eleven operators $\mathcal{O}^{0\nu}_{1,2,\ldots,11}$ at one-loop level by using the diagrammatic approach~\cite{Antusch:2008tz,Gavela:2008ra}.
%\sout{and all possible realizations through one-loop diagrams would be %identified} \textcolor{red}{and identify all possible realizations through %one-loop diagrams by using the diagrammatic %approach~\cite{Antusch:2008tz,Gavela:2008ra}}. \sout{We shall follow the %diagrammatic approach introduced in %Refs.}~\cite{Antusch:2008tz,Gavela:2008ra} \sout{to find all possible %one-loop realizations of the short-range $0\nu\beta\beta$ decay operators.}

%decompose the short distance operators of neutrinoless double beta decay

In this work, we consider the cases that the mediators are either scalar or
fermion fields. The topologies of the diagrams for scalar and vector are the same, nevertheless vector fields are expected to be the gauge bosons of some new symmetry and their masses are generated through the spontaneous breaking of that symmetry. That is to say the SM gauge group should be extended to accommodate the new vector bosons, consequently we only discuss scalar messengers here. Furthermore, we assume the new fermions are vector-like so that the resulting models are anomaly free.

This paper is organized as follows. In section~\ref{sec:generate}, the general procedures of decomposing a non-renormalizable operator are sketched, and we discuss how to generate the topologies, diagrams and models of the short-range $0\nu\beta\beta$ operators. In section~\ref{sec:relaton to mass}, the relation between $0\nu\beta\beta$ models and neutrino masses is discussed. In section~\ref{sec:example}, we give an example of one-loop model for short-range $0\nu\beta\beta$ decay, the leading contribution to neutrino masses arises at two-loop level, and the prediction for half-life is analyzed. Section~\ref{sec:conclusion} summarizes our main results. The complete set of $0\nu\beta\beta$ decay operators below the electroweak scale as well as the general formula for half-life are given in Appendix~\ref{sec:0nubb-eff-operators-LEFT}. We list the non-renormalizable topologies and finite non-genuine diagrams in Appendix~\ref{app:non-renorm-topology}. We present the possible one-loop models and the quantum number assignments of mediators for the $0\nu\beta\beta$ decay operator N4 with the diagram N-1-1-1 in Appendix~\ref{app:example-models}.

\section{\label{sec:generate}Generate genuine topologies, diagrams and models}

The diagrammatic decomposition approach~\cite{Antusch:2008tz,Gavela:2008ra} allows us to systematically identify the possible ultraviolet (UV) completion of any effective high dimensional operator $\mathcal{O}_{EF}$ in the SM effective theory, and it has been used to systematically classify the Majorana neutrino mass models generated at one-loop level~\cite{Bonnet:2012kz}, two-loop level~\cite{AristizabalSierra:2014wal} and three-loop level~\cite{Cepedello:2018rfh}. The classification of Dirac neutrino mass models have also been performed in this approach~\cite{Yao:2017vtm,Yao:2018ekp,CentellesChulia:2019xky,Jana:2019mgj}. The general procedure of decomposing a non-renormalizable operator $\mathcal{O}_{EF}$ at $L-$ loop level is as follows: \ding{172} Generate the $L-$loop connected topology with three-point and four-point vertices, and the number of the external lines should be exactly equal to the number of the fields contained in the considered operators. The topologies with tadpoles and self-energies are generally discarded. \ding{173} Assign the fields involved in the concerned operators to the external lines, and specify the Lorentz nature (spinor or scalar) of each internal line. Thus the topologies are promoted to diagrams. For the four-point vertex, the corresponding interaction is renormalizable if and only if all the four relevant legs are scalars, while the trilinear vertices can be the fermion-fermion-scalar and scalar-scalar-scalar interactions. We discard all the diagrams with non-renormalizable interactions. \ding{174} Construct the UV complete models by  specifying the quantum numbers of the internal fields, and each vertex should be invariant under the SM gauge symmetry.

Although the resulting UV models obtained in this way can generate the concerned effective operator $\mathcal{O}_{EF}$ at $L-$loop level, one should guarantee that it is the leading order contribution. It is called a genuine model if there is no lower order contribution to $\mathcal{O}_{EF}$ otherwise the model is non-genuine.
The topologies associated with the genuine models is called genuine topologies. If all the UV models generated from a topology are non-genuine, the corresponding topology is non-genuine. In a similar fashion, one can define genuine diagrams and non-genuine diagrams. A diagram is non-genuine if it is possible to compress one or more loops into a renormalizable vertex~\cite{Cepedello:2018rfh}, in this case, lower order contributions will be allowed. In general, any internal loop (or loops) with three external legs can be compressed into a three point vertex regardless of the field content of a model, thus the associated topology is non-genuine. It is only possible to compress a internal loop (or loops) with four external legs into a renormalizable vertex if all external lines are scalars.
%The renormalizable point interaction arising from the loop compression could %vanish for very specific choices of fields, usually repeated fields are %necessary. The relevant topologies (diagrams) are called special genuine.
In the present work, we shall apply the above procedures to study the one-loop decomposition of the neutrinoless double decay operators in Eq.~\eqref{eqn-basis-summary-dim9}.

\subsection{\label{sec:Gtopologies} Generate topologies}

\begin{figure}[htbp]
\centering
\includegraphics[width=\textwidth]{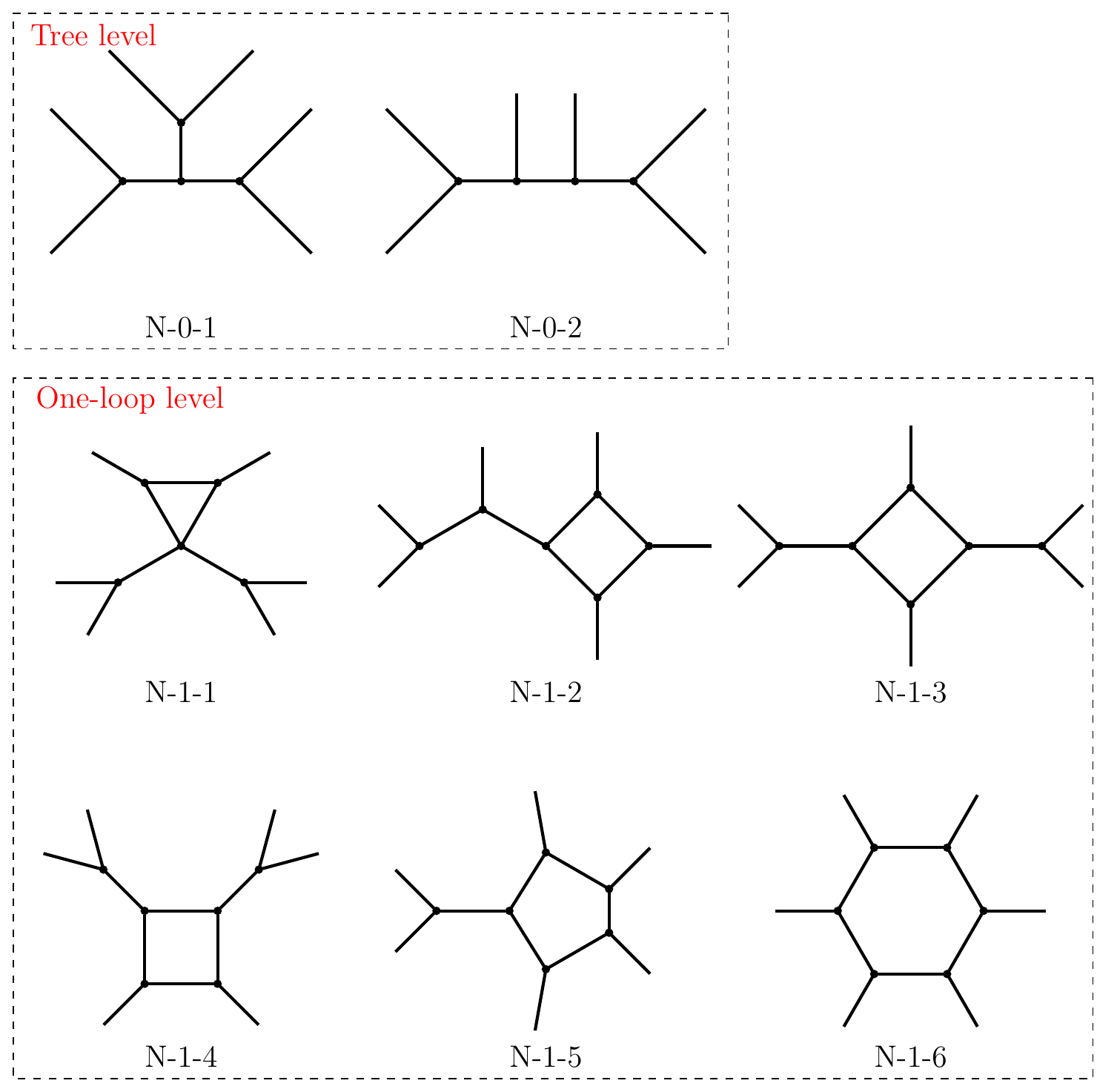}
%\caption{\label{fig:N0-1}List of topologies associated to genuine diagrams %up to one loops.}
\caption{\label{fig:N0-1}Genuine topologies which can lead to tree-level and one-loop genuine models for the short-range part of the $0\nu\beta\beta$ decay amplitude.}
\end{figure}

All the dimension 9 effective $0\nu\beta\beta$ decay operators involve six quark and lepton fields. We first use our own code to generate the connected one-loop topologies with six external legs. All the topologies with tadpoles and self-energy are excluded, because their loop integrals are divergent and thus lower order counter term is required in a consistent renormalization scheme. After excluding the topologies which contain an internal loop with three legs, we find there are totally 43 topologies shown in figure~\ref{fig:N0-1} and figure~\ref{fig:NNormalize}. Only the six one-loop topologies in figure~\ref{fig:N0-1} can lead to genuine diagrams (models). The topologies in figure~\ref{fig:NNormalize} need non-renormalizable interactions to accommodate six external fermion lines. For the sake of completeness, we show the non-genuine topologies which contain an internal loop with three legs in figure~\ref{fig:NGenuineT}. We see that all these non-genuine topologies can be regarded as extensions of the tree-level $0\nu\beta\beta$ decay models, where one of the vertices is generated at one loop. As regards the tree-level decomposition of the $0\nu\beta\beta$ decay operators, there are only two renormalizable topologies~\cite{Bonnet:2012kh} which are displayed in figure~\ref{fig:N0-1}. Generally three mediators are needed in the tree-level UV complete models and their quantum numbers are unambiguously fixed. We also provide the tree-level decomposition in the present work, since these results are essential when determining the genuineness of a one-loop $0\nu\beta\beta$ decay model.

\subsection{\label{sec:Gdiagram} Generate diagrams}
\begin{figure}[htbp]
\centering
\includegraphics[width=\textwidth]{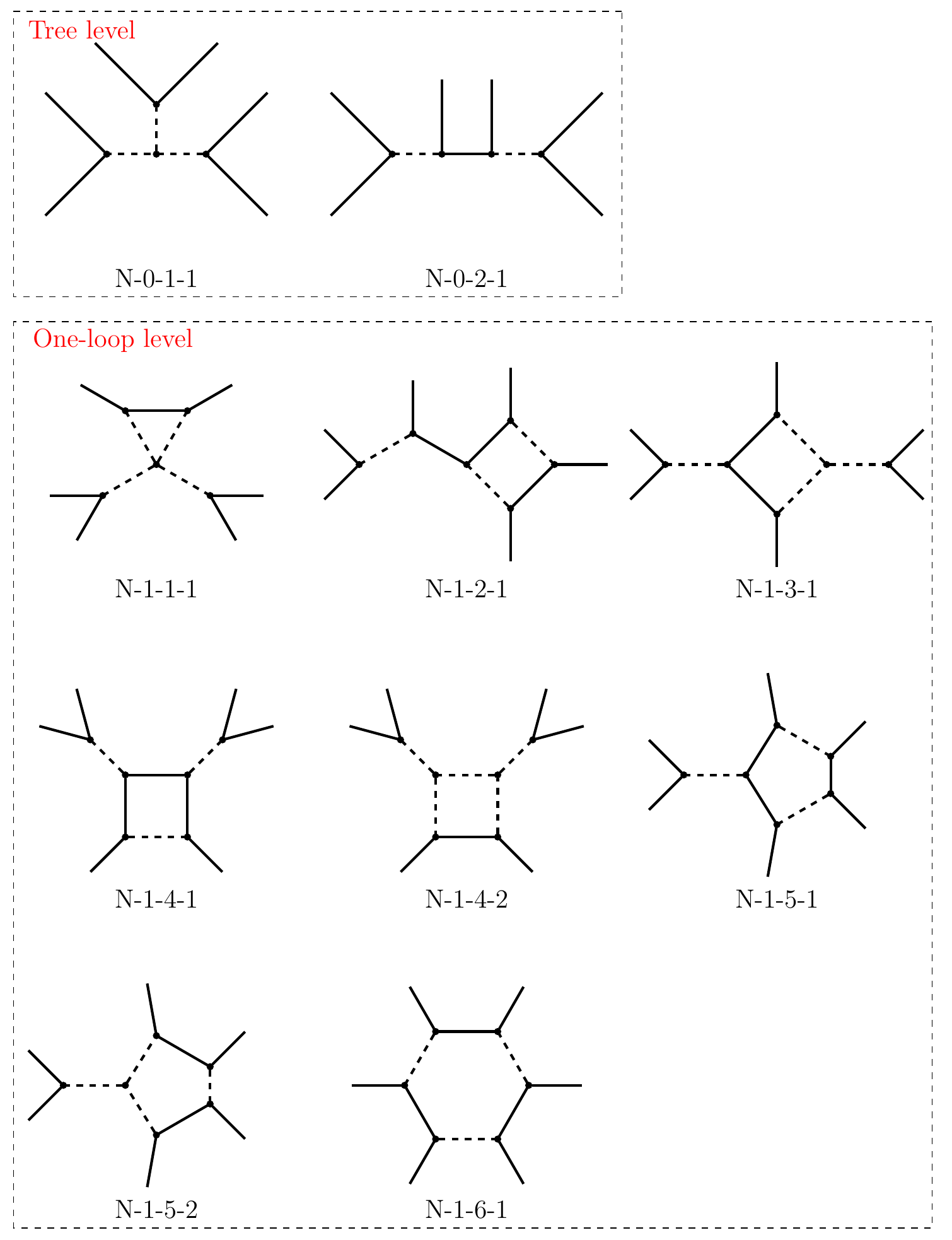}
\caption{\label{fig:DiaN0-1}List of genuine diagrams arising from the topologies in figure~\ref{fig:N0-1}, where the solid and dashed lines stand for fermion and scalar fields respectively.}
\end{figure}

All the six external legs for $0\nu\beta\beta$ decay are quark and lepton fields. We proceed to insert fermion or scalar into each internal line, and
Lorentz invariance requires that each vertex must contain even number of fermion. The UV complete models are assumed to be renormalizable, thus the interaction vertices should be dimension-four three-point and four-point vertices, and consequently the interaction terms in the Lagrangian are of the form $\overline{F}_1F_2S_1$, $S_1S_2S_3$ or $S_1S_2S_3S_4$, where $F_i$ and $S_i$ denote a generic fermion and scalar fields respectively. A topology is non-renormalizable if the diagrams for all possible fermionic and scalar insertions contain non-renormalizable vertex. We show the non-renormalizable topologies in figure~\ref{fig:NNormalize} for completeness. Some diagrams arising from a topology are possibly superfluous. One needs to consider all possible permutations of fermion and scalar lines and compare the couplings of the vertices. If the couplings of two diagrams match with each other, they are exactly the same diagrams. For the genuine topologies in figure~\ref{fig:N0-1}, we show the corresponding independent diagrams in figure~\ref{fig:DiaN0-1}. Both topologies N-1-4 and N-1-5 give rise to two diagrams, while there is only one diagram for each of the remaining four topologies N-1-1, N-1-2, N-1-3 and N-1-6. Because all the external legs of the short-range $0\nu\beta\beta$ decay are fermions, the four-point interaction $S_1S_2S_3S_4$ is only involved in the diagram N-1-1-1 and the interaction vertices are of the forms $\overline{F}_1F_2S_1$ and $S_1S_2S_3$ in other cases, as can be seen from figure~\ref{fig:DiaN0-1}. In particular, it is impossible to compress the loop into a renormalizable vertex, hence we expect all these diagrams in figure~\ref{fig:DiaN0-1} could give genuine models. We also show the tree-level diagrams which lead to $0\nu\beta\beta$ decay in figure~\ref{fig:DiaN0-1}, since they are essential when determining whether a loop model is genuine. There are only two diagrams at tree level, and the Lorentz natures of all mediators are completely fixed.

\subsection{\label{sec:Gmodel}Generate models}

The next step is to generate models for the diagrams displayed in figure~\ref{fig:DiaN0-1}, and we need to specify the $SU(3)_{C}$, $SU(2)_{L}$ and $U(1)_{Y}$ quantum numbers to each line in these diagrams. Firstly we attach the fields in the $0\nu\beta\beta$ decay operators to external lines of the diagram, subsequently gauge invariance allows us to determine the quantum numbers of each internal line. Notice that there are only finite  possibilities for the quantum numbers of the internal fermions and scalars in the tree-level models. However, any one-loop diagram in figure~\ref{fig:DiaN0-1} leads to an infinite number of genuine models. For illustration, we shall impose an upper limit on the dimensions of the $SU(3)_{C}$ and $SU(2)_{L}$ representations assigned to the mediator fields. The details of how to generate models are given in the following subsections.

\subsubsection{Attach external fields}
\begin{figure}[htbp]
\centering
\includegraphics[width=0.9\textwidth]{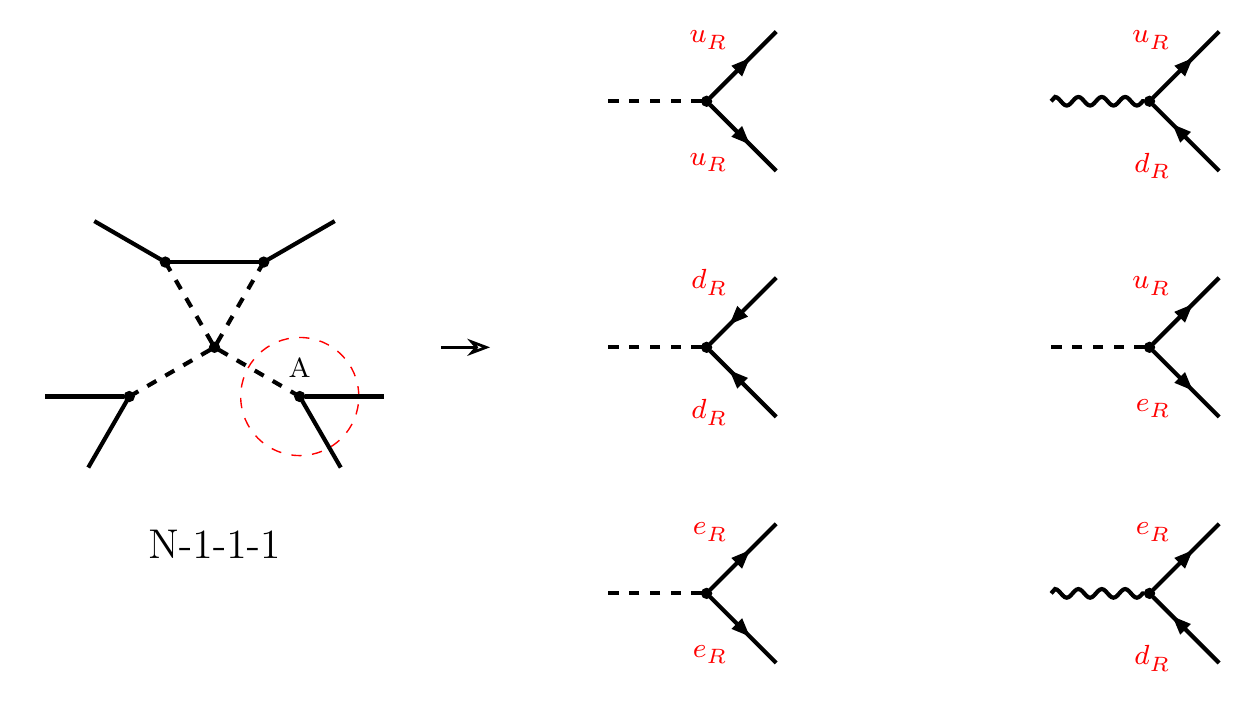}
\caption{\label{fig:external} The possible ways of attaching the fields of the operator N4 to the external legs of the vertex A in diagram N-1-1-1, where the dashed lines and wavy lines on the right side denote scalar and vector fields respectively. Notice that Lorentz invariance fixes the mediator to be scalar or vector because the external legs are fermions with definite chirality.  }
\end{figure}

The eleven $0\nu\beta\beta$ decay operators in Eq.~\eqref{eqn-basis-summary-dim9} can be divided into six categories as listed in table~\ref{tab:N1toN6}, and the operators of each category involve the same fields as well as the same UV completion. The vector fields should be gauge bosons of certain gauge symmetry and their masses are generated by the spontaneous breaking of symmetry. Thus it would be necessary to extend the SM gauge group if vectors are mediators. Therefore we focus on the scalar and fermion mediated cases in the present work, the cases of vectors as mediators would be not considered. The SM lepton and quark fields are chiral fields in weak basis, Lorentz invariance requires vector mediators for certain attachment of external fields, and these cases would be dropped. As an example, we shall consider the one-loop UV completion of the $0\nu\beta\beta$ decay operator N4 based on the topology N-1-1. Let us consider the diagram in figure~\ref{fig:external}, there are six possible ways to attach the fields of the operator N4 to the external lines of the vertex A. If the two external legs are assigned to be $u_R$, $d_R$ or $e_R$, $d_R$, the non-vanishing fermion bilinears are $\bar{u}_R\gamma^{\mu}d_R$ and $\bar{e}_R\gamma^{\mu}d_R$ respectively, consequently the internal leg has to be a vector field $X_\mu$ to form a Lorentz invariant interaction $\bar{u}_R\gamma^{\mu}d_RX_{\mu}$ and $\bar{e}_R\gamma^{\mu}d_R X_{\mu}$. As a result, these two cases will be discarded. Since we can freely attach the fields to the external lines, therefore there are generally multiple external line structures for a given diagram and some of them are superfluous. The redundant assignments of external lines must base on the same diagram. In order to find out the  redundant ones, we need to consider the possible permutations of vertices and compare the interlinkage between vertices. If several kinds of attachments have the same structure of interlinkage, they will be identified as one.

%\vskip0.3in

\begin{table}[t!]
\renewcommand{\tabcolsep}{0.5mm}
\renewcommand{\arraystretch}{1.3}
\centering
\begin{tabular}{|c|c|c|}\hline\hline
 \texttt{Notation}      ~&~ $0\nu\beta\beta$ \texttt{decay operators}
 %\cite{Graesser:2016bpz}
 ~&~    \texttt{External fields}\\ \hline
N1      &       $\mathcal{O}^{0\nu}_1,~\mathcal{O}^{0\nu}_2$  & $\overline{Q}, Q, \bar{u}_{R}, d_{R}, \overline{\ell}, \ell^{c}$\\ \hline
N2      &       $\mathcal{O}^{0\nu}_3,~\mathcal{O}^{0\nu}_4$   & $Q, Q, \bar{u}_{R}, \bar{u}_{R}, \overline{\ell}, \ell^{c}$\\ \hline
N3      & $\mathcal{O}^{0\nu}_5,~\mathcal{O}^{0\nu}_6$  & $\overline{Q}, \overline{Q}, d_{R}, d_{R}, \overline{\ell}, \ell^{c}$\\ \hline
N4 & $\mathcal{O}^{0\nu}_7$  & $\bar{u}_{R}, \bar{u}_{R}, d_{R}, d_{R}, \bar{e}_{R}, e_{R}^{c} $\\ \hline
N5 & $\mathcal{O}^{0\nu}_8,~\mathcal{O}^{0\nu}_9$  & $\overline{u}_{R}, \overline{Q}, d_{R}, d_{R}, \overline{\ell}, e_{R}^{c}$\\\hline
N6 & $\mathcal{O}^{0\nu}_{10},~\mathcal{O}^{0\nu}_{11}$  & $\bar{u}_{R}, \bar{u}_{R}, Q, d_{R}, \overline{\ell}, e_{R}^{c}$\\ \hline\hline
\end{tabular}
\caption{Summary of fields involved in the effective $0\nu\beta\beta$ decay operators of dimension 9.}
\label{tab:N1toN6}
\end{table}

\begin{figure}[htbp]
\centering
\includegraphics[width=0.9\textwidth]{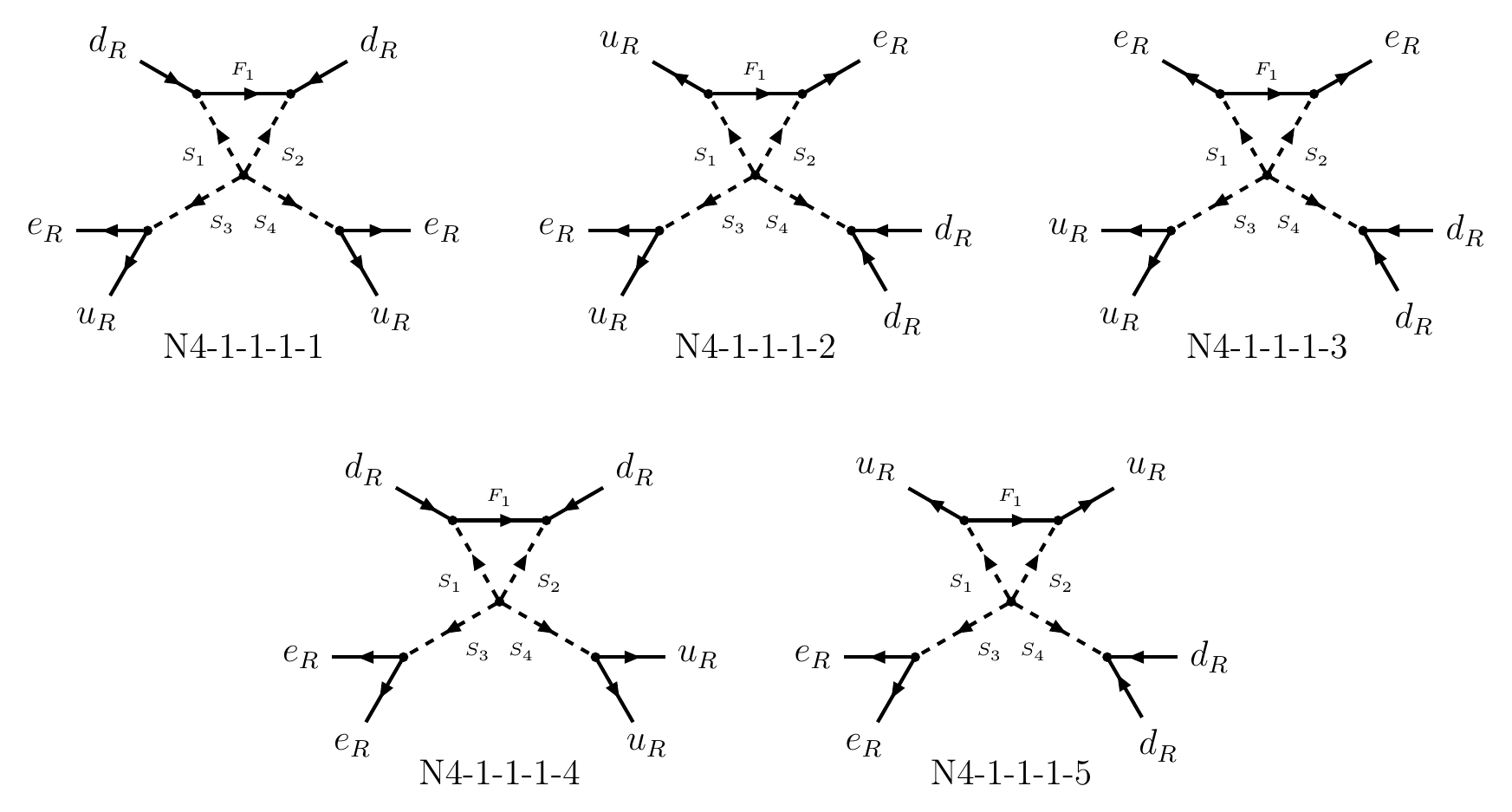}
\caption{\label{fig:external-N4} The possible ways to attach the fields of the operator N4 to the diagram N-1-1-1.   }
\end{figure}
%

%\subsubsection{Assign $U(1)_{Y}$ charge}
\subsubsection{Determine $U(1)_{Y}$ charge}

Once the external legs are specified,
the hypercharges fulfilling of the internal lines can be determined by imposing the hypercharge conservation vertex by vertex. Note that the hypercharge of a field is related to its electric charge via the Gell-Mann-Nishijima formula $Q=T_3+Y$, where $T_3$ is the third component of the weak isospin. For example, the equations of hypercharge conservation for the diagram N4-1-1-1-1 read as
\begin{eqnarray}
\nonumber&&\qquad\qquad Y_{F_1}-Y_{d_R}-Y_{S_1}=0,~~~~ Y_{F_1}+Y_{S_2}+Y_{d_R}=0,\\
&&Y_{S_1}+Y_{S_2}+Y_{S_3}+Y_{S_4}=0,~~~Y_{e_R}+Y_{u_R}-Y_{S_1}=0,~~~ Y_{e_R}+Y_{u_R}-Y_{S_4}=0\,,
\end{eqnarray}
with $Y_{e_R}=-1$, $Y_{u_R}=\frac{2}{3}$ and $Y_{d_R}=-\frac{1}{3}$. The solution to the above equations is given by
\begin{eqnarray}
Y_{F_1}=-\frac{1}{3}+\alpha,~~~Y_{S_1}=\alpha,~~Y_{S_2}=-\frac{2}{3}-\alpha,~~~Y_{S_3}=Y_{S_4}=-\frac{1}{3}\,,
\end{eqnarray}
where $\alpha$ is arbitrary real parameter. A definite value of $\alpha$ should be taken in a concrete model. In a similar way, we can determine the hypercharges for the other diagrams in figure~\ref{fig:external-N4} as follows,
\begin{eqnarray}
\nonumber&&\text{N4-1-1-1-2}~:~ Y_{F_1}=-\frac{2}{3}+\alpha,~~Y_{S_1}=\alpha,~~Y_{S_2}=-\frac{1}{3}-\alpha,~~Y_{S_3}=-\frac{1}{3},~~Y_{S_4}=\frac{2}{3}\,,\\
\nonumber&&\text{N4-1-1-1-3}~:~ Y_{F_1}=1+\alpha,~~Y_{S_1}=\alpha,~~Y_{S_2}=-2-\alpha,~~Y_{S_3}=\frac{4}{3},~~Y_{S_4}=\frac{2}{3}\,,\\
\nonumber&&\text{N4-1-1-1-4}~:~ Y_{F_1}=-\frac{1}{3}+\alpha,~~Y_{S_1}=\alpha,~~Y_{S_2}=\frac{2}{3}-\alpha,~~Y_{S_3}=-2,~~Y_{S_4}=\frac{4}{3}\,,\\
&&\text{N4-1-1-1-5}~:~Y_{F_1}=-\frac{2}{3}+\alpha,~~Y_{S_1}=\alpha,~~Y_{S_2}=\frac{4}{3}-\alpha,~~Y_{S_3}=-2,~~Y_{S_4}=\frac{2}{3}\,.
\end{eqnarray}
We see that a hypercharge flow encoded in the parameter $\alpha$ in the loop is allowed.

\vskip0.4in

\subsubsection{Assign $SU(2)_{L}$ transformation}

\begin{figure}[htbp]
\centering
\includegraphics[width=0.9\textwidth]{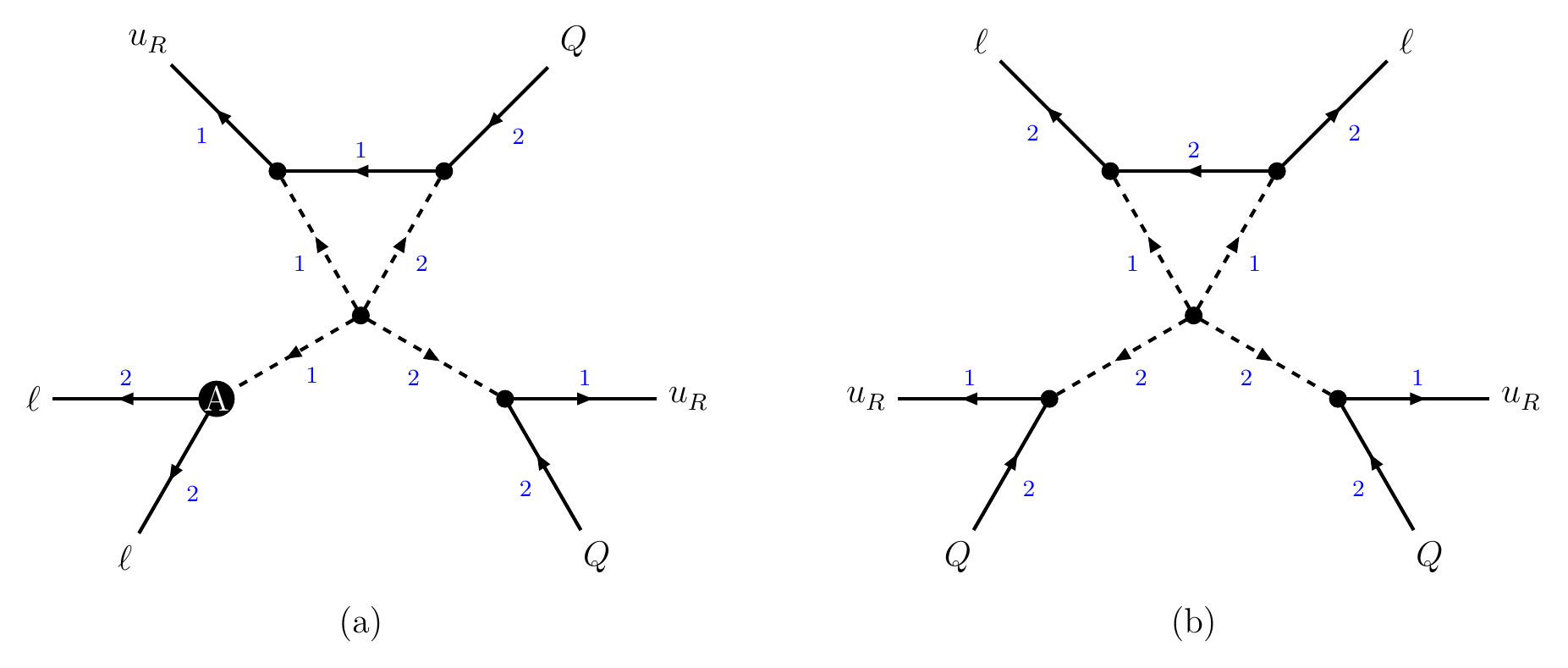}
\caption{\label{fig:SU2} Both models shown in panels (a) and (b) fulfill the $SU(2)_{L}$ invariance, where the numbers 1 and 2 denote the $SU(2)$ transformations of the fields. For the diagram (a), there is a vanishing point interaction $\bar{\ell}\ell^cS$ at the vertex $A$. The model in panel (b) doesn't contribute to the $0\nu\beta\beta$ decay.}
\end{figure}

We proceed to assign the $SU(2)_{L}$ transformations to the internal lines, and they are denoted by the dimensions of the $SU(2)_{L}$ representations. For the concerned renormalizable interactions, the invariance under $SU(2)_{L}$ gauge symmetry requires the following conditions should be fulfilled,
\begin{eqnarray}
\nonumber\overline{F}_1F_2S_1~&:&~ n_{F_1}\otimes n_{F_2}\otimes n_{S_1}\supset1\,, \\
\nonumber S_1S_2S_3~&:&~ n_{S_1}\otimes n_{S_2}\otimes n_{S_3}\supset1\,,\\
\label{eq:SU2-invariance}S_1S_2S_3S_4~&:&~ n_{S_1}\otimes n_{S_2}\otimes n_{S_3}\otimes n_{S_4}\supset1\,,
\end{eqnarray}
where $n_{F_i}$ and $n_{S_i}$ denote the $SU(2)_L$ representations to which the fields $F_i$ and $S_i$ are assigned. Notice that each $SU(2)$ representation and its complex conjugate are equivalent. Moreover, it is necessary to check that some $0\nu\beta\beta$ decay operators can really be generated after integrating out the mediators. For example, let us consider the models in figure~\ref{fig:SU2}, the external legs are $Q, Q, \bar{u}_{R}, \bar{u}_{R}, \overline{\ell}, \ell^{c}$ which coincide with those of the operators $\mathcal{O}^{0\nu}_3$ and $\mathcal{O}^{0\nu}_4$. In the left panel, the renormalizable interaction $\bar{\ell}\ell^cS$ at the vertex $A$ is vanishing because the $SU(2)$ singlet contraction of two doublets is antisymmetric. The model in the right panel of figure~\ref{fig:SU2} leads to the effective operator $\bar{e}_{L}\nu^c_{L}\bar{u}_{R}u_{L}\bar{u}_Rd_{L}$ which can not induce the $0\nu\beta\beta$ decay process, although its interaction vertices are invariant under SM gauge symmetry.
As a consequence, we need to explicitly expand the $SU(2)_L$ index of the interaction at each vertex, contract the internal lines and see whether the $0\nu\beta\beta$ decay operator can be produced. For illustration, we shall consider the $SU(2)_L$ singlet, doublet and triplet assignments for the messenger fields in the present work, and higher dimensional representations can be discussed in the same fashion.

\vskip0.2in

The SM quantum number assignments for the internal fields are independent of their Lorentz property, therefore this step can be performed at the topology level. In table~\ref{tab:N11SU2}, we list the possible $SU(2)_L$ assignments of fields for the topology N-1-1. For the set of operators N1, N2 and N3, four of the external fields are $SU(2)_L$ doublets and the other two are  singlets, consequently there are four different assignments of external legs, as shown in table~\ref{tab:N11SU2}. The external lines of the operator N4 are all singlets under $SU(2)_L$. The fields of the operators N5 and N6 include  two doublets and four singlets of $SU(2)$, and there are four different assignments of external fields as well. The possible $SU(2)_L$ quantum numbers of the internal fields can be fixed by the invariance condition in Eq.~\eqref{eq:SU2-invariance}, and it is necessary to check that the $0\nu\beta\beta$ decay operators can really be produced, as explained above.

\begin{figure}[htbp]
\centering
\includegraphics[width=0.5\textwidth]{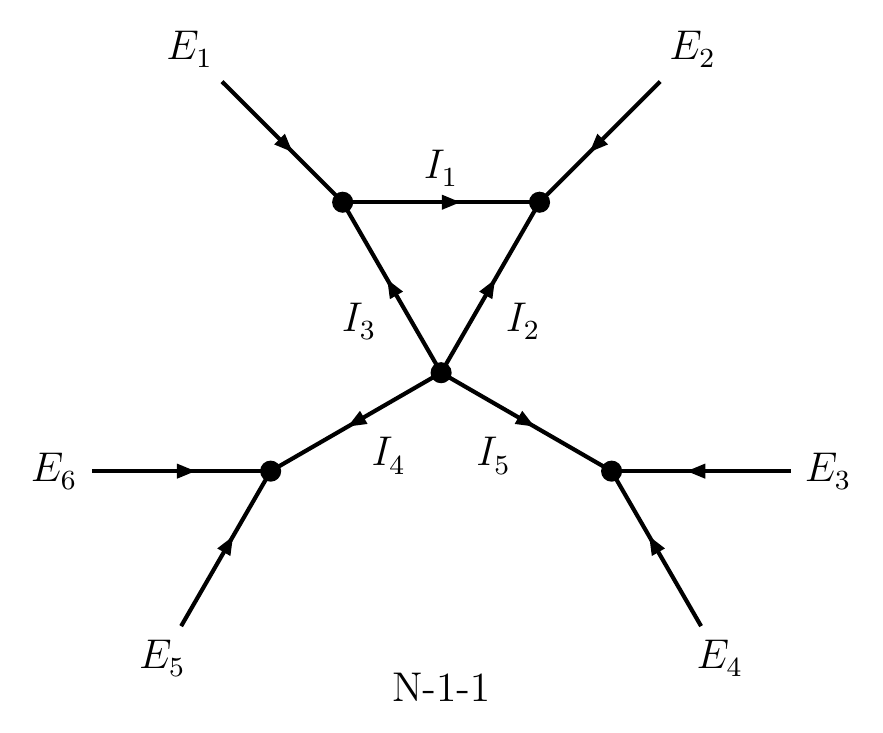}
\caption{\label{fig:N11SU2SU3}Field assignments and color flow for the topology N-1-1, where $E_i$ and $I_i$ denote the external and internal fields respectively.
%line label and arrows.
}
\end{figure}
\begin{table}[htbp]
        \renewcommand{\tabcolsep}{0.5mm}
        \renewcommand{\arraystretch}{1.3}
        \centering
        \begin{tabular}{|c|c|c|c|c|c|c|c|c|c|c|c||c|c|c|c|c|c|c|c|c|c|c|c|}
                \hline\hline
                \texttt{Operator} &
                $E_{1}$ & $E_{2}$ & $E_{3}$ & $E_{4}$ & $E_{5}$ & $E_{6}$ &
                $I_{1}$ & $I_{2}$ & $I_{3}$ & $I_{4}$ & $I_{5}$ &
                \texttt{Operator} &
                $E_{1}$ & $E_{2}$ & $E_{3}$ & $E_{4}$ & $E_{5}$ & $E_{6}$ &
                $I_{1}$ & $I_{2}$ & $I_{3}$ & $I_{4}$ & $I_{5}$\\ \hline
                \multirow{16}{*}{N1,N2,N3} & \multirow{8}{*}{$1$} & \multirow{8}{*}{$2$} & \multirow{8}{*}{$2$} & \multirow{8}{*}{$2$} & \multirow{8}{*}{$1$} & \multirow{8}{*}{$2$} &
                $2$ & $1$ & $2$ & $2$ & $1$ &
                \multirow{13}{*}{N1,N2,N3} & \multirow{8}{*}{$2$} & \multirow{8}{*}{$2$} & \multirow{8}{*}{$1$} & \multirow{8}{*}{$1$} & \multirow{8}{*}{$2$} & \multirow{8}{*}{$2$} &
                $2$ & $1$ & $1$ & $1$ & $1$  \\ \cline{8-12} \cline{20-24}
            & & & & & & &
                $2$ & $1$ & $2$ & $2$ & $3$ &
                & & & & & & &
                $2$ & $1$ & $3$ & $3$ & $1$ \\ \cline{8-12} \cline{20-24}
                & & & & & & &
                $1$ & $2$ & $1$ & $2$ & $1$  &
                & & & & & & &
                $1$ & $2$ & $2$ & $1$ & $1$  \\ \cline{8-12} \cline{20-24}
                & & & & & & &
                $1$ & $2$ & $1$ & $2$ & $3$ &
                & & & & & & &
                $1$ & $2$ & $2$ & $3$ & $1$  \\ \cline{8-12} \cline{20-24}
                & & & & & & &
                $3$ & $2$ & $3$ & $2$ & $1$ &
                & & & & & & &
                $3$ & $2$ & $2$ & $1$ & $1$   \\ \cline{8-12} \cline{20-24}
                & & & & & & &
                $3$ & $2$ & $3$ & $2$ & $3$ &
                & & & & & & &
                $3$ & $2$ & $2$ & $3$ & $1$  \\ \cline{8-12} \cline{20-24}
                & & & & & & &
                $2$ & $3$ & $2$ & $2$ & $1$ &
                & & & & & & &
                $2$ & $3$ & $3$ & $1$ & $1$   \\ \cline{8-12} \cline{20-24}
                & & & & & & &
                $2$ & $3$ & $2$ & $2$ & $3$ &
                & & & & & & &
                $2$ & $3$ & $3$ & $3$ & $1$ \\ \cline{2-12} \cline{14-24}
                & \multirow{8}{*}{$1$} & \multirow{8}{*}{$1$} & \multirow{8}{*}{$2$} & \multirow{8}{*}{$2$} & \multirow{8}{*}{$2$} & \multirow{8}{*}{$2$} &
                $1$ & $1$ & $1$ & $1$ & $1$ &
                & \multirow{5}{*}{$2$} & \multirow{5}{*}{$2$} & \multirow{5}{*}{$1$} & \multirow{5}{*}{$2$} & \multirow{5}{*}{$2$} & \multirow{5}{*}{$1$} &
                $2$ & $1$ & $1$ & $2$ & $2$ \\ \cline{8-12} \cline{20-24}
                & & & & & & &
                $1$ & $1$ & $1$ & $3$ & $3$ &
                & & & & & & &
                $2$ & $1$ & $3$ & $2$ & $2$\\ \cline{8-12} \cline{20-24}
                & & & & & & &
                $2$ & $2$ & $2$ & $1$ & $1$ &
                & & & & & & &
                $1$ & $2$ & $2$ & $2$ & $2$\\ \cline{8-12} \cline{20-24}
                & & & & & & &
                $2$ & $2$ & $2$ & $3$ & $1$ &
                & & & & & & &
                $3$ & $2$ & $2$ & $2$ & $2$ \\ \cline{8-12} \cline{20-24}
                & & & & & & &
                $2$ & $2$ & $2$ & $3$ & $3$ &
                & & & & & & &
                $2$ & $3$ & $3$ & $2$ & $2$ \\ \cline{8-12} \cline{12-24}
                & & & & & & &
                $3$ & $3$ & $3$ & $1$ & $1$ &
                \multirow{3}{*}{N4} & \multirow{3}{*}{$1$} & \multirow{3}{*}{$1$} & \multirow{3}{*}{$1$} & \multirow{3}{*}{$1$} & \multirow{3}{*}{$1$} & \multirow{3}{*}{$1$} &
                $1$ & $1$ & $1$ & $1$ & $1$ \\ \cline{8-12} \cline{20-24}
                & & & & & & &
                $3$ & $3$ & $3$ & $3$ & $1$ &
                & & & & & & &
                $2$ & $2$ & $2$ & $1$ & $1$ \\ \cline{8-12} \cline{20-24}
                & & & & & & &
                $3$ & $3$ & $3$ & $3$ & $3$ &
                & & & & & & &
                $3$ & $3$ & $3$ & $1$ & $1$ \\ \hline
                \multirow{8}{*}{N5,N6} & \multirow{5}{*}{$1$} & \multirow{5}{*}{$1$} & \multirow{5}{*}{$1$} & \multirow{5}{*}{$1$} & \multirow{5}{*}{$2$} & \multirow{5}{*}{$2$} &
                $1$ & $1$ & $1$ & $1$ & $1$ &
                \multirow{8}{*}{N5,N6} & \multirow{4}{*}{$2$} & \multirow{4}{*}{$1$} & \multirow{4}{*}{$1$} & \multirow{4}{*}{$1$} & \multirow{4}{*}{$2$} & \multirow{4}{*}{$1$} &
                $2$ & $2$ & $1$ & $2$ & $1$ \\ \cline{8-12} \cline{20-24}
                & & & & & & &
                $2$ & $2$ & $2$ & $1$ & $1$ &
                & & & & & & &
                $1$ & $1$ & $2$ & $2$ & $1$\\ \cline{8-12} \cline{20-24}
                & & & & & & &
                $2$ & $2$ & $2$ & $3$ & $1$ &
                & & & & & & &
                $3$ & $3$ & $2$ & $2$ & $1$\\ \cline{8-12} \cline{20-24}
                & & & & & & &
                $3$ & $3$ & $3$ & $1$ & $1$ &
                & & & & & & &
                $2$ & $2$ & $3$ & $2$ & $1$\\ \cline{8-12} \cline{14-24}
                & & & & & & &
                $3$ & $3$ & $3$ & $3$ & $1$ &
                & \multirow{4}{*}{$2$} & \multirow{4}{*}{$2$} & \multirow{4}{*}{$1$} & \multirow{4}{*}{$1$} & \multirow{4}{*}{$1$} & \multirow{4}{*}{$1$} &
                $2$ & $1$ & $1$ & $1$ & $1$ \\ \cline{2-12} \cline{20-24}
                & \multirow{3}{*}{$1$} & \multirow{3}{*}{$1$} & \multirow{3}{*}{$1$} & \multirow{3}{*}{$2$} & \multirow{3}{*}{$2$} & \multirow{3}{*}{$1$} &
                $1$ & $1$ & $1$ & $2$ & $2$ &
                & & & & & & &
                $1$ & $2$ & $2$ & $1$ & $1$\\ \cline{8-12} \cline{20-24}
                & & & & & & &
                $2$ & $2$ & $2$ & $2$ & $2$ &
                & & & & & & &
                $3$ & $2$ & $2$ & $1$ & $1$\\ \cline{8-12} \cline{20-24}
                & & & & & & &
                $3$ & $3$ & $3$ & $2$ & $2$ &
                & & & & & & &
                $2$ & $3$ & $3$ & $1$ & $1$\\ \hline \hline
        \end{tabular}
\caption{\label{tab:N11SU2}Assign the $SU(2)_{L}$ quantum numbers to fields of the topology N-1-1 in figure~\ref{fig:N0-1}. The labels $E_i$ and $I_i$ of the external and internal lines are showed in figure~\ref{fig:N11SU2SU3}.}
\end{table}

\subsubsection{Assign $SU(3)_{C}$ transformation}
\begin{table}[hbpt]
        \renewcommand{\tabcolsep}{0.5mm}
        \renewcommand{\arraystretch}{1.2}
        \small
        \centering
        \begin{tabular}{|c|c|c|c|c|c|c|c|c|c|c||c|c|c|c|c|c|c|c|c|c|c|}
                \hline
                $E_{1}$ & $E_{2}$ & $E_{3}$ & $E_{4}$ & $E_{5}$ & $E_{6}$ &
                $I_{1}$ & $I_{2}$ & $I_{3}$ & $I_{4}$ & $I_{5}$ &
                $E_{1}$ & $E_{2}$ & $E_{3}$ & $E_{4}$ & $E_{5}$ & $E_{6}$ &
                $I_{1}$ & $I_{2}$ & $I_{3}$ & $I_{4}$ & $I_{5}$\\ \hline
                \multirow{5}{*}{$1$} & \multirow{5}{*}{$\bar{3}$} & \multirow{5}{*}{$3$} & \multirow{5}{*}{$\bar{3}$} & \multirow{5}{*}{$1$} & \multirow{5}{*}{$3$} &
                $1$ & $3$ & $1$ & $\bar{3}$ & $1$ &
                \multirow{5}{*}{$1$} & \multirow{5}{*}{$3$} & \multirow{5}{*}{$\bar{3}$} & \multirow{5}{*}{$3$} & \multirow{5}{*}{$\bar{3}$} & \multirow{5}{*}{$1$} &
                $1$ & $\bar{3}$ & $1$ & $3$ & $1$ \\ \cline{7-11} \cline{18-22}
                & & & & & &
                $3$ & $1$ & $3$ & $\bar{3}$ & $1$ &
                & & & & & &
                $3$ & $3$ & $3$ & $3$ & $1$\\ \cline{7-11} \cline{18-22}
                & & & & & &
                $\bar{3}$ & $\bar{3}$ & $\bar{3}$ & $\bar{3}$ & $1$  &
                & & & & & &
                $3$ & $\bar{6}$ & $3$ & $3$ & $1$ \\ \cline{7-11} \cline{18-22}
                & & & & & &
                $\bar{3}$ & $6$ & $\bar{3}$ & $\bar{3}$ & $1$ &
                & & & & & &
                $\bar{3}$ & $1$ & $\bar{3}$ & $3$ & $1$ \\ \cline{7-11} \cline{18-22}
                & & & & & &
                $6$ & $\bar{3}$ & $6$ & $\bar{3}$ & $1$ &
                & & & & & &
                $\bar{6}$ & $3$ & $\bar{6}$ & $3$ & $1$ \\ \cline{1-11} \cline{12-22}
                \multirow{3}{*}{$\bar{3}$} & \multirow{3}{*}{$\bar{3}$} & \multirow{3}{*}{$3$} & \multirow{3}{*}{$1$} & \multirow{3}{*}{$1$} & \multirow{3}{*}{$3$} &
                $\bar{3}$ & $\bar{3}$ & $1$ & $\bar{3}$ & $\bar{3}$ &
                \multirow{3}{*}{$1$} & \multirow{3}{*}{$1$} & \multirow{3}{*}{$3$} & \multirow{3}{*}{$\bar{3}$} & \multirow{3}{*}{$\bar{3}$} & \multirow{3}{*}{$3$} &
                $1$ & $1$ & $1$ & $1$ & $1$ \\ \cline{7-11} \cline{18-22}
                & & & & & &
                $\bar{3}$ & $6$ & $1$ & $\bar{3}$ & $\bar{3}$ &
                & & & & & &
                $3$ & $\bar{3}$ & $3$ & $1$ & $1$ \\ \cline{7-11} \cline{18-22}
                & & & & & &
                $1$ & $3$ & $3$ & $\bar{3}$ & $\bar{3}$ &
                & & & & & &
                $6$ & $\bar{6}$ & $6$ & $1$ & $1$ \\ \hline
                \multirow{10}{*}{$1$} & \multirow{10}{*}{$3$} & \multirow{10}{*}{$1$} & \multirow{10}{*}{$3$} & \multirow{10}{*}{$\bar{3}$} & \multirow{10}{*}{$\bar{3}$} &
                $1$ & $\bar{3}$ & $1$ & $\bar{3}$ & $\bar{3}$ &
                \multirow{10}{*}{$1$} & \multirow{10}{*}{$\bar{3}$} & \multirow{10}{*}{$3$} & \multirow{10}{*}{$3$} & \multirow{10}{*}{$\bar{3}$} & \multirow{10}{*}{$1$} &
                $1$ & $3$ & $1$ & $3$ & $3$ \\ \cline{7-11} \cline{18-22}
                & & & & & &
                $1$ & $\bar{3}$ & $1$ & $6$ & $\bar{3}$ &
                & & & & & &
                $1$ & $3$ & $1$ & $3$ & $\bar{6}$\\ \cline{7-11} \cline{18-22}
                & & & & & &
                $3$ & $3$ & $3$ & $\bar{3}$ & $\bar{3}$ &
                & & & & & &
                $3$ & $1$ & $3$ & $3$ & $3$\\ \cline{7-11} \cline{18-22}
                & & & & & &
                $3$ & $3$ & $3$ & $6$ & $\bar{3}$ &
                & & & & & &
                $3$ & $1$ & $3$ & $3$ & $\bar{6}$\\ \cline{7-11} \cline{18-22}
                & & & & & &
                $3$ & $\bar{6}$ & $3$ & $\bar{3}$ & $\bar{3}$ &
                & & & & & &
                $\bar{3}$ & $\bar{3}$ & $\bar{3}$ & $3$ & $3$\\ \cline{7-11} \cline{18-22}
                & & & & & &
                $3$ & $\bar{6}$ & $3$ & $6$ & $\bar{3}$ &
                & & & & & &
                $\bar{3}$ & $\bar{3}$ & $\bar{3}$ & $3$ & $\bar{6}$ \\ \cline{7-11} \cline{18-22}
                & & & & & &
                $\bar{3}$ & $1$ & $\bar{3}$ & $\bar{3}$ & $\bar{3}$ &
                & & & & & &
                $\bar{3}$ & $6$ & $\bar{3}$ & $3$ & $3$\\ \cline{7-11} \cline{18-22}
                & & & & & &
                $\bar{3}$ & $1$ & $\bar{3}$ & $6$ & $\bar{3}$ &
                & & & & & &
                $\bar{3}$ & $6$ & $\bar{3}$ & $3$ & $\bar{6}$\\
                \cline{7-11} \cline{18-22}
                & & & & & &
                $\bar{6}$ & $3$ & $\bar{6}$ & $\bar{3}$ & $\bar{3}$ &
                & & & & & &
                $6$ & $\bar{3}$ & $6$ & $3$ & $3$ \\ \cline{7-11} \cline{18-22}
                & & & & & &
                $\bar{6}$ & $3$ & $\bar{6}$ & $6$ & $\bar{3}$ &
                & & & & & &
                $6$ & $\bar{3}$ & $6$ & $3$ & $\bar{6}$\\ \hline
                \multirow{7}{*}{$\bar{3}$} & \multirow{7}{*}{$3$} & \multirow{7}{*}{$1$} & \multirow{7}{*}{$3$} & \multirow{7}{*}{$\bar{3}$} & \multirow{7}{*}{$1$} &
                $\bar{3}$ & $1$ & $1$ & $3$ & $\bar{3}$ &
                \multirow{3}{*}{$3$} & \multirow{3}{*}{$3$} & \multirow{3}{*}{$1$} & \multirow{3}{*}{$\bar{3}$} & \multirow{3}{*}{$\bar{3}$} & \multirow{3}{*}{$1$} &
                $\bar{3}$ & $1$ & $3$ & $3$ & $3$ \\ \cline{7-11} \cline{18-22}
                & & & & & &
                $1$ & $\bar{3}$ & $3$ & $3$ & $\bar{3}$ &
                & & & & & &
                $\bar{3}$ & $1$ & $\bar{6}$ & $3$ & $3$\\ \cline{7-11} \cline{18-22}
                & & & & & &
                $3$ & $3$ & $\bar{3}$ & $3$ & $\bar{3}$ &
                & & & & & &
                $1$ & $\bar{3}$ & $\bar{3}$ & $3$ & $3$\\ \cline{7-11} \cline{12-22}
                & & & & & &
                $3$ & $\bar{6}$ & $\bar{3}$ & $3$ & $\bar{3}$  &
                \multirow{4}{*}{$\bar{3}$} & \multirow{4}{*}{$\bar{3}$} & \multirow{4}{*}{$3$} & \multirow{4}{*}{$3$} & \multirow{4}{*}{$1$} & \multirow{4}{*}{$1$} &
                $\bar{3}$ & $\bar{3}$ & $1$ & $1$ & $3$ \\ \cline{7-11} \cline{18-22}
                & & & & & &
                $\bar{6}$ & $3$ & $\bar{3}$ & $3$ & $\bar{3}$ &
                & & & & & &
                $\bar{3}$ & $6$ & $1$ & $1$ & $\bar{6}$ \\ \cline{7-11} \cline{18-22}
                & & & & & &
                $3$ & $3$ & $6$ & $3$ & $\bar{3}$ &
                & & & & & &
                $1$ & $3$ & $3$ & $1$ & $3$ \\ \cline{7-11} \cline{18-22}
                & & & & & &
                $3$ & $\bar{6}$ & $6$ & $3$ & $\bar{3}$ &
                & & & & & &
                $1$ & $3$ & $3$ & $1$ & $\bar{6}$ \\ \hline
                \multirow{10}{*}{$1$} & \multirow{10}{*}{$1$} & \multirow{10}{*}{$3$} & \multirow{10}{*}{$3$} & \multirow{10}{*}{$\bar{3}$} & \multirow{10}{*}{$\bar{3}$} &
                $1$ & $1$ & $1$ & $\bar{3}$ & $3$ &
                \multirow{5}{*}{$\bar{3}$} & \multirow{5}{*}{$3$} & \multirow{5}{*}{$\bar{3}$} & \multirow{5}{*}{$3$} & \multirow{5}{*}{$1$} & \multirow{5}{*}{$1$} &
                $\bar{3}$ & $1$ & $1$ & $1$ & $1$ \\ \cline{7-11} \cline{18-22}
                & & & & & &
                $1$ & $1$ & $1$ & $6$ & $\bar{6}$ &
                & & & & & &
                $1$ & $\bar{3}$ & $3$ & $1$ & $1$\\ \cline{7-11} \cline{18-22}
                & & & & & &
                $3$ & $\bar{3}$ & $3$ & $\bar{3}$ & $3$ &
                & & & & & &
                $3$ & $3$ & $\bar{3}$ & $1$ & $1$\\ \cline{7-11} \cline{18-22}
                & & & & & &
                $3$ & $\bar{3}$ & $3$ & $\bar{3}$ & $\bar{6}$ &
                & & & & & &
                $\bar{6}$ & $3$ & $\bar{3}$ & $1$ & $1$\\ \cline{7-11} \cline{18-22}
                & & & & & &
                $3$ & $\bar{3}$ & $3$ & $6$ & $3$ &
                & & & & & &
                $3$ & $\bar{6}$ & $6$ & $1$ & $1$\\ \cline{7-11} \cline{12-22}
                & & & & & &
                $3$ & $\bar{3}$ & $3$ & $6$ & $\bar{6}$ &
                \multirow{4}{*}{$3$} & \multirow{4}{*}{$3$} & \multirow{4}{*}{$\bar{3}$} & \multirow{4}{*}{$\bar{3}$} & \multirow{4}{*}{$1$} & \multirow{4}{*}{$1$} &
                $\bar{3}$ & $1$ & $3$ & $1$ & $\bar{3}$ \\ \cline{7-11} \cline{18-22}
                & & & & & &
                $6$ & $\bar{6}$ & $6$ & $\bar{3}$ & $3$ &
                & & & & & &
                $\bar{3}$ & $1$ & $\bar{6}$ & $1$ & $6$\\ \cline{7-11} \cline{18-22}
                & & & & & &
                $6$ & $\bar{6}$ & $6$ & $\bar{3}$ & $\bar{6}$ &
                & & & & & &
                $1$ & $\bar{3}$ & $\bar{3}$ & $1$ & $\bar{3}$\\
                \cline{7-11} \cline{18-22}
                & & & & & &
                $6$ & $\bar{6}$ & $6$ & $6$ & $3$ &
                & & & & & &
                $1$ & $\bar{3}$ & $\bar{3}$ & $1$ & $6$ \\ \cline{7-11} \cline{12-22}
                & & & & & &
                $6$ & $\bar{6}$ & $6$ & $6$ & $\bar{6}$ \\ \cline{1-11}
        \end{tabular}
\caption{\label{tab:N11SU3} Assign the $SU(3)_{C}$ quantum numbers to fields of the topology N-1-1 in figure~\ref{fig:N0-1}.  The labels $E_i$ and $I_i$ of the external and internal fields are showed in figure~\ref{fig:N11SU2SU3}.}
\end{table}

\begin{figure}[hptb]
\centering
\includegraphics[width=0.40\textwidth]{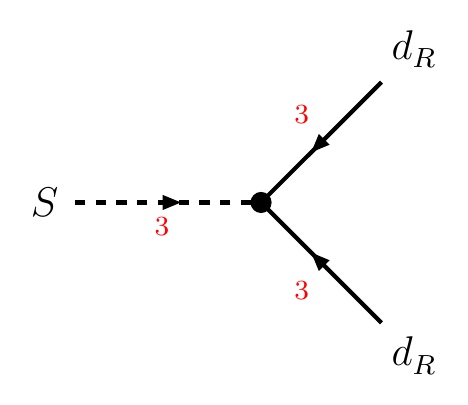}
\caption{\label{fig:SU3} An example of $SU(3)_{C}$ invariant vertex which is vanishing because of antisymmetric color contractions. }
\end{figure}

Then we discuss the possible transformation properties of the internal fields under $SU(3)_C$, we consider the $SU(3)$ representation up to dimension 8, namely, $1$, 3, $\bar{3}$, 6, $\bar{6}$ and $8$, high dimensional representations can be discussed analogously. All the $0\nu\beta\beta$ decay operators in Eq.~\eqref{eqn-basis-summary-dim9} involve two quarks, two anti-quarks and two lepton fields which are in the irreducible representations 3, $\bar{3}$ and $1$ of the $SU(3)_C$ group respectively. We assign $SU(3)_C$ quantum numbers to each internal line, then consider whether $SU(3)_C$ invariant contractions can be formed at the vertices by using the tensor product reduction rules such as
\begin{small}
\begin{eqnarray}
\nonumber&&3\otimes3=\bar{3}\oplus6,~~~3\otimes\bar{3}=1\oplus8,~~~3\otimes 6=8\otimes 10,~~~3\otimes \bar{6} =\bar{3} \oplus \overline{15}\,,\\
\nonumber&&3 \otimes 8= 3 \oplus \bar{6} \oplus 15,~~~3 \otimes 10= 15 \oplus 15,~~~3 \otimes \overline{10} = \overline{6} \oplus \overline{24}\,,\\
\nonumber&&6\otimes 6=\bar{6}\oplus 15\oplus 15,~~~6\otimes \bar{6}=1\oplus 8\oplus 27\,,~~~6\otimes 8=\bar{3} \oplus 6 \oplus \overline{15} \oplus 24,\\
\nonumber&&6\otimes 10= \overline{15}\oplus 21 \oplus 24,~~~6\otimes \overline{10} = \bar{3} \oplus \overline{15}\oplus \overline{42}\,,~~~8\otimes 8=1\oplus 8\oplus 8 \oplus 10\oplus \overline{10} \oplus 27\,,\\
&&8\otimes 10= 8 \oplus 10 \oplus 27 \oplus 35,~~~10 \otimes 10 =\overline{10} \oplus 27 \oplus 28 \oplus 35,~~~10 \otimes \overline{10} = 1 \oplus 8 \oplus 27 \oplus 64\,.
\end{eqnarray}
\end{small}
We shall use the Mathematica package \textbf{GroupMath}~\cite{Fonseca:2020vke} to determine the $SU(3)_C$ gauge invariance of the interaction vertices. The renormalizable interaction compatible with $SU(3)_C$ gauge symmetry could be vanishing, in particular, when the repeated fields are involved. We take the interaction displayed in figure~\ref{fig:SU3} as an example, which is of the form $\overline{d_{R}^{c}}d_{R}S$, with $S$ being a color triplet. Expanding the color index of the down quark explicitly, we obtain
\begin{eqnarray}
\nonumber\overline{d_{R}^{c}}d_{R}S&=&
\overline{d_{Rr}^{c}}d_{Rg}S_{b}-\overline{d_{Rg}^{c}}d_{Rr}S_{b}-
\overline{d_{Rr}^{c}}d_{Rb}S_{g}+\overline{d_{Rb}^{c}}d_{Rr}S_{g}+ \overline{d_{Rg}^{c}}d_{Rb}S_{r}-\overline{d_{Rb}^{c}}d_{Rg}S_{r}\\
\nonumber&=&\overline{d_{Rr}^{c}}d_{Rg}S_{b}-\overline{d_{Rr}^{c}}d_{Rg}S_{b}
-\overline{d_{Rr}^{c}}d_{Rb}S_{g}+\overline{d_{Rr}^{c}}d_{Rb}S_{g}+
\overline{d_{Rg}^{c}}d_{Rb}S_{r}-\overline{d_{Rg}^{c}}d_{Rb}S_{r}\\
&=&0\,,
\end{eqnarray}
where $r$, $g$, $b$ denote the three different colors of quark fields. Hence it is necessary to explicitly expand the $SU(3)_C$ indices of each interaction vertex to see whether point interactions are vanishing. The $SU(3)_C$ quantum numbers of the internal fields can be discussed at the topology level as well. We list the possible color charge assignments of the topology N-1-1 in table~\ref{tab:N11SU3}, where the $SU(3)_C$ representations up to dimension 6 are considered due to limited space and the results including octet assignment are given in the attachment~\cite{Chen:2021sup}.
The labels of the fields and the color flow of the topology N-1-1 are shown in figure~\ref{fig:N11SU2SU3}.

As explained above, we can start from a diagram in figure~\ref{fig:DiaN0-1}, attach the fields of the $0\nu\beta\beta$ decay operators to the external legs, subsequently determine the possible $U(1)_Y$, $SU(2)_L$ and $SU(3)_C$ quantum numbers of the internal fields by imposing gauge invariance at each vertex, then the one-loop UV complete models for the effective $0\nu\beta\beta$ decay operators can be obtained. The different models are named as Ni-a-b-c-d-e-f, where Ni with i=1, 2, 3, 4, 5, 6 denotes the set of $0\nu\beta\beta$ decay operators, the second index ``a'' refer to tree (a=0) and one-loop (a=1) models, the symbols ``b'' and ``c'' stand for the topology and diagram respectively, the index ``d'' indicates the attachment of external fields, and the last two indices ``e'' and ``f'' denote the assignments of the $SU(2)_L$ and $SU(3)_C$ respectively. For instance, combining figure~\ref{fig:external-N4} with the SM electroweak-sector quantum number assignments in table \ref{tab:N11SU2} and table \ref{tab:N11SU3}, we can obtain all the models arising from the diagram N-1-1-1. The possible UV completions for the operator N4 are listed in table~\ref{tab:N41111}, table~\ref{tab:N41112}, table~\ref{tab:N41113}, table~\ref{tab:N41114} and table~\ref{tab:N41115}.

\vskip0.5in

\subsubsection{\label{sec:DetermineGe}Determine genuineness}

\begin{figure}[hptb]
\centering
\includegraphics[width=0.95\textwidth]{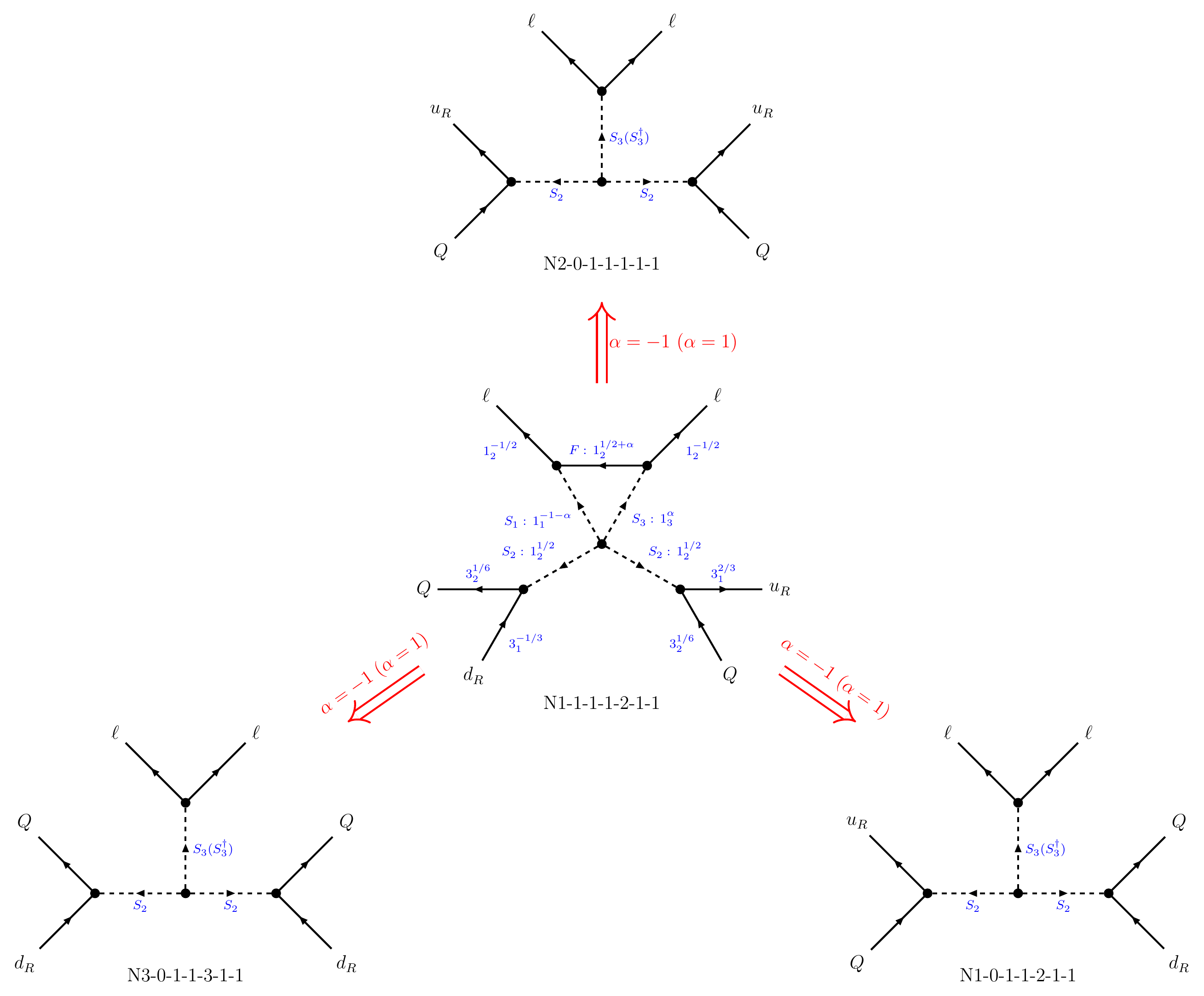}
\caption{\label{fig:N2111311} One example of a non-genuine model N1-1-1-1-2-1-1, the mediators $S_2$ and $S_3$ will also lead to tree-level contributions N1-0-1-1-2-1-1, N2-0-1-1-1-1-1, N3-0-1-1-3-1-1 when the hypercharge parameter $\alpha=\pm1$. The quantum numbers are given in the notation $C_{L}^{Y}$, where $C$ refers to the $SU(3)_{C}$ transformation, $L$ refers to the $SU(2)_{L}$ transformation, and $Y$ stands for $U(1)_{Y}$ charge. }
\end{figure}

Numerous one-loop models for short-range $0\nu\beta\beta$ decays can be generated through the algorithm described in previous sections, in particular all the tree-level $0\nu\beta\beta$ decay models are generated. A one-loop model is the dominant contribution if and only if the new messenger fields are insufficient to mediate the more important tree-level diagram, then such model would be called genuine.
In other words, the fields which generate the tree-level $0\nu\beta\beta$ decays should be absent in a genuine one-loop model. If the lower order contributions can not be forbidden without extra symmetry, the model would be non-genuine. We determine the genuineness of each model by comparing its field content with that of tree-level models one by one. Since the quantum numbers of the internal particles in the tree-level $0\nu\beta\beta$ decay models are unambiguously fixed, genuineness of a one-loop model generally excludes certain value of the hypercharge parameter $\alpha$.
We take the model N1-1-1-1-2-1-1 for illustration, see figure~\ref{fig:N2111311}, we have defined the notation $C_{L}^{Y}$ for the quantum numbers in the diagram, where $C$ refers to the $SU(3)_{C}$ transformation, $L$ refers to the $SU(2)_{L}$ transformation, and $Y$ stands for $U(1)_{Y}$ charge. When the hypercharge parameter $\alpha=\pm1$, the mediators $S_2$ and $S_3$ and the associated interactions allow to generate the tree-level $0\nu\beta\beta$ decay models N1-0-1-1-2-1-1, N2-0-1-1-1-1-1, N3-0-1-1-3-1-1 which are the more important ones, as shown in figure~\ref{fig:N2111311}. Therefore the genuineness of the one-loop model N1-1-1-1-2-1-1 requires $\alpha\neq\pm1$. The condition of genuineness has been considered for each possible one-loop decomposition of the $0\nu\beta\beta$ decay operators, see the attachment for the full results~\cite{Chen:2021sup}.

\section{\label{sec:relaton to mass}Neutrino masses in $0\nu\beta\beta$ decay models }

We proceed to discuss the neutrino masses which are generated in the above models of short-range $0\nu\beta\beta$ decay. The famous black box theorem establishes that the models which can lead to $0\nu\beta\beta$ decay must produce Majorana neutrino masses at or below four-loop order~\cite{Schechter:1981bd}. The four-loop diagram guaranteed by the black box theorem produces a very tiny neutrino mass of order $m_{\nu}\sim 10^{-24}$ eV which is too small to explain the current neutrino experimental data~\cite{Duerr:2011zd}. However, generally the neutrino masses are generated at a lower loop order in a specific $0\nu\beta\beta$ decay model so that the correct order of neutrino masses could be obtained. It is well known that the  exchange of a light neutrino between  two SM charged current interactions could lead to $0\nu\beta\beta$ decay, and it is usually called mass mechanism which is characterized by the effective neutrino mass $m_{\beta\beta}$ in Eq.~\eqref{eq:mbb-eff}. For models with tree and one-loop neutrino masses, the mass mechanism is generally expected to dominate unless the model parameters are fine-tuned. The short-range contribution could be comparable with the mass mechanism for models with two-loop and three-loop neutrino masses~\cite{Helo:2015fba}. One can construct neutrino mass diagrams by using the SM fields and the mediators which appear in one-loop renormalizable $0\nu\beta\beta$ decay models.
%We are mainly interested in UV completions in which where new physics %contributions to $0\nu\beta\beta$ can compete with the mass mechanism.
In order to identify the neutrino mass diagrams, one need to compare the messenger fields of the $0\nu\beta\beta$ decay models with those of neutrino mass models. In the present work, we are concerned with the neutrino mass models up to two-loop level which can be generated from $0\nu\beta\beta$ decay models.

\begin{figure}[hptb]
\centering
\includegraphics[width=\textwidth]{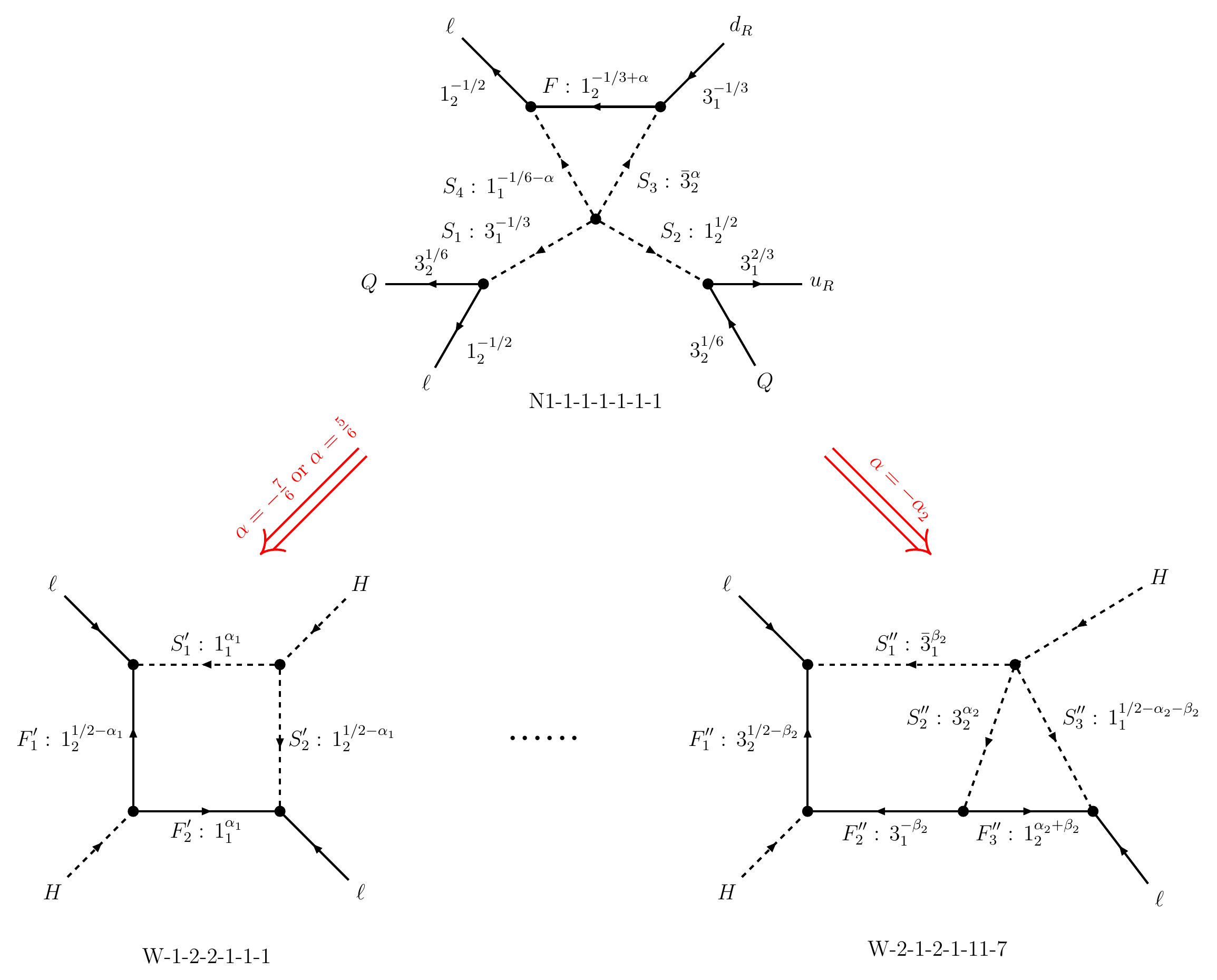}
\caption{\label{fig:RelatedToMass} The neutrino mass models generated from the $0\nu\beta\beta$ decay model N1-1-1-1-1-1-1. The one-loop neutrino mass diagram W-1-2-2-1-1-1 can be generated by the mediators of N1-1-1-1-1-1-1 when the hypercharge $\alpha=-7/6, 5/6$, and the two-loop neutrino mass diagram W-2-1-2-1-11-7 can always be generated.}
\end{figure}

%The method used in this section is the same as Sec.\ref{sec:DetermineGe}.
%\sout{We take model N1-1-1-1-1-1-1 for example.}
We take model N1-1-1-1-1-1-1 for example. The new mediators could possibly give rise to the one-loop neutrino mass diagram W-1-2-2-1-1-1, as shown in figure~\ref{fig:RelatedToMass}. By examining the $SU(3)_C$ and $SU(2)_L$ quantum numbers of the mediators, one could possibly identify the fields in the mass model W-1-2-2-1-1-1 with those of the $0\nu\beta\beta$ decay model as follows,
\begin{equation}
S_1'\sim S_4^{(\dagger)}\, \quad S_2'\sim S_{2}^{(\dagger)}, H\, \quad F_1'\sim F^{(c)}, \ell_{L}^{(c)}\, \quad F_2'\sim e_{R}^{(c)}\,.
\end{equation}
%Therefore we have the following relations.
Furthermore, the identification of the hypercharge requires the following relations should be fulfilled,
\begin{eqnarray}
\nonumber&&S_1': \alpha_{1}=-\frac{1}{6}-\alpha~~~\text{or}~~~\alpha_{1}=\frac{1}{6}+\alpha\,,\\
\nonumber&& S_2':\frac{1}{2}-\alpha_{1}=\frac{1}{2}
~~~ \text{or}~~~\frac{1}{2}-\alpha_{1}=-\frac{1}{2}\,,\\
\nonumber&& F_1': \frac{1}{2}-\alpha_{1}=-\frac{1}{3}+\alpha
~~\text{or}~~   \frac{1}{2}-\alpha_{1}=\frac{1}{3}-\alpha~~ \text{or} ~~\frac{1}{2}-\alpha_{1}=\frac{1}{2}
~~ \text{or} ~~ \frac{1}{2}-\alpha_{1}=-\frac{1}{2}\,,\\
%\nonumber&&~~~ \text{or} ~~~\frac{1}{2}-\alpha_{1}=\frac{1}{2}
%~~~ \text{or} ~~~ \frac{1}{2}-\alpha_{1}=-\frac{1}{2}\,,\\
&&F_2': \alpha_{1}=-1 ~~~ \text{or} ~~~ \alpha_{1}=1\,.
\end{eqnarray}
%The solution is
The solutions to these equations are given by
\begin{equation}
\left\{\begin{array}{cl}
\alpha&=-7/6,\\
\alpha_{1}&=1,
\end{array} \right.
\quad \text{or} \quad
\left\{\begin{array}{cl}
\alpha&=5/6,\\
\alpha_{1}&=1.
\end{array} \right.
\end{equation}
Hence the interactions in the $0\nu\beta\beta$ decay model N1-1-1-1-1-1-1 lead us to the one-loop neutrino mass model W-1-2-2-1-1-1 when $\alpha=-7/6$ or $\alpha=5/6$. Furthermore, by matching with the mediators of the one-loop decomposition of the Weinberg operator~\cite{Bonnet:2012kz}, we find that the absence of one-loop neutrino masses excludes the values of $\alpha=$$-7/6,-1/6,5/6$. Then we proceed to show that the two-loop neutrino mass diagram W-2-1-2-1-11-7 given in figure~\ref{fig:RelatedToMass} can be constructed without introducing new interactions. All the mediators of the model W-2-1-2-1-11-7 should be contained in the model N1-1-1-1-1-1-1, their transformations under $SU(3)_C\times SU(2)_L$ implies the following possible identifications,
\begin{equation}
S_1''\sim S_1^{\dagger}\,, ~~ S_2''\sim S_{3}^{\dagger}\,, ~~ S_3''\sim S_{4}^{(\dagger)}\,, ~~ F_1''\sim Q_{L}\,, ~~  F_2''\sim u_{R}, d_{R}\,, ~~ F_3''\sim F^{(c)}, \ell_{L}^{(c)}\,.
\end{equation}
The equality of hypercharges imposes the following constraints:
\begin{eqnarray}
\nonumber&& S_1'': \beta_{2}=\frac{1}{3}\,, \\
\nonumber&& S_2'': \alpha_{2}=-\alpha\,,\\
\nonumber&& S_3'': \frac{1}{2}-\alpha_{2}-\beta_{2}=-\frac{1}{6}-\alpha
\quad \text{or} \quad \frac{1}{2}-\alpha_{2}-\beta_{2}=\frac{1}{6}+\alpha\,,\\
\nonumber&& F_1'': \frac{1}{2}-\beta_{2}=\frac{1}{6}\,,\\
\nonumber&& F_2'': -\beta_{2}=\frac{2}{3} \quad \text{or} \quad -\beta_{2}=-\frac{1}{3}\,,\\
&&F_3'': \alpha_{2}+\beta_{2}=-\frac{1}{3}+\alpha ~~  \text{or} ~~ \alpha_{2}+\beta_{2}=\frac{1}{3}-\alpha ~~ \text{or}~~ \alpha_{2}+\beta_{2}=-\frac{1}{2} ~~ \text{or} ~~\alpha_{2}+\beta_{2}=\frac{1}{2}\,,
\end{eqnarray}
whose solution is
\begin{equation}
\alpha=-\alpha_{2},~~~\beta_{2}=1/3\,.
\end{equation}
This implies that the $0\nu\beta\beta$ decay model N1-1-1-1-1-1-1 always leads to the two-loop neutrino mass diagram W-2-1-2-1-11-7 independent of the value of $\alpha$. After applying the method to all the mass models up to two-loop level, one can find that the model N1-1-1-1-1-1-1 can generate neutrino masses when $\alpha=-7/6, -1/6, 1/3, 5/6, -\alpha_{2}$. In a similar way, we consider the neutrino mass diagrams for each $0\nu\beta\beta$ decay model.
%\textcolor{green}{and we find that the neutrino masses can be generated at a %level lower than three-loop.  }

\section{\label{sec:example} An example model of one-loop $0\nu\beta\beta$ decay }

\begin{figure}[htbp]
\centering
\includegraphics[width=\textwidth]{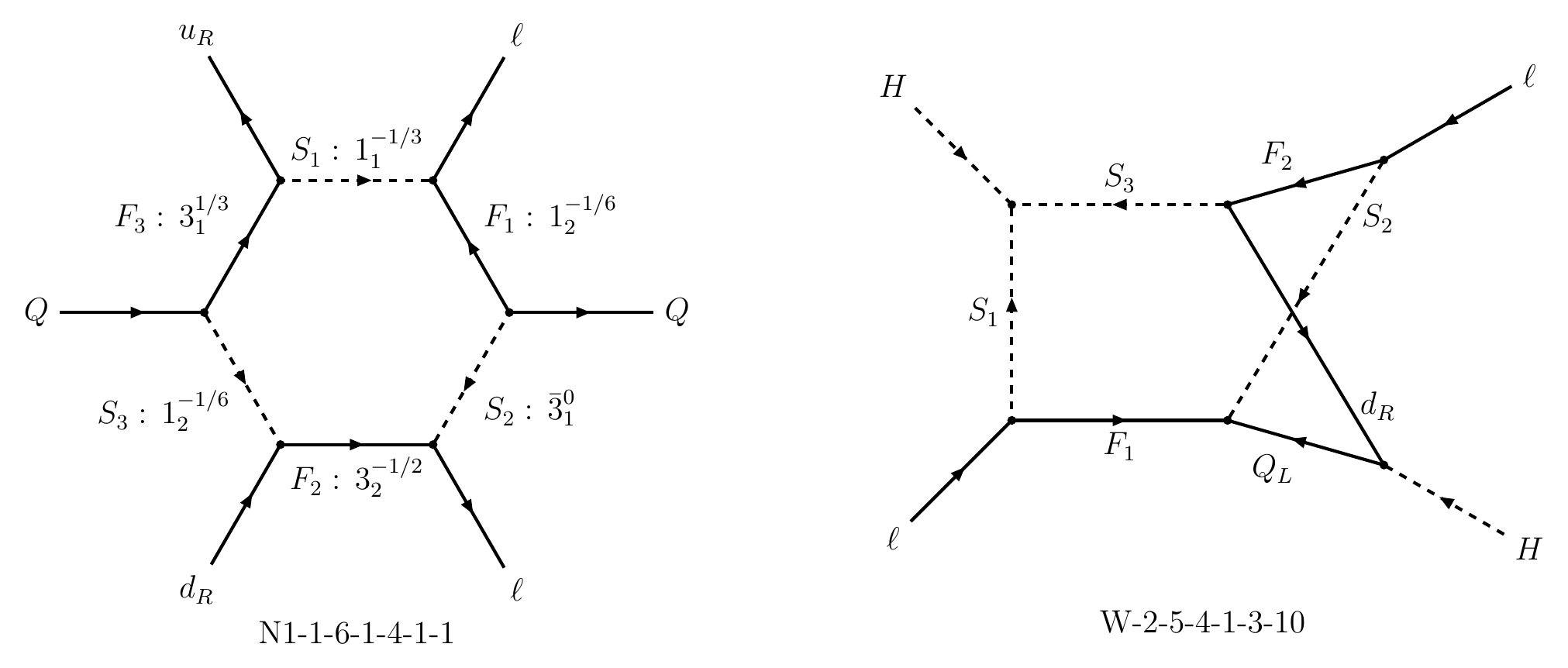}
\caption{\label{fig:N1161411-mass} An example of $0\nu\beta\beta$ decay model N1-1-6-1-4-1-1~(left panel) with the hypercharge parameter $\alpha=0$, which is a UV completion of the operator N1 based on the diagram N-1-6-1. The neutrino masses arise from the two-loop diagram W-2-5-4-1-3-10 (right panel) at leading order, and it is mediated by the messenger fields of the model N1-1-6-1-4-1-1. }
\end{figure}

\begin{table}[h]
\renewcommand{\tabcolsep}{0.5mm}
\renewcommand{\arraystretch}{1.3}
\centering
\begin{tabular}{|c|c|c|c|c|c|c|}\hline \hline
        & \multicolumn{6}{c|}{N1-1-6-1-4-1-1} \\ \hline
 Fields & $F_{1}$ & $F_{2}$ & $F_{3}$ & $S_{1}$ & $S_{2}$ & $S_{3}$\\ \hline
$C_{L}^{Y}$  & $ 1_{2}^{-1/6}$ & $ 3_{2}^{-1/2}$ & $3_{1}^{1/3}$ & $1_{1}^{-1/3}$ & $\bar{3}_{1}^{0}$ & $1_{2}^{-1/6}$\\ \hline \hline
\end{tabular}
\caption{\label{tab:N1161411} The mediators of the $0\nu\beta\beta$ decay model shown in the left panel of figure~\ref{fig:N1161411-mass} as well as their transformation properties under the SM gauge group. The quantum numbers are given in the notation $C_{L}^{Y}$, where $C$ refers to the $SU(3)_{C}$ transformation, $L$ refers to the $SU(2)_{L}$ transformation, and $Y$ stands for $U(1)_{Y}$ charge. }
\end{table}

In this section, we present an example of model N1-1-6-1-4-1-1 which is a decomposition of the $0\nu\beta\beta$ decay operator N1 based on the diagram N-1-6-1 in figure~\ref{fig:DiaN0-1}. We add six new fields to the SM particle content in this model, and their quantum number assignments are listed in table~\ref{tab:N1161411}. The interaction Lagrangian for these fields should respect the SM gauge symmetry, and it is of the following form
\begin{eqnarray}
\nonumber\mathcal{L}_{\text{int}} &=& y_{I\alpha}\overline{Q_{L}}_{,\alpha}i\tau^{2}F_{1}^{c} S_{2}^{*}+
y_{II\alpha}\overline{\ell}_{\alpha}F_{1}S_{1}+y_{III\alpha}\overline{u_{R}}_{,\alpha}F_{3}S_{1}^{*}+y_{IV\alpha}S_{3}^{\dagger}\overline{F_{3}}Q_{L,\alpha}\\
\label{eq:interaction-example-model}&&+y_{V\alpha}\overline{F_{2}}d_{R,\alpha}S_{3}
+y_{VI\alpha}\overline{\ell}_{\alpha}F_{2}S_{2}+\sum_{i=1}^{3}m_{F_{i}}\overline{F_{i}}F_{i}+\mathrm{H.c.}\,,
\end{eqnarray}
where $\tau^2$ is the Pauli matrix, and $i\tau^2=\epsilon$ is the 2-component antisymmetric tensor. This Lagrangian can generate the one-loop diagram for $0\nu\beta\beta$ decay shown in figure~\ref{fig:N1161411-mass}. The gauge invariant scalar potential terms involving new scalars are given by
\begin{equation}
\label{eq:V-potential}V\supset
\left({\mu}S_{1}S_{3}^{T}i\tau^{2}H + \mathrm{H.c.}\right) +                 \sum_{i=1}^{3}{\lambda}_{S_{i}H}S_{i}^{\dagger}S_{i}H^{\dagger}H+
\sum_{i,j=1}^{3}{\lambda}_{S_{i}S_{j}}S_{i}^{\dagger}S_{i}S_{j}^{\dagger}S_{j}
+\sum_{i=1}^{3}m_{S_{i}}^{2}S_{i}^{\dagger}S_{i}\,.
\end{equation}
We denote $S_3=(S_{3,1}, S_{3,2})^T$, the first term in Eq.~\eqref{eq:V-potential} can induce mixing between the two states $S_1$ and $S_{3,1}$ after electroweak symmetry breaking. The bilinear terms of $S_1$ and $S_{3,1}$ read as
\begin{equation}
\mu v_{H}S_{1}S_{3,1}+\mu^{*} v_{H}S^{\dagger}_{1}S^{\dagger}_{3,1}
+m_{S_{1}}^{2}S_{1}^{\dagger}S_{1}+m_{S_{3}}^{2}S_{3,1}^{\dagger}S_{3,1}
=\left(S_{1}^{\dagger} \quad S_{3,1}\right)
\left(\begin{array}{cc}
m_{S_{1}}^{2} ~& \mu^{*}\,v_{H}\\
{\mu}\,v_{H} ~& m_{S_{3}}^{2}
\end{array}\right)
\left(\begin{array}{c}
S_{1}\\
S_{3,1}^{\dagger}
\end{array}
\right)\,,
\end{equation}
where $v_H$ is the vacuum expectation value of the Higgs field with $\langle H^0\rangle=v_H$, we have redefined $m_{S_{1}}^{2}$ and $m_{S_{3}}^{2}$ to absorb the contributions of ${\lambda}_{S_{1}H}v^2_H$ and ${\lambda}_{S_{3}H}v^2_H$ respectively. Assuming the mixing parameter $\mu$ to be real, the above mass matrix can be diagonalized by a two-dimensional rotation matrix
\begin{equation}
\left(
\begin{array}{cc}
\cos\theta ~& \sin\theta \\
-\sin\theta ~& \cos\theta
\end{array}\right)
\left(\begin{array}{cc}
m_{S_{1}}^{2} ~& \mu\,v_{H}\\
{\mu}\,v_{H} ~& m_{S_{3}}^{2}
\end{array}\right)
\left(
\begin{array}{cc}
\cos\theta ~& -\sin\theta \\
\sin\theta ~& \cos\theta
\end{array}\right)=\text{diag}(m^2_{S_{n1}}, m^2_{S_{n2}})\,,
\end{equation}
where
\begin{eqnarray}
\nonumber\tan2\theta&=&\frac{2{\mu}\,v_{H}}{m_{S_{1}}^{2}-m_{S_{3}}^{2}}\,,\\
\nonumber m^2_{S_{n1}}&=&\frac{1}{2}\left[ m_{S_{1}}^{2}+m_{S_{3}}^{2}+\sqrt{(m_{S_{1}}^{2}- m_{S_{3}}^{2})^2+4\mu^2v^2_{H}}\right]\,,\\
\label{eq:scalar-mass}m^2_{S_{n2}}&=&\frac{1}{2}\left[ m_{S_{1}}^{2}+ m_{S_{3}}^{2}-\sqrt{(m_{S_{1}}^{2}- m_{S_{3}}^{2})^2+4\mu^2v^2_{H}}\right]\,.
\end{eqnarray}
Accordingly the mass eigenstates are related to the interaction basis by
\begin{equation}
\left(\begin{array}{c}
S_{n1}\\
S_{n2}
\end{array}
\right)=\left(
\begin{array}{cc}
\cos\theta ~& \sin\theta \\
-\sin\theta ~& \cos\theta
\end{array}\right)
\left(\begin{array}{c}
S_{1}\\
%S_{3,1}^{\dagger}
S_{3,1}^{*}
\end{array}
\right)\,.
\end{equation}
Rotating the interaction in Eq.~\eqref{eq:interaction-example-model} to the scalar mass eigenstate basis, the terms relevant to $0\nu\beta\beta$ decay are given by
\begin{align}
\nonumber
\mathcal{L}_{\text{int}}=&y_{I1}\overline{u_{L}}F_{1,2}^{c} S_{2}^{*}
+y_{IIe}\overline{e_{L}}F_{1,2}(S_{n1}\cos\theta-S_{n2}\sin\theta)
+y_{III1}\overline{u_{R}}F_{3}(S_{n1}^{*}\cos\theta-S_{n2}^{*}\sin\theta)\\
&+y_{IV1}\overline{F_{3}}d_{L}S_{3,2}^{*}+y_{V1}\overline{F_{2,2}}d_{R}S_{3,2}+y_{VIe}\overline{e_{L}}F_{2,2}S_{2}+\mathrm{H.c.}+\ldots\,.
  \label{eq:Lint}
\end{align}
The one-loop diagram in figure~\ref{fig:N1161411-mass} can be straightforwardly calculated, and we find the following effective operator is generated
\begin{equation}
\label{eq:SR-oper-model}\frac{G_{F}^{2}\cos^{2}\theta_{C}}{2m_{P}}\epsilon_{1}^{LRL}J_{L}J_{R}j_{L}\,,
\end{equation}
where $J_{R,L}=\bar{u}(1\pm\gamma_{5})d$, $j_{L}=\bar{e}(1+\gamma_{5})e^{c}$ are defined in Eq.~\eqref{eq:current}, and the coefficient $\epsilon_{1}^{LRL}$ is given by
%\textcolor{green}{what is the corresponding $SU(2)_L\times U(1)_Y$ invariant operator? }
\begin{eqnarray}
\nonumber&&\epsilon_{1}^{LRL}=\frac{m_{F_{2}}m_{F_{3}}m_{P}}{128\pi^2m^{7}_{F_{1}}G_{F}^{2}\cos^{2}\theta_{C}} y_{I1}y_{IIe}y_{III1}y_{IV1}y_{V1}y_{VIe}\Bigg[\cos^{2}\theta\, \mathcal{F}\left(1,\frac{m_{F_{2}}^{2}}{m_{F_{1}}^{2}},\frac{m_{F_{3}}^{2}}{m_{F_{1}}^{2}},\frac{m_{S_{n1}}^{2}}{m_{F_{1}}^{2}},\frac{m_{S_{2}}^{2}}{m_{F_{1}}^{2}},\frac{m_{S_{3}}^{2}}{m_{F_{1}}^{2}}\right)\\
\label{eq:epsilon1-LRL}&&\hskip0.45in +\sin^{2}\theta\,\mathcal{F}\left(1,\frac{m_{F_{2}}^{2}}{m_{F_{1}}^{2}},\frac{m_{F_{3}}^{2}}{m_{F_{1}}^{2}},\frac{m_{S_{n2}}^{2}}{m_{F_{1}}^{2}},\frac{m_{S_{2}}^{2}}{m_{F_{1}}^{2}},\frac{m_{S_{3}}^{2}}{m_{F_{1}}^{2}}\right)\Bigg]\,.
\end{eqnarray}
The function $\mathcal{F}(x_{1},x_{2},x_{3},x_{4},x_{5},x_{6})$ is the loop integral and it is given by
\begin{eqnarray}
\nonumber\mathcal{F}(x_{1},x_{2},x_{3},x_{4},x_{5},x_{6})&=&\frac{1}{i\pi^{2}}\int d^{4}p\frac{1}{\left(p^{2}-x_{1}\right)\left(p^{2}-x_{2}\right)\left(p^{2}-x_{3}\right)\left(p^{2}-x_{4}\right)\left(p^{2}-x_{5}\right)\left(p^{2}-x_{6}\right)}\\
%\nonumber&&\frac{A_{0}(x_{6})}{\left(x_{6}-x_{1}\right)\left(x_{6}-x_{2}\right)\left(x_{6}-x_{3}\right)\left(x_{6}-x_{4}\right)\left(x_{6}-x_{5}\right)}\\
\nonumber &=&\frac{x_{1}\left(1-\ln x_{1}\right)}{\left(x_{1}-x_{2}\right)\left(x_{1}-x_{3}\right)\left(x_{1}-x_{4}\right)\left(x_{1}-x_{5}\right)\left(x_{1}-x_{6}\right)}\\
\nonumber&&+\frac{x_{2}\left(1-\ln x_{2}\right)}{\left(x_{2}-x_{1}\right)\left(x_{2}-x_{3}\right)\left(x_{2}-x_{4}\right)\left(x_{2}-x_{5}\right)\left(x_{2}-x_{6}\right)}\\
\nonumber&&+\frac{x_{3}\left(1-\ln x_{3}\right)}{\left(x_{3}-x_{1}\right)\left(x_{3}-x_{2}\right)\left(x_{3}-x_{4}\right)\left(x_{3}-x_{5}\right)\left(x_{3}-x_{6}\right)}\\
\nonumber&&+\frac{x_{4}\left(1-\ln x_{4}\right)}{\left(x_{4}-x_{1}\right)\left(x_{4}-x_{2}\right)\left(x_{4}-x_{3}\right)\left(x_{4}-x_{5}\right)\left(x_{4}-x_{6}\right)}\\
\nonumber&&+\frac{x_{5}\left(1-\ln x_{5}\right)}{\left(x_{5}-x_{1}\right)\left(x_{5}-x_{2}\right)\left(x_{5}-x_{3}\right)\left(x_{5}-x_{4}\right)\left(x_{5}-x_{6}\right)}\\
\label{eq:integral-F}&&+\frac{x_{6}\left(1-\ln x_{6}\right)}{\left(x_{6}-x_{1}\right)\left(x_{6}-x_{2}\right)\left(x_{6}-x_{3}\right)\left(x_{6}-x_{4}\right)\left(x_{6}-x_{5}\right)}\,.
\end{eqnarray}
In the limit $m_{F_{1}}=m_{F_{2}}=m_{F_{3}}=m_{S_{1}}=m_{S_{2}}=m_{S_{3}}\equiv \bar{m}$, we have $x_1=x_2=x_3=x_5=x_6=1$, $x_4=m_{S_{n1}}^{2}/m_{F_{1}}^{2}=1+\mu v_H/\bar{m}^2$ or $x_4=m_{S_{n2}}^{2}/m_{F_{1}}^{2}=1-\mu v_H/\bar{m}^2$, Eq.~\eqref{eq:integral-F} is not valid anymore. In this case, we can use the following formula instead,
\begin{equation}
\mathcal{F}(1, 1, 1, x_{4}, 1, 1) =\frac{3+10x_{4}-18x_{4}^{2}+6x_{4}^{3}-x_{4}^{4}+12x_{4}\ln x_{4}}{12\left(1-x_{4}\right)^{5}}\,.
\end{equation}
At present, the most stringent experimental limits on the $0\nu\beta\beta$ decay half-life of Germanium and Xenon are $T_{1/2}(^{76}\text{Ge})>1.8\times 10^{26}$ yr~\cite{GERDA:2020xhi} and $T_{1/2}(^{136}\text{Xe})>1.07\times 10^{26}$ yr~\cite{KamLAND-Zen:2016pfg} respectively at $90\%$ confidence level, which lead to the upper limits on the short-range coupling $|\epsilon_{1}^{LRL}|<2.90\times10^{-10}$ and $|\epsilon_{1}^{LRL}|<2.50\times10^{-10}$ respectively~\cite{Deppisch:2020ztt} if only the short-range mechanism is considered.

\subsection{Prediction for neutrino masses }

\begin{figure}[hptb]
\centering
\includegraphics[]{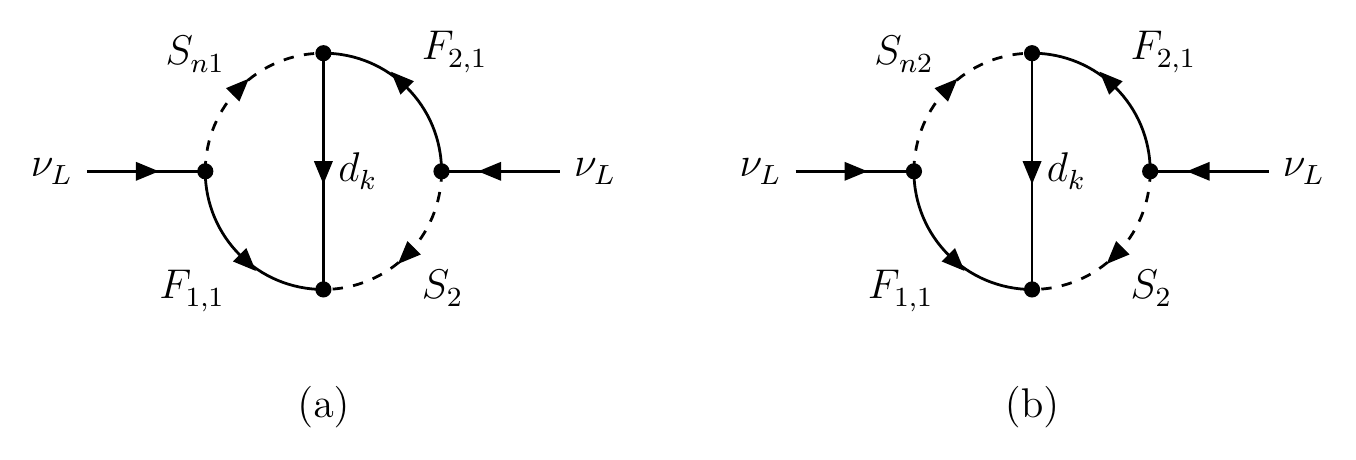}
\caption{\label{fig:neutrino-mass-example-model}
The diagrams for the neutrino mass generation in the model W-2-5-4-1-3-10, which is mediated by the fields of the $0\nu\beta\beta$ decay model N1-1-6-1-4-1-1, where $d_{k}$ with $k=1, 2, 3$ stand for the down quark, strange quark and bottom quark respectively.}
\end{figure}

We have checked that for this model the leading order neutrino masses arise from a two-loop model W-2-5-4-1-3-10, as shown in figure~\ref{fig:N1161411-mass}. The relevant interaction Lagrangian is given by
\begin{eqnarray}
\nonumber\mathcal{L}^{\nu}_{\text{int}}&=& y_{I\alpha}\overline{Q_{L}}_{,\alpha}i\tau^{2}F_{1}^{c}S_{2}^{*}
+
y_{II\alpha}\overline{\ell}_{\alpha}F_{1}S_{1}
+
y_{d,\alpha\beta}\overline{Q_{L}}_{,\alpha}d_{R,\beta}H\\
&&
+y_{V\alpha}\overline{F_{2}}d_{R,\alpha}S_{3}
+y_{VI\alpha}\overline{\ell}_{\alpha}F_{2}S_{2}
+{\mu}S_{1}S_{3}^{T}i\tau^{2}H+\mathrm{H.c.}\,.
\end{eqnarray}
The lepton number is violated by the simultaneous presence of the above terms. After the electroweak symmetry breaking, the above interaction can be written in the mass eigenstate basis as follows,
\begin{eqnarray}
\mathcal{L}^{\nu}_{\text{int}}&=&\nonumber y_{I\alpha}\overline{d_{L}}_{,\alpha}F^{c}_{1,1}S_{2}^{*}
+
y_{II\alpha}\overline{\nu_{L}}_{,\alpha}F_{1,1}(S_{n1}\cos\theta-S_{n2}\sin\theta)\\
%\nonumber&+Y^{*}_{1i}\overline{F^{c}}_{1,1}d_{L,i}S_{2}\\
&&+y_{V\alpha}\overline{F_{2}}_{,1}d_{R,\alpha}(S_{n1}^{*}\sin\theta+S_{n2}^{*}\cos\theta)
+y_{VI\alpha}\overline{\nu_{L}}_{,\alpha}F_{2,1}S_{2}+\mathrm{H.c.}+\ldots\,,
\end{eqnarray}
where the unitary rotations of the down type quarks to mass eigenstates have been absorbed into the couplings $y_{I\alpha}$ and $y_{V\alpha}$. The neutrino mass matrix arises from the two-loop diagrams of figure~\ref{fig:neutrino-mass-example-model} in the mass basis and it can be straightforwardly calculated as
\begin{eqnarray}
\nonumber (m_{\nu})_{\alpha\beta}&=&\frac{3}{512\pi^4}\sin2\theta\,(y^{*}_{II\alpha}y^{*}_{VI\beta}+y^{*}_{II\beta}y^{*}_{VI\alpha})y^{*}_{Ik}y^{*}_{Vk}
\frac{m_{F_{1}}m_{F_{2}}m_{d_{k}}}{m^2_{S_{2}}}\bigg[\mathcal{I}\left(
\frac{m_{F_{1}}^{2}}{m^2_{S_{2}}},
\frac{m_{S_{n2}}^{2}}{m^2_{S_{2}}},
\frac{m_{F_{2}}^{2}}{m^2_{S_{2}}}, 1,
\frac{m_{d_{k}}^{2}}{m^2_{S_{2}}}
\right)\\
\label{eq:nu-mass-model}&&-\mathcal{I}
\left(\frac{m_{F_{1}}^{2}}{m^2_{S_{2}}},
\frac{m_{S_{n1}}^{2}}{m^2_{S_{2}}},
\frac{m_{F_{2}}^{2}}{m^2_{S_{2}}}, 1,
\frac{m_{d_{k}}^{2}}{m^2_{S_{2}}}
\right)\bigg]\,.
\end{eqnarray}
Here the loop integral $\mathcal{I}\left(r_{a}, t_{\alpha}, r_{b}, t_{\beta}, r_{c} \right)$ is
\begin{small}
\begin{eqnarray}
\nonumber &&\mathcal{I}\left(r_{a}, t_{\alpha}, r_{b}, t_{\beta}, r_{c} \right)=\left(\frac{1}{i\pi^2}\right)^2\int d^{4}k \int d^{4}q
\frac{1}{\left(k^{2}-r_{a}\right)\left(k^{2}-t_{\alpha}\right)\left(q^{2}-r_{b}\right)\left(q^{2}-t_{\beta}\right)\left[\left(q+k\right)^{2}-r_{c}\right]}\\
&&\label{eq:func-I}\qquad\qquad\qquad =\frac{1}{r_{a}-t_{\alpha}}\frac{1}{r_{b}-t_{\beta}}\left[ T_{3}(r_{a},r_{b},r_{c})-T_{3}(r_{a},t_{\beta},r_{c})-T_{3}(t_{\alpha},r_{b},r_{c})
+T_{3}(t_{\alpha},t_{\beta},r_{c})\right] \,,
\end{eqnarray}
\end{small}
with
\begin{align}
T_{3}(x_{1},x_{2},x_{3})=\left(\frac{1}{i\pi^2}\right)^2
\int d^{4}k\int d^{4}q\frac{1}{\left(k^{2}-x_{1}\right)\left(q^{2}-x_{2}\right)\left[\left(q+k\right)^{2}-x_{3}\right]}\,.
\end{align}
The momentum space integration $T_{3}(x_{1},x_{2},x_{3})$ can be performed through the introduction of Feynman parameters, and it has been calculated in Refs.~\cite{vanderBij:1983bw,McDonald:2003zj,Angel:2013hla,Freitas:2016zmy}. The analytical expression of $\mathcal{I}\left(r_{a}, t_{\alpha}, r_{b}, t_{\beta}, r_{c} \right)$ reads as
\begin{eqnarray}
\nonumber \mathcal{I}\left(r_{a}, t_{\alpha}, r_{b}, t_{\beta}, r_{c} \right)&=&\frac{r_c}{(r_{a}-t_{\alpha})(r_{b}-t_{\beta})}\big[g(r_{a}/r_{c}, r_{b}/r_{c})-g(r_{a}/r_{c}, t_{\beta}/r_{c})\\
&&-g(t_{\alpha}/r_{c}, r_{b}/r_{c})
+g(t_{\alpha}/r_{c}, t_{\beta}/r_{c})\Big]\,,
\end{eqnarray}
where
\begin{eqnarray}
\nonumber
g(s,t)&=&\frac{s}{2}\ln s \ln t +\sum_{\pm}\pm\frac{s(1-s)+3st+2(1-t)x_{\pm}}{2\omega}\\
&&\times \left[\mathrm{Li}_{2}\left( \frac{x_{\pm}}{x_{\pm}-s}\right)
-\mathrm{Li}_{2}\left( \frac{x_{\pm}-s}{x_{\pm}}\right)
+\mathrm{Li}_{2}\left( \frac{t-1}{x_{\pm}}\right)
-\mathrm{Li}_{2}\left( \frac{t-1}{x_{\pm}-s}\right)\right]
\end{eqnarray}
with $\mathrm{Li}_{2}(x)$ being the dilogarithm function
\begin{equation}
\mathrm{Li}_{2}(x)=-\int_{0}^{x}\frac{\ln(1-y)}{y}dy
\end{equation}
and
\begin{equation}
x_{\pm}=\frac{1}{2}(-1+s+t\pm\omega)\,,\quad \omega=\sqrt{1+s^{2}+t^{2}-2(s+t+st)}~\,.
\end{equation}
The neutrino mass in Eq.~\eqref{eq:nu-mass-model} is dominated by the contribution of the bottom quark because the hierarchical mass pattern $m_{b}\gg m_s\gg m_d$. Only one generation for every new fermions and scalars are introduced in our model, therefore the neutrino mass matrix of Eq.~\eqref{eq:nu-mass-model} is of rank two so that the lightest neutrino is massless. If one allows for more than one generation of the new fermions and scalars, the corresponding neutrino mass matrix can be read out analogously and the mass of the lightest neutrino could be non-zero.

\subsection{Half-life time of $0\nu\beta\beta$ decay }

As shown in previous subsection, the mediators of the $0\nu\beta\beta$ decay model can generate non-vanishing light neutrino masses at two-loop level. Therefore both the short-range operator in Eq.~\eqref{eq:SR-oper-model} and the mass mechanism of the light neutrino exchange contribute to the $0\nu\beta\beta$ decay, and the two contributions should be added coherently. Using the general formula of Eq.~\eqref{eq:inverse-half-life} for the inverse half-life, we have
\begin{eqnarray}
\nonumber T_{1/2}^{-1}&=&G_{11+}^{(0)}\abs{\frac{m_{\beta\beta}}{m_{e}}\mathcal{M}_{\nu}+\epsilon_{1}^{LRL}\mathcal{M}^{LR}_{1}}^2\\
&=&G_{11+}^{(0)}\left[\abs{\mathcal{M}_{\nu}}^2\frac{\abs{m_{\beta\beta}}^2}{m^2_e}+
\abs{\mathcal{M}^{LR}_{1}}^2\abs{\epsilon_{1}^{LRL}}^2+
2\abs{\frac{m_{\beta\beta}}{m_{e}}\mathcal{M}_{\nu}}\abs{\epsilon_{1}^{LRL}\mathcal{M}^{LR}_{1}}\cos(\alpha-\beta)
\right]\,,
\end{eqnarray}
where $\alpha$ and $\beta$ denote the complex phases of $m_{\beta\beta}$ and $\epsilon_{1}^{LRL}$ respectively, note that the nuclear matrix elements are conventionally defined to be real. We see that the interference between the short-range contribution and mass mechanism is constructive for $\alpha-\beta=0$ and destructive for $\alpha-\beta=\pi$. It is remarkable that the two contributions would exactly cancel each other if $\epsilon_{1}^{LRL}=-m_{\beta\beta}\mathcal{M}_{\nu}/(m_{e}\mathcal{M}^{LR}_{1})$.

\begin{figure}[htbp]
\centering
\includegraphics[width=0.95\textwidth]{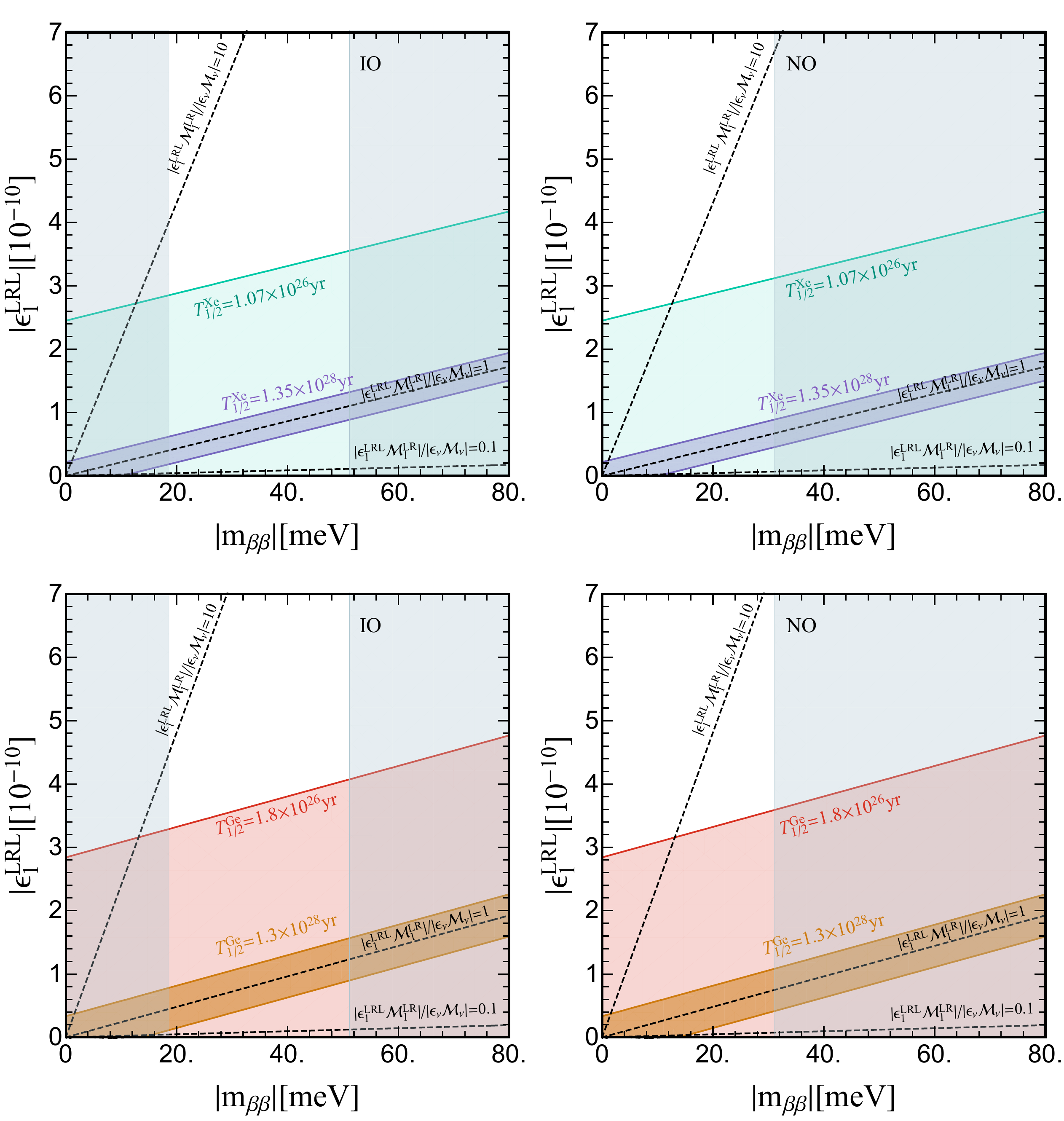}
\caption{\label{fig:paraConstraints} Constraints on the effective neutrino mass $|m_{\beta\beta}|$ and the short-range coupling $\epsilon_{1}^{LRL}$ for normal ordering and inverted ordering neutrino masses. The black dashed lines from top to bottom correspond to $\abs{\epsilon_{1}^{LRL}\mathcal{M}^{LR}_{1}}/\abs{\epsilon_{\nu}\mathcal{M}_{\nu}}=10,\; 1,\; 0.1$ respectively. The highlighted regions represent the parameter space allowed by the current bounds $T_{1/2}(^{76}\text{Ge})>1.8\times 10^{26}$ yr~\cite{GERDA:2020xhi}, $T_{1/2}(^{136}\text{Xe})>1.07\times 10^{26}$ yr~\cite{KamLAND-Zen:2016pfg} and the future sensitivities $T_{1/2}(^{76}\text{Ge})>1.3\times 10^{28}$ yr~\cite{LEGEND:2021bnm}, $T_{1/2}(^{136}\text{Xe})>1.35\times 10^{28}$ yr~\cite{nEXO:2021ujk}.
The gray bands denote the regions excluded by the cosmological data $\sum_{i=1}^{3}m_{\nu_{i}}<0.12$ eV at $95\%$ confidence level from the Planck Collaboration~\cite{Planck:2018vyg}, and the oscillation parameters are fixed to the current best fit values~\cite{Esteban:2020cvm}.
}
\end{figure}

We show the constraints on the effective Majorana neutrino mass $\abs{m_{\beta\beta}}$ and the short-range coupling $|\epsilon_{1}^{LRL}|$ in figure~\ref{fig:paraConstraints}, where the values of phase space factor and nuclear matrix elements are adopted from Ref.~\cite{Deppisch:2020ztt}. The highlighted areas denote the allowed regions by the current limits and future sensitivities of the $0\nu\beta\beta$ decay half-life of the isotopes $^{76}\text{Ge}$ and $^{136}\text{Xe}$, where the phase $\alpha-\beta$ freely varies between $0$ to $\pi$. The half-life sensitivity would be increased by a factor of approximately one hundred in upcoming generation of tonne-scale experiments, therefore the constraint on parameter space would be improved considerably, as can be seen from figure~\ref{fig:paraConstraints}. Moreover, we notice that the constraint imposed by $^{136}\text{Xe}$ is a bit more stringent than that of $^{76}\text{Ge}$. Cosmology has played an important role in constraining neutrino properties, and the analysis of cosmological observables can provide an upper bound on a total neutrino mass. Within the standard minimal $\Lambda$CDM model, the present cosmological bound on neutrino mass is $\sum_{i}m_{\nu_i}<0.12$ eV for Planck TT, TE, EE+lowE+lensing+BAO~\cite{Planck:2018vyg}. Taking into account the values of the neutrino mass squared differences and mixing angles measured by the neutrino oscillation experiments~\cite{Esteban:2020cvm}, one can obtain that the effective Majorana neutrino mass is in the region: $|m_{\beta\beta}|\leq31.12$ meV for normal ordering (NO) and  $18.62\,\text{meV}\leq|m_{\beta\beta}|\leq51.14$ meV for inverted ordering (IO). With this constraint the current bounds $T_{1/2}(^{76}\text{Ge})>1.8\times 10^{26}$ yr~\cite{GERDA:2020xhi} and $T_{1/2}(^{136}\text{Xe})>1.07\times 10^{26}$ yr~\cite{KamLAND-Zen:2016pfg} require the coupling $|\epsilon_{1}^{LRL}|$ smaller than $4\times10^{-10}$. It is remarkable that the forthcoming tonne-scale searches for $0\nu\beta\beta$ decays would constrain the parameter space into a narrow band where the short-range mechanism is comparable with the mass mechanism.

\begin{figure}[h]
\centering
\includegraphics[width=\textwidth]{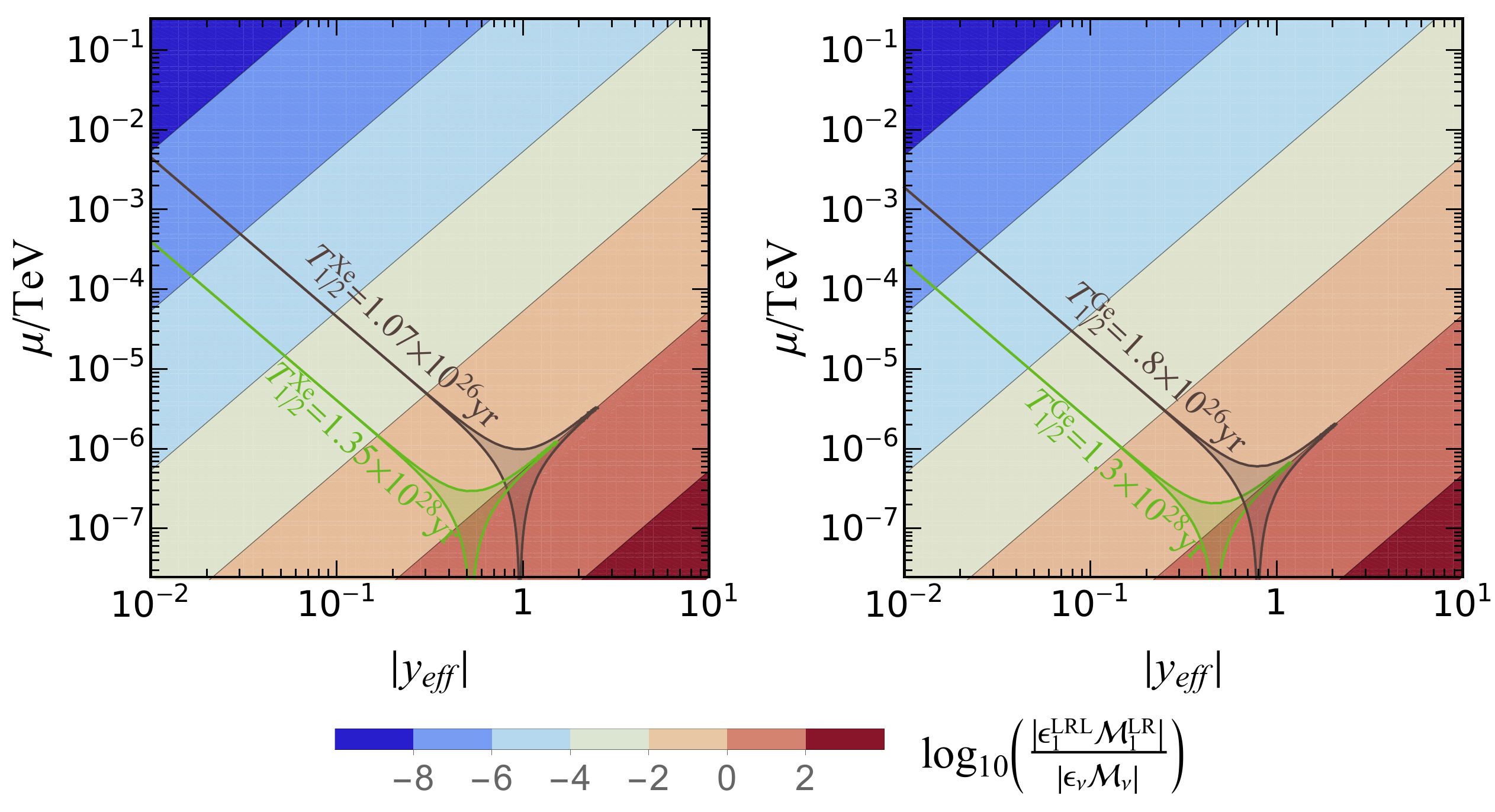}
\caption{\label{fig:frac} The relative size of the short-range mechanism and mass mechanism to the decay rate in the plane $\mu$ versus the couplings $y_{eff}$. The mass mechanism is dominant in the blue region while the short-range mechanism is dominant in the dark red region. The shaded regions denote the contours of the current limit and future sensitivities in ${}^{136}\text{Xe}$ (left) and ${}^{76}\text{Ge}$ (right) $0\nu\beta\beta$ decay search, where the phase $\alpha-\beta$ freely varies between $0$ and $\pi$.}
\end{figure}

We proceed to discuss the relative size of the short-range mechanism and mass mechanism to the decay rate. We take $m_{S_{1}}=m_{S_{2}}=m_{S_{3}}=m_{F_{1}}=m_{F_{2}}=m_{F_{3}}=1$ TeV and the coupling constants  $y_{Ik}=y_{III1}=y_{IIe}=y_{IV1}=y_{Vk}=y_{VIe}\equiv y_{eff}$ to be real. We display the contour region of $\log_{10}\big(\big|\frac{\epsilon^{\text{LRL}}_1\mathcal{M}^{\text{LR}}_1}{\epsilon_{\nu}\mathcal{M}_{\nu}}\big|\big)$ in the plane $\mu$ versus $y_{eff}$ in figure~\ref{fig:frac}, where $\epsilon_{\nu}\equiv m_{\beta\beta}/m_e$. The parameter $\mu$ dictates the mass splitting of $m^2_{S_{n1}}$ and $m^2_{S_{n2}}$, as shown in Eq.~\eqref{eq:scalar-mass}. If $\mu$ is quite small with $\mu/\text{TeV}\ll1$, $m^2_{S_{n1}}$ and $m^2_{S_{n2}}$ tend to be degenerate so that cancellation between two terms of the light neutrino mass matrix in Eq.~\eqref{eq:nu-mass-model} occurs. As a result, the mass mechanism is generally sub-dominant to the short-range contribution for $\mu/\text{TeV}\ll1$. Furthermore, from Eq.~\eqref{eq:epsilon1-LRL} and Eq.~\eqref{eq:nu-mass-model} we see $\epsilon_{1}^{LRL}\propto Y^6_{eff}$ and $(m_{\nu})_{\alpha\beta}\propto Y^4_{eff}$. Therefore the mass mechanism is dominant in the region $y_{eff}\ll1$ while the short-range mechanism is dominant in the region $y_{eff}>1$.
%\textcolor{red}{(The relation between $\epsilon_{\nu}$, $m_{\beta\beta}$ and %$(m_{\nu})_{ee}$?)}


\section{\label{sec:conclusion}Conclusion}

The $0\nu\beta\beta$ decay is the most sensitive experiments to establish that the massive neutrinos are Majorana or Dirac particles. If the signal of $0\nu\beta\beta$ decay is observed in the future, lepton number would be violated by two units and neutrinos must be Majorana particles. The rare $0\nu\beta\beta$ decay process might also provide important information on the neutrino mass spectrum, the Majorana phases and the absolute scale of neutrino masses. On the other hand, if neutrinos are Majorana particles, $0\nu\beta\beta$ decay can be mediated by exchange of a light neutrino between two charged current vertices, and it is the so-called mass mechanism. Additional contributions from short-range mechanism is also possible, where the exchanged particles are much heavier than the $0\nu\beta\beta$ decay energy scale typically 100 MeV, and they can be represented by non-renormalizable contact interactions with six external fermions.

In the present work, we have systematically studied the possible one-loop UV completions of dimension-9 short-range $0\nu\beta\beta$ decay operators in Eq.~\eqref{eqn-basis-summary-dim9}. The general procedure to be followed involve three steps: firstly generate one-loop topologies with six legs, then promote topology to diagram by specifying the Lorentz nature of fields, finally attach fields to external lines and determine the SM quantum numbers of internal particles by gauge invariance. Notice that there are infinite possibilities for the quantum numbers of the internal fermions and scalars, thus a given diagram can lead to an infinite number of UV completions. To be more specific, excluding the topologies with tadpole and self-energy, there are only 48 one-loop connected topologies with three-point and four-point vertices and six external lines, as shown in figure~\ref{fig:N0-1}, figure~\ref{fig:NNormalize} and figure~\ref{fig:NGenuineT}. Only 11 of them in figure~\ref{fig:N0-1} and figure~\ref{fig:NGenuineT} can accommodate 6 external fermion lines using renormalizable interactions. The 5 topologies in figure~\ref{fig:NGenuineT} contain a one-loop sub-diagram which can be compressed into a three-point vertex so that the more important tree-level contributions can not be forbidden. As a result, it is sufficient to only consider the 6 one-loop topologies in figure~\ref{fig:N0-1} which lead to 8 one-loop genuine diagrams shown in figure~\ref{fig:DiaN0-1}. One can determine the transformations of the internal fields under SM gauge group by imposing gauge invariance at each vertex,
the $SU(3)_C$, $SU(2)_L$ representations and hypercharge $U(1)_Y$ assignments can be unambiguously fixed in tree-level realizations while there are an infinite number of UV completions at one-loop level. The complete list of the models together with the internal fields and their quantum numbers are given in the supplementary file~\cite{Chen:2021sup}, where the $SU(2)_L$ and $SU(3)_C$ representations with dimensions less than 4 and 10 respectively are considered.

In order to guarantee the dominance of the one-loop contribution to $0\nu\beta\beta$ decay, the mediators of the tree diagrams should be absent in the one-loop models. As a consequence, certain values of the hypercharge parameter $\alpha$ are excluded. The $0\nu\beta\beta$ decay models violate lepton number, thus they can also generate Majorana neutrino masses, as is required by the black box theorem~\cite{Schechter:1981bd}. Therefore both short-range contribution and mass mechanism are present in these models. We have discussed the Majorana neutrino masses up to two-loop level which can be generated from the mediators of the $0\nu\beta\beta$ decay models.
Furthermore, we give a concrete one-loop $0\nu\beta\beta$ decay model in which the neutrino masses are generated at 2-loop level. The predictions for the neutrino mass matrix and the half-life of $0\nu\beta\beta$ decay are presented. The short-range contribution can be as important as the mass mechanism in some region of parameter space.

There are many experiments that are searching for $0\nu\beta\beta$ decay or are in various stages of planning and construction over the world. We expect it would become an important probe for physics beyond the SM. If the observation of $0\nu\beta\beta$ decay experiments is in conflict with neutrino mass ordering measured by neutrino oscillation experiments and the neutrino mass bounds from $\beta$ decay and cosmology in the future, the new contributions beyond the mass mechanism would be necessary. Then our results can be used to construct short-range $0\nu\beta\beta$ decay models. Furthermore, generally a few messenger fields are required in the one-loop $0\nu\beta\beta$ decay models, it is interesting to study the phenomenological implications of these new fields in collider and cosmology, and this is left for future work.

\section*{Acknowledgements}

PTC and GJD are supported by the National Natural Science Foundation of China under Grant Nos 11975224, 11835013 and 11947301 and the Key Research Program of the Chinese Academy of Sciences under Grant NO. XDPB15. CYY is supported in part by the Grants No.~NSFC-11975130, No.~NSFC-12035008, No.~NSFC-12047533, by the National Key Research and Development Program of China under Grant No. 2017YFA0402200 and the China Post-doctoral Science Foundation under Grant No. 2018M641621.

\newpage

\section*{Appendix}

\begin{appendix}

\section{\label{sec:0nubb-eff-operators-LEFT} $0\nu\beta\beta$ decay operators below the electroweak scale and half-life }

The SM gauge symmetry $SU(3)_C\times SU(2)_L\times U(1)_Y$ is spontaneous broken down to $SU(3)_C\times U(1)_{EM}$. The leading operators which contribute to $0\nu\beta\beta$ decay at short distance involve four quark fields and two charged leptons, they respect $SU(3)_{C}\times U(1)_{EM}$ symmetry and can be written as the product of three fermion currents~\cite{Pas:2000vn,Deppisch:2020ztt}
%in terms of leptonic and hadronic currents as~\cite{Pas:2000vn}
\begin{align}
\nonumber&(\mathcal{O}_{1})_{\{LL\}L}= J_{L}J_{L}j_{L},& &(\mathcal{O}_{1})_{\{LL\}R}=J_{L}J_{L}j_{R}\,,\\
\nonumber&(\mathcal{O}_{1})_{\{RR\}L}= J_{R}J_{R}j_{L},& &(\mathcal{O}_{1})_{\{RR\}R}=J_{R}J_{R}j_{R}\,,\\
\nonumber&(\mathcal{O}_{1})_{\{LR\}L}= J_{L}J_{R}j_{L},& &(\mathcal{O}_{1})_{\{LR\}R}=J_{L}J_{R}j_{R}\,,\\
\nonumber&(\mathcal{O}_{2})_{\{LL\}L}=J^{\mu\nu}_{L}J_{L\mu\nu}j_{L},&  &(\mathcal{O}_{2})_{\{LL\}R} =J^{\mu\nu}_{L}J_{L\mu\nu}j_{R}\,,\\
\nonumber&(\mathcal{O}_{2})_{\{RR\}L}=J^{\mu\nu}_{R}J_{R\mu\nu}j_{L},&  &(\mathcal{O}_{2})_{\{RR\}R} =J^{\mu\nu}_{R}J_{R\mu\nu}j_{R}\,,\\
\nonumber&(\mathcal{O}_{3})_{\{LL\}L}= J^{\mu}_{L}J_{L\mu}j_{L},& &(\mathcal{O}_{3})_{\{LL\}R}=J^{\mu}_{L}J_{L\mu}j_{R}\,,\\
\nonumber&(\mathcal{O}_{3})_{\{RR\}L}= J^{\mu}_{R}J_{R\mu}j_{L},& &(\mathcal{O}_{3})_{\{RR\}R}=J^{\mu}_{R}J_{R\mu}j_{R}\,,\\
\nonumber&(\mathcal{O}_{3})_{\{LR\}L}= J^{\mu}_{L}J_{R\mu}j_{L},& &(\mathcal{O}_{3})_{\{LR\}R}=J^{\mu}_{L}J_{R\mu}j_{R}\,,\\
\nonumber&(\mathcal{O}_{4})_{LL}=J_{L}^{\mu}J_{L\mu\nu}j^{\nu},& &(\mathcal{O}_{4})_{RR}=J_{R}^{\mu}J_{R\mu\nu}j^{\nu}\,,\\
\nonumber&(\mathcal{O}_{4})_{LR}=J_{L}^{\mu}J_{R\mu\nu}j^{\nu},& &(\mathcal{O}_{4})_{RL}=J_{R}^{\mu}J_{L\mu\nu}j^{\nu}\,, \\
\nonumber&(\mathcal{O}_{5})_{LL}=J_{L}J_{L}^{\mu}j_{\mu},& &(\mathcal{O}_{5})_{RR}=J_{R}J_{R}^{\mu}j_{\mu}\,,\\
\label{eq:0nubb-LEFT}&(\mathcal{O}_{5})_{LR}=J_{L}J_{R}^{\mu}j_{\mu},& &(\mathcal{O}_{5})_{RL}=J_{R}J_{L}^{\mu}j_{\mu}\,.
\end{align}
Here the normalization factor $\frac{G^2_F\cos^2\theta_C}{2m_P}$ has been omitted, where $m_P$ is the proton mass and $\theta_C$ is the Cabibbo angle. The factor $\cos^2\theta_C$ is included so that the phase space factor of short-range contribution can be defined in the same way as that for the standard light neutrino exchange~\cite{Deppisch:2020ztt}.  The hadronic current $J$ and leptonic current $j$ are bilinears of quarks and electrons,
\begin{eqnarray}
\nonumber J_{R,L}&=&\bar{u}_{a}(1\pm\gamma_{5})d_{a},\quad
J_{R,L}^{\mu}=\bar{u}_{a}\gamma^{\mu}(1\pm\gamma_{5})d_{a},\quad
J_{R,L}^{\mu\nu}=\bar{u}_{a}\sigma^{\mu\nu}(1\pm\gamma_{5})d_{a},\\
&&\qquad \qquad j_{R,L}=\bar{e}(1\mp\gamma_{5})e^{c},\quad j^{\mu}=\bar{e}\gamma^{\mu}\gamma_{5}e^{c}\,,
   \label{eq:current}
\end{eqnarray}
where $a=1, 2, 3$ is $SU(3)_{C}$ color indices and $\sigma^{\mu\nu}=\frac{i}{2}[\gamma^{\mu},\gamma^{\nu}]$.
The chirality assignment in the charged lepton current $j_{R,L}$ is flipped, and the subscripts $L$ and $R$ are associated with $1+\gamma_5$ and $1-\gamma_5$ respectively. Thus the operators $j_R$ and $j_L$ describes the creation of two right-handed and left-handed electrons respectively. Notice that using the Fierz identity one can show that both operators $(\mathcal{O}_{2})_{\{LR\}L}=J^{\mu\nu}_{L}J_{R\mu\nu}j_{L}$ and $(\mathcal{O}_{2})_{\{LR\}R} =J^{\mu\nu}_{L}J_{R\mu\nu}j_{R}$ are vanishing. Hence the most general short-range interaction Lagrangian responsible for $0\nu\beta\beta$ decay can be written as
\begin{equation}
\mathcal{L}_\text{short}=\frac{G^2_F\cos^2\theta_C}{2m_P}\sum_{\chi_1, \chi_2, \chi}\left[\sum^3_{i=1}\epsilon_i^{\chi_1\chi_2\chi}(\mathcal{O}_{i})_{\{\chi_1\chi_2\}\chi}+
\sum^{5}_{i=4}\epsilon_i^{\chi_1\chi_2}(\mathcal{O}_{i})_{\chi_1\chi_2}\right]+ \mathrm{H.c.}\,,
\end{equation}
where the indices $\chi_1, \chi_2, \chi$ are $L$ or $R$ and they denote the chiralities of the quark and electron currents in the order of appearance in Eq.~\eqref{eq:0nubb-LEFT}. Under the presence of both standard mass mechanism and the short-range mechanism, the inverse $0\nu\beta\beta$ decay half-life time can be expressed as~\cite{Deppisch:2020ztt}
\begin{eqnarray}
\nonumber T_{1/2}^{-1} &=& G_{11+}^{(0)}\abs{
\sum_{I=1}^{3}\epsilon_{I}^{L}\mathcal{M}_{I}
+\epsilon_{\nu}\mathcal{M}_{\nu}}^{2}+G_{11+}^{(0)}\abs{\sum_{I=1}^{3}\epsilon_{I}^{R}\mathcal{M}_{I}}^{2}
+G_{66}^{(0)}\abs{\sum_{I=4}^{5}\epsilon_{I}\mathcal{M}_{I}}^{2}\\
\nonumber&&+G_{11-}^{(0)}\times2\Re\left[\left(\sum_{I=1}^{3}\epsilon_{I}^{L}\mathcal{M}_{I}
+\epsilon_{\nu}\mathcal{M}_{\nu}\right)\left(\sum_{I=1}^{3}\epsilon_{I}^{R}\mathcal{M}_{I}\right)^{*}\right]\\
\label{eq:inverse-half-life}&&+G_{16}^{(0)}\times2\Re\left[\left(\sum_{I=1}^{3}\epsilon_{I}^{L}\mathcal{M}_{I}-\sum_{I=1}^{3}\epsilon_{I}^{R}\mathcal{M}_{I}
+\epsilon_{\nu}\mathcal{M}_{\nu}\right)
\left(\sum_{I=4}^{5}\epsilon_{I}\mathcal{M}_{I}\right)^{*}\right]\,,
\end{eqnarray}
where $\epsilon_{\nu}\equiv m_{\beta\beta}/m_e$, the summation is over the different types of short-range contributions including different chiralities with $I=(i, \chi_1\chi_2)$, $i=1, \ldots, 5$ and $\chi_1, \chi_2\in\left\{L, R\right\}$. The notations $\epsilon_{I}^{L}$ and $\epsilon_{I}^{R}$ stand for the coefficients $\epsilon^{\chi_1\chi_2L}_{i}$ and $\epsilon^{\chi_1\chi_2R}_{i}$ respectively when $i=1, 2, 3$. Moreover, $G_{11+}^{(0)}$, $G_{66}^{(0)}$, $G_{11-}^{(0)}$, $G_{16}^{(0)}$ are the phase space factors, $\mathcal{M}_{\nu}$ and $\mathcal{M}_{I}$ are the nuclear matrix elements for  the mass mechanism and the short-range operators respectively, their numerical values within the interacting boson model are listed in Ref.~\cite{Deppisch:2020ztt}.

%and $\epsilon^{L}_{I}$($\epsilon^{R}_{I}$) stand for %$\epsilon^{XYL}_{I}$($\epsilon^{XYR}_{I}$).

As shown in Eq.~\eqref{eqn-basis-summary-dim9}, there are eleven $0\nu\beta\beta$ decay operators invariant under the SM gauge symmetry  $SU(3)_C \times  SU(2)_L \times U(1)_Y$ and twenty-four $SU(3)_C \times U(1)_{EM}$ invariant operators at dimension $d=9$. The matching between the $0\nu\beta\beta$ decay operators at the electroweak scale is listed in table~\ref{tab:SMEFT-to-LEFT1}. Multiplying the dimension-9 operators in Eq.~\eqref{eqn-basis-summary-dim9} by $H^{\dagger}H$ and $(H^{\dagger}H)^2$,
one can obtain the remaining thirteen low-energy dimension-9 operators after electroweak symmetry breaking. Following the general procedures listed in section~\ref{sec:generate}, the possible UV completions of the above effective operators in Eq.~\eqref{eq:0nubb-LEFT} can be straightforwardly found out.

\begin{table}[hptb]
\renewcommand{\tabcolsep}{0.5mm}
\renewcommand{\arraystretch}{1.3}
\centering
\begin{tabular}{|c|c|c|}\hline\hline
%$0\nu\beta\beta$ operators above the electroweak scale
%&      Low energy operator in \cite{Graesser:2016bpz}
%&      $0\nu\beta\beta$ operators below the electroweak scale \\ %\cite{Bonnet:2012kh}\\
%\hline
Above the electroweak scale     &  Below the electroweak scale \\ \hline
$\mathcal{O}^{0\nu}_{1}$        & $\frac{1}{8}(\mathcal{O}_{3})_{\{LR\}L}$\\ \hline
$\mathcal{O}^{0\nu}_{2}$  &                     $-\frac{1}{4}(\mathcal{O}_{1})_{\{LR\}L} -\frac{1}{24}(\mathcal{O}_{3})_{\{LR\}L}$\\ \hline
$\mathcal{O}^{0\nu}_{3}$  & $\frac{1}{8}(\mathcal{O}_{1})_{\{LL\}L}$\\ \hline
$\mathcal{O}^{0\nu}_{4}$        &       $-\frac{5}{24}(\mathcal{O}_{1})_{\{LL\}L}
-\frac{1}{32}(\mathcal{O}_{2})_{\{LL\}L}$\\ \hline
$\mathcal{O}^{0\nu}_{5}$                & $\frac{1}{8}(\mathcal{O}_{1})_{\{RR\}L}$\\ \hline
$\mathcal{O}^{0\nu}_{6}$        & $-\frac{5}{24}(\mathcal{O}_{1})_{\{RR\}L}
-\frac{1}{32}(\mathcal{O}_{2})_{\{RR\}L}$\\ \hline
$\mathcal{O}^{0\nu}_{7}$        & $\frac{1}{8}(\mathcal{O}_{3})_{\{RR\}R}$\\ \hline
$\mathcal{O}^{0\nu}_{8}$  & $\frac{1}{8}(\mathcal{O}_{5})_{RR}$\\ \hline
$\mathcal{O}^{0\nu}_{9}$        & $\frac{i}{8}(\mathcal{O}_{4})_{RR} -\frac{5}{24}(\mathcal{O}_{5})_{RR}$ \\    \hline
$\mathcal{O}^{0\nu}_{10}$       &       $\frac{1}{8}(\mathcal{O}_{5})_{RL}$\\ \hline
$\mathcal{O}^{0\nu}_{11}$       & $-\frac{i}{8}(\mathcal{O}_{4})_{RL} -\frac{5}{24}(\mathcal{O}_{5})_{RL}$\\ \hline\hline
\end{tabular}
\caption{\label{tab:SMEFT-to-LEFT1} Matching between the $0\nu\beta\beta$ decay operators above the electroweak scale and those below the electroweak scale. }
\end{table}

\section{\label{app:non-renorm-topology}Non-renormalizable topologies and finite non-genuine diagrams}

\begin{figure}[hptb]
\centering
\includegraphics[width=\textwidth]{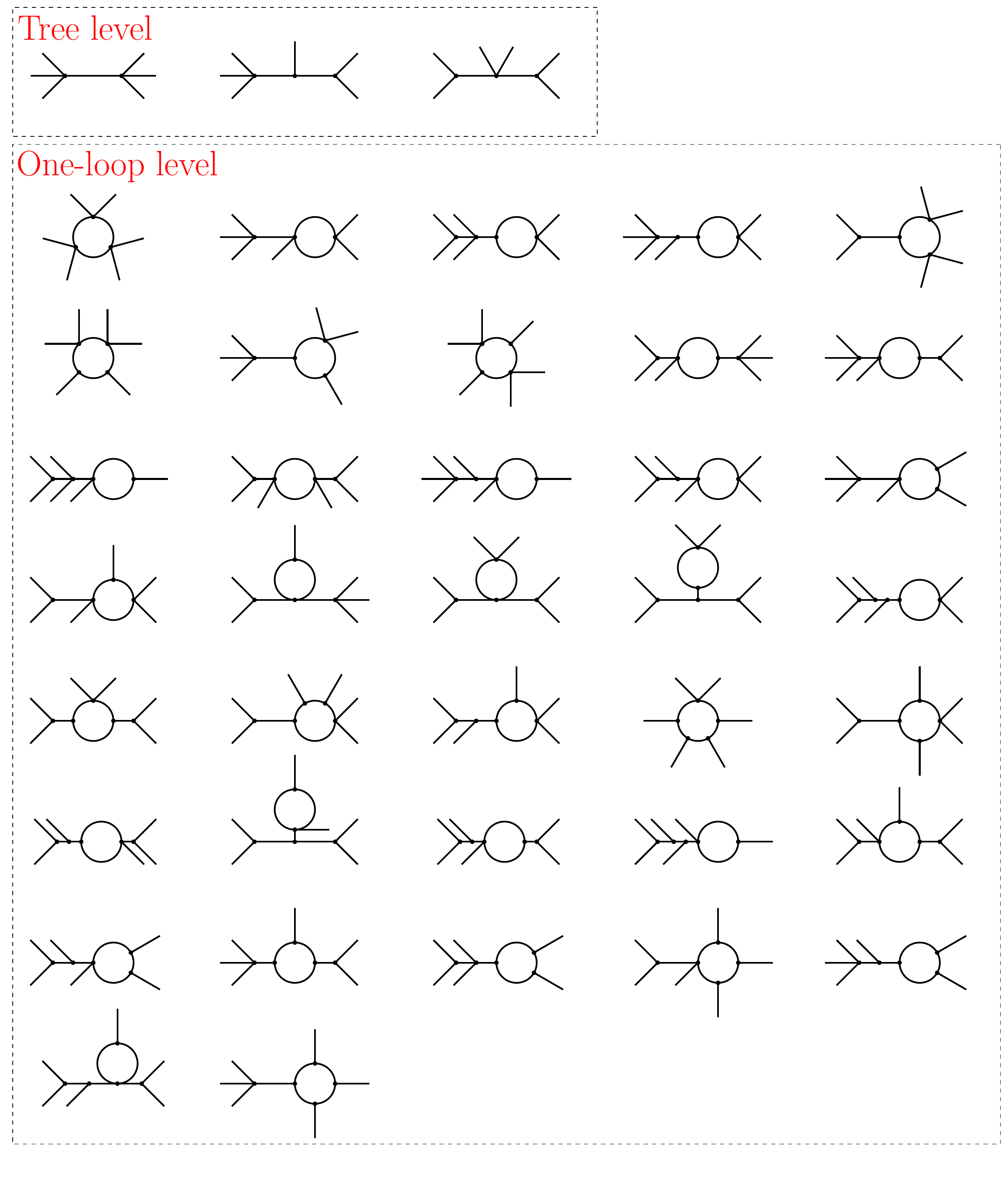}
\caption{\label{fig:NNormalize}The non-renormalizable tree-level and one-loop topologies.}
\end{figure}

\begin{figure}[hptb]
\centering
\includegraphics[width=0.65\textwidth]{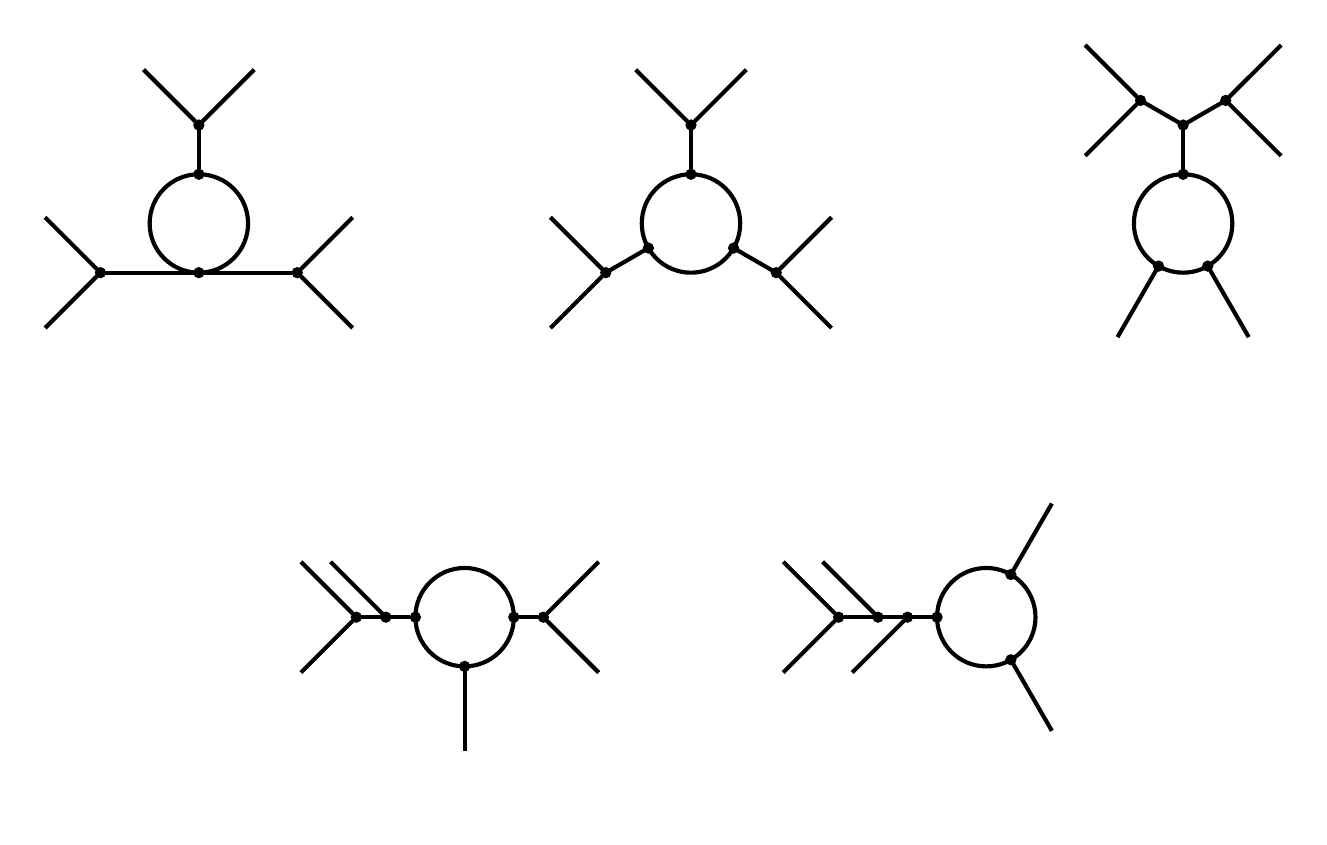}
\caption{\label{fig:NGenuineT}The one-loop topologies leading to non-genuine diagrams.}
\end{figure}

\begin{figure}[hptb]
\centering
\includegraphics[width=0.85\textwidth]{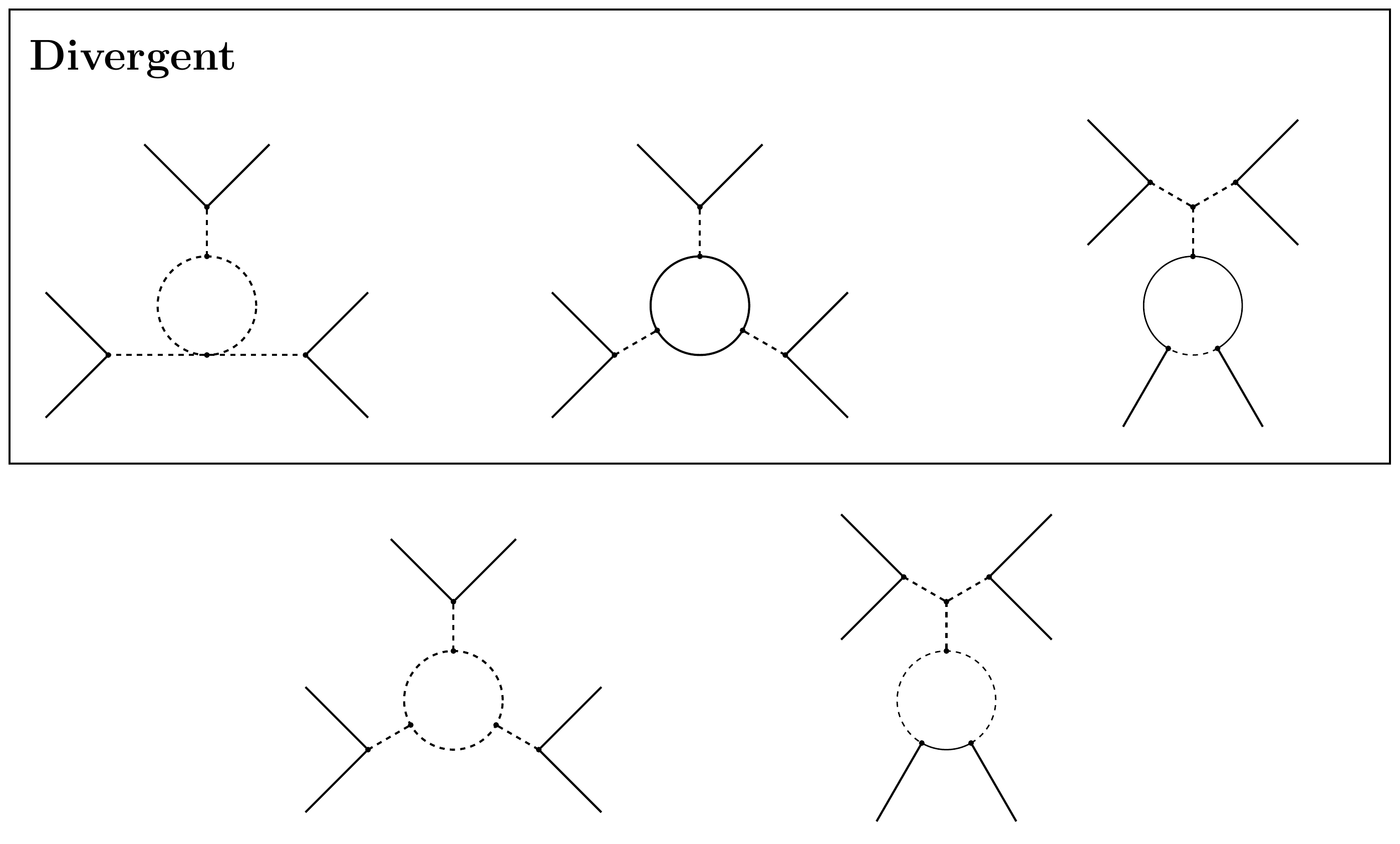}
\caption{\label{fig:NGenuineD1} Non-genuine one-loop diagrams which correspond to the one-loop generation of one of the couplings in the tree-level diagram N-0-1-1(see figure \ref{fig:DiaN0-1}). The diagrams in the box are divergent. }
\end{figure}

\begin{figure}[hptb]
\centering
\includegraphics[width=0.65\textwidth]{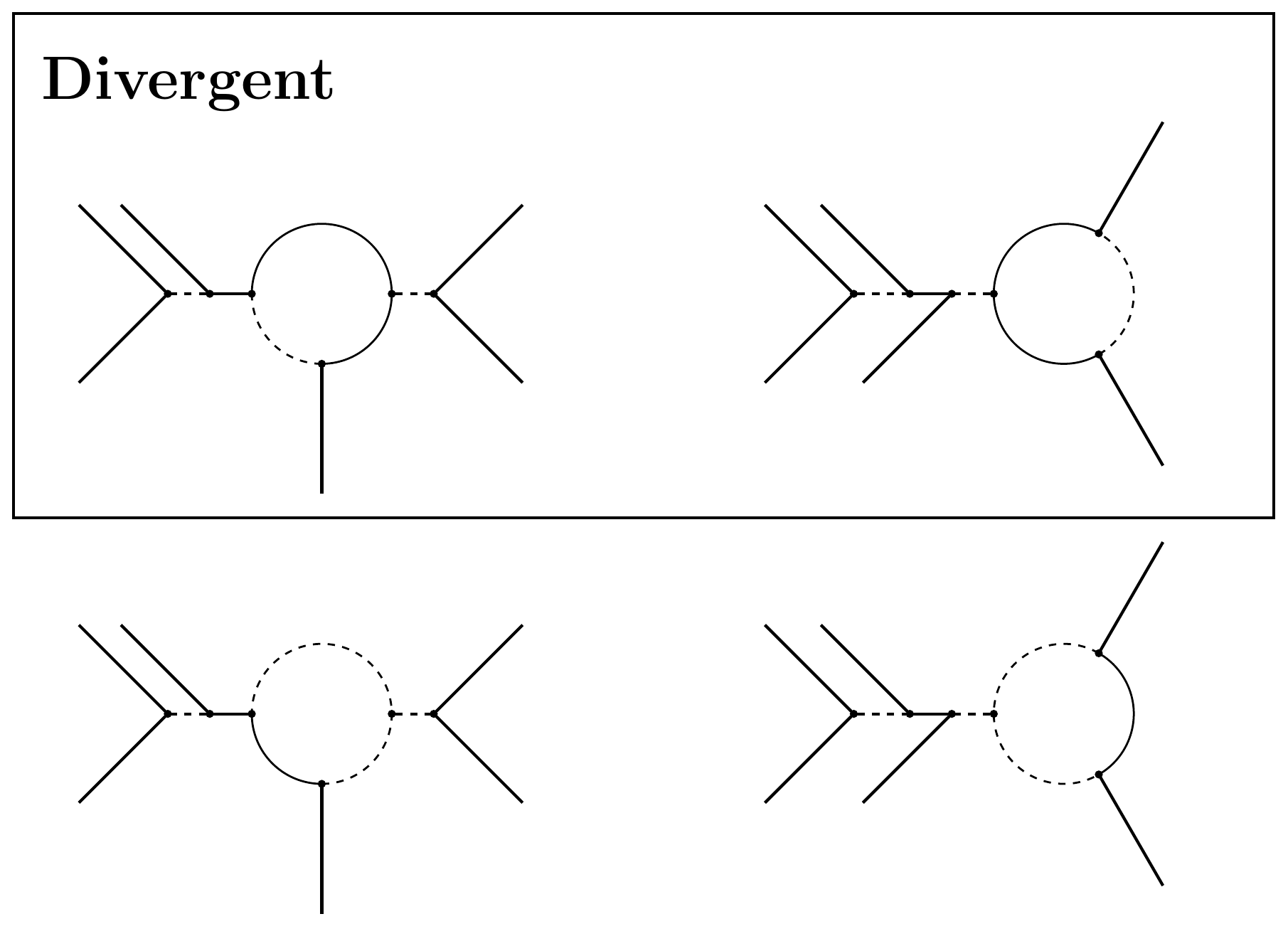}
\caption{\label{fig:NGenuineD2} Non-genuine one-loop diagrams which correspond to the one-loop generation of one of the couplings in the tree-level diagram N-0-2-1 (see figure~\ref{fig:DiaN0-1}). The diagrams in the box are divergent.}
\end{figure}

For completeness, we list the topologies which always lead to the non-renormalizable diagrams in figure~\ref{fig:NNormalize} as well as the non-genuine topologies in figure~\ref{fig:NGenuineT}. The non-genuine model arising from the topologies in figure~\ref{fig:NGenuineT} can be regarded as extensions of the tree-level $0\nu\beta\beta$ decay models, and one of the vertices is generated at one-loop level, as illustrated in figure~\ref{fig:NGenuineD1} and figure~\ref{fig:NGenuineD2}.


\section{\label{app:example-models}Possible UV completions for the $0\nu\beta\beta$ decay operator N4 based on the diagram N-1-1-1  }

In this appendix, we present the possible one-loop models which generate the $0\nu\beta\beta$ decay operator N4, and the models are based on the diagram N-1-1-1. The SM quantum numbers of the messenger fields are listed in tables~\ref{tab:N41111}-\ref{tab:N41115}.

\begin{table}[htbp]
        \renewcommand{\tabcolsep}{0.5mm}
        \renewcommand{\arraystretch}{1.3}
        \centering
        \begin{tabular}{|c|c|c|c|c|c||c|c|c|c|c||c|c|c|c|c|}
        \hline\hline
        \multirow{2}{*}{Model Name} &
        \multicolumn{5}{c||}{$U(1)_{Y}$}& \multicolumn{5}{c||}{$SU(2)_{L}$} & \multicolumn{5}{c|}{$SU(3)_{C}$}\\ \cline{2-16}
        & $F_{1}$ & $S_{1}$ & $S_{2}$ & $S_{3}$ & $S_{4}$ &
        $F_{1}$ & $S_{1}$ & $S_{2}$ & $S_{3}$ & $S_{4}$ &
        $F_{1}$ & $S_{1}$ & $S_{2}$ & $S_{3}$ & $S_{4}$\\ \hline
        N4-1-1-1-1-1-1 & \multirow{27}{*}{$-\frac{1}{3}+\alpha$} & \multirow{27}{*}{$\alpha$} & \multirow{27}{*}{$\frac{2}{3}-\alpha$} & \multirow{27}{*}{$-\frac{1}{3}$} & \multirow{27}{*}{$-\frac{1}{3}$} &
        \multirow{9}{*}{$1$} & \multirow{9}{*}{$1$} & \multirow{9}{*}{$1$} & \multirow{9}{*}{$1$} & \multirow{9}{*}{$1$} &
        $\bar{3}$ & $3$ & $1$ & $3$ & $3$\\ \cline{1-1}\cline{12-16}
        N4-1-1-1-1-1-2 &  &  &  &  &  &  &  &  &  &  &
        $\bar{3}$ & $\bar{6}$ & $1$ & $3$ & $3$\\ \cline{1-1}\cline{12-16}
        N4-1-1-1-1-1-3 &  &  &  &  &  &  &  &  &  &  &
        $3$ & $8$ & $3$ & $3$ & $3$\\ \cline{1-1}\cline{12-16}
        N4-1-1-1-1-1-4 &  &  &  &  &  &  &  &  &  &  &
        $\bar{6}$ & $8$ & $3$ & $3$ & $3$\\ \cline{1-1}\cline{12-16}
        N4-1-1-1-1-1-5 &  &  &  &  &  &  &  &  &  &  &
        $1$ & $\bar{3}$ & $\bar{3}$ & $3$ & $3$\\ \cline{1-1}\cline{12-16}
        N4-1-1-1-1-1-6 &  &  &  &  &  &  &  &  &  &  &
        $8$ & $\bar{3}$ & $\bar{3}$ & $3$ & $3$\\ \cline{1-1}\cline{12-16}
        N4-1-1-1-1-1-7 &  &  &  &  &  &  &  &  &  &  &
        $8$ & $6$ & $\bar{3}$ & $3$ & $3$\\ \cline{1-1}\cline{12-16}
        N4-1-1-1-1-1-8 &  &  &  &  &  &  &  &  &  &  &
        $3$ & $8$ & $\bar{6}$ & $3$ & $3$\\ \cline{1-1}\cline{12-16}
        N4-1-1-1-1-1-9 &  &  &  &  &  &  &  &  &  &  &
        $8$ & $6$ & $6$ & $3$ & $3$\\ \cline{1-1}\cline{7-16}
        N4-1-1-1-1-2-1 &  &  &  &  &  &
        \multirow{9}{*}{$2$} & \multirow{9}{*}{$2$} & \multirow{9}{*}{$2$} & \multirow{9}{*}{$1$} & \multirow{9}{*}{$1$} &
        $\bar{3}$ & $3$ & $1$ & $3$ & $3$\\ \cline{1-1}\cline{12-16}
        N4-1-1-1-1-2-2 &  &  &  &  &  &  &  &  &  &  &
        $\bar{3}$ & $\bar{6}$ & $1$ & $3$ & $3$\\ \cline{12-16}
        N4-1-1-1-1-2-3 &  &  &  &  &  &  &  &  &  &  &
        $3$ & $8$ & $3$ & $3$ & $3$\\ \cline{1-1}\cline{12-16}
        N4-1-1-1-1-2-4 &  &  &  &  &  &  &  &  &  &  &
        $\bar{6}$ & $8$ & $3$ & $3$ & $3$\\ \cline{1-1}\cline{12-16}
        N4-1-1-1-1-2-5 &  &  &  &  &  &  &  &  &  &  &
        $1$ & $\bar{3}$ & $\bar{3}$ & $3$ & $3$\\ \cline{1-1}\cline{12-16}
        N4-1-1-1-1-2-6 &  &  &  &  &  &  &  &  &  &  &
        $8$ & $\bar{3}$ & $\bar{3}$ & $3$ & $3$\\ \cline{1-1}\cline{12-16}
        N4-1-1-1-1-2-7 &  &  &  &  &  &  &  &  &  &  &
        $8$ & $6$ & $\bar{3}$ & $3$ & $3$\\ \cline{1-1}\cline{12-16}
        N4-1-1-1-1-2-8 &  &  &  &  &  &  &  &  &  &  &
        $3$ & $8$ & $\bar{6}$ & $3$ & $3$\\ \cline{1-1}\cline{12-16}
        N4-1-1-1-1-2-9 &  &  &  &  &  &  &  &  &  &  &
        $8$ & $6$ & $6$ & $3$ & $3$\\ \cline{1-1}\cline{7-16}
        N4-1-1-1-1-3-1 &  &  &  &  &  &
        \multirow{9}{*}{$3$} & \multirow{9}{*}{$3$} & \multirow{9}{*}{$3$} & \multirow{9}{*}{$1$} & \multirow{9}{*}{$1$} &
        $\bar{3}$ & $3$ & $1$ & $3$ & $3$\\ \cline{1-1}\cline{12-16}
        N4-1-1-1-1-3-2 &  &  &  &  &  &  &  &  &  &  &
        $\bar{3}$ & $\bar{6}$ & $1$ & $3$ & $3$\\ \cline{1-1}\cline{12-16}
        N4-1-1-1-1-3-3 &  &  &  &  &  &  &  &  &  &  &
        $3$ & $8$ & $3$ & $3$ & $3$\\ \cline{1-1}\cline{12-16}
        N4-1-1-1-1-3-4 &  &  &  &  &  &  &  &  &  &  &
        $\bar{6}$ & $8$ & $3$ & $3$ & $3$\\ \cline{1-1}\cline{12-16}
        N4-1-1-1-1-3-5 &  &  &  &  &  &  &  &  &  &  &
        $1$ & $\bar{3}$ & $\bar{3}$ & $3$ & $3$\\ \cline{1-1}\cline{12-16}
        N4-1-1-1-1-3-6 &  &  &  &  &  &  &  &  &  &  &
        $8$ & $\bar{3}$ & $\bar{3}$ & $3$ & $3$\\ \cline{1-1}\cline{12-16}
        N4-1-1-1-1-3-7 &  &  &  &  &  &  &  &  &  &  &
        $8$ & $6$ & $\bar{3}$ & $3$ & $3$\\ \cline{1-1}\cline{12-16}
        N4-1-1-1-1-3-8 &  &  &  &  &  &  &  &  &  &  &
        $3$ & $8$ & $\bar{6}$ & $3$ & $3$\\ \cline{1-1}\cline{12-16}
        N4-1-1-1-1-3-9 &  &  &  &  &  &  &  &  &  &  &
        $8$ & $6$ & $6$ & $3$ & $3$\\ \hline\hline %\cline{7-16}
        \end{tabular}
\caption{\label{tab:N41111} The one-loop models for the $0\nu\beta\beta$ decay operator N4 arising from the diagram N4-1-1-1-1 shown in figure~\ref{fig:external-N4}. }
\end{table}
\begin{table}[htbp]
        \renewcommand{\tabcolsep}{0.5mm}
        \renewcommand{\arraystretch}{1.3}
        \centering
        \begin{tabular}{|c|c|c|c|c|c||c|c|c|c|c||c|c|c|c|c|}
                \hline\hline
                \multirow{2}{*}{Model Name} &
                \multicolumn{5}{c||}{$U(1)_{Y}$}& \multicolumn{5}{c||}{$SU(2)_{L}$} & \multicolumn{5}{c|}{$SU(3)_{C}$}\\ \cline{2-16}
                & $F_{1}$ & $S_{1}$ & $S_{2}$ & $S_{3}$ & $S_{4}$ &
                $F_{1}$ & $S_{1}$ & $S_{2}$ & $S_{3}$ & $S_{4}$ &
                $F_{1}$ & $S_{1}$ & $S_{2}$ & $S_{3}$ & $S_{4}$\\ \hline
                N4-1-1-1-2-1-1 & \multirow{27}{*}{$-\frac{2}{3}+\alpha$} & \multirow{27}{*}{$\alpha$} & \multirow{27}{*}{$-\frac{1}{3}-\alpha$} & \multirow{27}{*}{$-\frac{1}{3}$} & \multirow{27}{*}{$\frac{2}{3}$} &
                \multirow{9}{*}{$1$} & \multirow{9}{*}{$1$} & \multirow{9}{*}{$1$} & \multirow{9}{*}{$1$} & \multirow{9}{*}{$1$} &
                $1$ & $3$ & $1$ & $3$ & $\bar{6}$\\ \cline{1-1}\cline{12-16}
                N4-1-1-1-2-1-2 &  &  &  &  &  &  &  &  &  &  &
                $\bar{3}$  & $1$ & $3$ & $3$& $\bar{6}$\\ \cline{1-1}\cline{12-16}
                N4-1-1-1-2-1-3 &  &  &  &  &  &  &  &  &  &  &
                $\bar{3}$ & $8$ & $3$ & $3$ & $\bar{6}$\\ \cline{1-1}\cline{12-16}
                N4-1-1-1-2-1-4 &  &  &  &  &  &  &  &  &  &  &
                $3$ & $\bar{3}$ & $\bar{3}$ & $3$ & $\bar{6}$\\ \cline{1-1}\cline{12-16}
                N4-1-1-1-2-1-5 &  &  &  &  &  &  &  &  &  &  &
                $3$ & $6$ & $\bar{3}$ & $3$ & $\bar{6}$\\ \cline{1-1}\cline{12-16}
                N4-1-1-1-2-1-6 &  &  &  &  &  &  &  &  &  &  &
                $6$ & $8$ & $\bar{6}$ & $3$ & $\bar{6}$\\ \cline{1-1}\cline{12-16}
                N4-1-1-1-2-1-7 &  &  &  &  &  &  &  &  &  &  &
                $\bar{6}$ & $\bar{3}$ & $6$ & $3$ & $\bar{6}$\\ \cline{1-1}\cline{12-16}
                N4-1-1-1-2-1-8 &  &  &  &  &  &  &  &  &  &  &
                $8$ & $3$ & $8$ & $3$ & $\bar{6}$\\ \cline{1-1}\cline{12-16}
                N4-1-1-1-2-1-9 &  &  &  &  &  &  &  &  &  &  &
                $8$ & $\bar{6}$ & $8$ & $3$ & $\bar{6}$\\ \cline{1-1}\cline{7-16}
                N4-1-1-1-2-2-1 &  &  &  &  &  &
                \multirow{9}{*}{$2$} & \multirow{9}{*}{$2$} & \multirow{9}{*}{$2$} & \multirow{9}{*}{$1$} & \multirow{9}{*}{$1$} &
                $1$ & $3$ & $1$ & $3$ & $\bar{6}$\\ \cline{1-1}\cline{12-16}
                N4-1-1-1-2-2-2 &  &  &  &  &  &  &  &  &  &  &
                $\bar{3}$  & $1$ & $3$ & $3$& $\bar{6}$\\ \cline{1-1}\cline{12-16}
                N4-1-1-1-2-2-3 &  &  &  &  &  &  &  &  &  &  &
                $\bar{3}$ & $8$ & $3$ & $3$ & $\bar{6}$\\ \cline{1-1}\cline{12-16}
                N4-1-1-1-2-2-4 &  &  &  &  &  &  &  &  &  &  &
                $3$ & $\bar{3}$ & $\bar{3}$ & $3$ & $\bar{6}$\\ \cline{1-1}\cline{12-16}
                N4-1-1-1-2-2-5 &  &  &  &  &  &  &  &  &  &  &
                $3$ & $6$ & $\bar{3}$ & $3$ & $\bar{6}$\\ \cline{1-1}\cline{12-16}
                N4-1-1-1-2-2-6 &  &  &  &  &  &  &  &  &  &  &
                $6$ & $8$ & $\bar{6}$ & $3$ & $\bar{6}$\\ \cline{1-1}\cline{12-16}
                N4-1-1-1-2-2-7 &  &  &  &  &  &  &  &  &  &  &
                $\bar{6}$ & $\bar{3}$ & $6$ & $3$ & $\bar{6}$\\ \cline{1-1}\cline{12-16}
                N4-1-1-1-2-2-8 &  &  &  &  &  &  &  &  &  &  &
                $8$ & $3$ & $8$ & $3$ & $\bar{6}$\\ \cline{1-1}\cline{12-16}
                N4-1-1-1-2-2-9 &  &  &  &  &  &  &  &  &  &  &
                $8$ & $\bar{6}$ & $8$ & $3$ & $\bar{6}$\\ \cline{1-1}\cline{7-16}
                N4-1-1-1-2-3-1 &  &  &  &  &  &
                \multirow{9}{*}{$3$} & \multirow{9}{*}{$3$} & \multirow{9}{*}{$3$} & \multirow{9}{*}{$1$} & \multirow{9}{*}{$1$} &
                $1$ & $3$ & $1$ & $3$ & $\bar{6}$\\ \cline{1-1}\cline{12-16}
                N4-1-1-1-2-3-2 &  &  &  &  &  &  &  &  &  &  &
                $\bar{3}$  & $1$ & $3$ & $3$& $\bar{6}$\\ \cline{1-1}\cline{12-16}
                N4-1-1-1-2-3-3 &  &  &  &  &  &  &  &  &  &  &
                $\bar{3}$ & $8$ & $3$ & $3$ & $\bar{6}$\\ \cline{1-1}\cline{12-16}
                N4-1-1-1-2-3-4 &  &  &  &  &  &  &  &  &  &  &
                $3$ & $\bar{3}$ & $\bar{3}$ & $3$ & $\bar{6}$\\ \cline{1-1}\cline{12-16}
                N4-1-1-1-2-3-5 &  &  &  &  &  &  &  &  &  &  &
                $3$ & $6$ & $\bar{3}$ & $3$ & $\bar{6}$\\ \cline{1-1}\cline{12-16}
                N4-1-1-1-2-3-6 &  &  &  &  &  &  &  &  &  &  &
                $6$ & $8$ & $\bar{6}$ & $3$ & $\bar{6}$\\ \cline{1-1}\cline{12-16}
                N4-1-1-1-2-3-7 &  &  &  &  &  &  &  &  &  &  &
                $\bar{6}$ & $\bar{3}$ & $6$ & $3$ & $\bar{6}$\\ \cline{1-1}\cline{12-16}
                N4-1-1-1-2-3-8 &  &  &  &  &  &  &  &  &  &  &
                $8$ & $3$ & $8$ & $3$ & $\bar{6}$\\ \cline{1-1}\cline{12-16}
                N4-1-1-1-2-3-9 &  &  &  &  &  &  &  &  &  &  &
                $8$ & $\bar{6}$ & $8$ & $3$ & $\bar{6}$\\ \hline\hline %\cline{7-16}
        \end{tabular}
\caption{\label{tab:N41112} The one-loop models for the $0\nu\beta\beta$ decay operator N4 arising from the diagram N4-1-1-1-2 shown in figure~\ref{fig:external-N4}.}
\end{table}
\begin{table}[htbp]
        \renewcommand{\tabcolsep}{0.5mm}
        \renewcommand{\arraystretch}{1.3}
        \centering
        \begin{tabular}{|c|c|c|c|c|c||c|c|c|c|c||c|c|c|c|c|}
                \hline\hline
                \multirow{2}{*}{Model Name} &
                \multicolumn{5}{c||}{$U(1)_{Y}$}& \multicolumn{5}{c||}{$SU(2)_{L}$} & \multicolumn{5}{c|}{$SU(3)_{C}$}\\ \cline{2-16}
                & $F_{1}$ & $S_{1}$ & $S_{2}$ & $S_{3}$ & $S_{4}$ &
                $F_{1}$ & $S_{1}$ & $S_{2}$ & $S_{3}$ & $S_{4}$ &
                $F_{1}$ & $S_{1}$ & $S_{2}$ & $S_{3}$ & $S_{4}$\\ \hline
                N4-1-1-1-3-1-1 & \multirow{12}{*}{$1+\alpha$} & \multirow{12}{*}{$\alpha$} & \multirow{12}{*}{$-2-\alpha$} & \multirow{12}{*}{$\frac{4}{3}$} & \multirow{12}{*}{$\frac{2}{3}$} &
                \multirow{4}{*}{$1$} & \multirow{4}{*}{$1$} & \multirow{4}{*}{$1$} & \multirow{4}{*}{$1$} & \multirow{4}{*}{$1$} &
                $1$ & $1$ & $1$ & $6$ & $\bar{6}$\\ \cline{1-1}\cline{12-16}
                N4-1-1-1-3-1-2&  &  &  &  &  &  &  &  &  &  &
                $\bar{3}$  & $\bar{3}$ & $3$ & $6$ & $\bar{6}$\\ \cline{1-1}\cline{12-16}
                N4-1-1-1-3-1-3&  &  &  &  &  &  &  &  &  &  &
                $6$ & $6$ & $\bar{6}$ & $6$ & $\bar{6}$\\ \cline{1-1}\cline{12-16}
                N4-1-1-1-3-1-4&  &  &  &  &  &  &  &  &  &  &
                $8$ & $8$ & $8$ & $6$ & $\bar{6}$\\ \cline{1-1}\cline{7-16}
                N4-1-1-1-3-2-1& &  &  &  &  &
                \multirow{4}{*}{$2$} & \multirow{4}{*}{$2$} & \multirow{4}{*}{$2$} & \multirow{4}{*}{$1$} & \multirow{4}{*}{$1$} &
                $1$ & $1$ & $1$ & $6$ & $\bar{6}$\\ \cline{1-1}\cline{12-16}
                N4-1-1-1-3-2-2&  &  &  &  &  &  &  &  &  &  &
                $\bar{3}$  & $\bar{3}$ & $3$ & $6$ & $\bar{6}$\\ \cline{1-1}\cline{12-16}
                N4-1-1-1-3-2-3&  &  &  &  &  &  &  &  &  &  &
                $6$ & $6$ & $\bar{6}$ & $6$ & $\bar{6}$\\ \cline{1-1}\cline{12-16}
                N4-1-1-1-3-2-4&  &  &  &  &  &  &  &  &  &  &
                $8$ & $8$ & $8$ & $6$ & $\bar{6}$\\ \cline{1-1}\cline{7-16}
                N4-1-1-1-3-3-1& &  &  &  &  &
                \multirow{4}{*}{$3$} & \multirow{4}{*}{$3$} & \multirow{4}{*}{$3$} & \multirow{4}{*}{$1$} & \multirow{4}{*}{$1$} &
                $1$ & $1$ & $1$ & $6$ & $\bar{6}$\\ \cline{1-1}\cline{12-16}
                N4-1-1-1-3-3-2&  &  &  &  &  &  &  &  &  &  &
                $\bar{3}$  & $\bar{3}$ & $3$ & $6$ & $\bar{6}$\\ \cline{1-1}\cline{12-16}
                N4-1-1-1-3-3-3&  &  &  &  &  &  &  &  &  &  &
                $6$ & $6$ & $\bar{6}$ & $6$ & $\bar{6}$\\ \cline{1-1}\cline{12-16}
                N4-1-1-1-3-3-4&  &  &  &  &  &  &  &  &  &  &
                $8$ & $8$ & $8$ & $6$ & $\bar{6}$\\ \hline \hline
        \end{tabular}
\caption{\label{tab:N41113} The one-loop models for the $0\nu\beta\beta$ decay operator N4 arising from the diagram N4-1-1-1-3 shown in figure~\ref{fig:external-N4}.}
\end{table}
\begin{table}[htbp]
        \renewcommand{\tabcolsep}{0.5mm}
        \renewcommand{\arraystretch}{1.3}
        \centering
        \begin{tabular}{|c|c|c|c|c|c||c|c|c|c|c||c|c|c|c|c|}
                \hline\hline
                \multirow{2}{*}{Model Name} &
                \multicolumn{5}{c||}{$U(1)_{Y}$}& \multicolumn{5}{c||}{$SU(2)_{L}$} & \multicolumn{5}{c|}{$SU(3)_{C}$}\\ \cline{2-16}
                & $F_{1}$ & $S_{1}$ & $S_{2}$ & $S_{3}$ & $S_{4}$ &
                $F_{1}$ & $S_{1}$ & $S_{2}$ & $S_{3}$ & $S_{4}$ &
                $F_{1}$ & $S_{1}$ & $S_{2}$ & $S_{3}$ & $S_{4}$\\ \hline
                N4-1-1-1-4-1-1 & \multirow{21}{*}{$-\frac{1}{3}+\alpha$} & \multirow{21}{*}{$\alpha$} & \multirow{21}{*}{$\frac{2}{3}-\alpha$} & \multirow{21}{*}{$-2$} & \multirow{21}{*}{$\frac{4}{3}$} &
                \multirow{7}{*}{$1$} & \multirow{7}{*}{$1$} & \multirow{7}{*}{$1$} & \multirow{7}{*}{$1$} & \multirow{7}{*}{$1$} &
                $\bar{3}$ & $\bar{6}$ & $1$ & $1$ & $6$\\ \cline{1-1}\cline{12-16}
                N4-1-1-1-4-1-2&  &  &  &  &  &  &  &  &  &  &
                $3$  & $8$ & $3$ & $1$& $6$\\ \cline{1-1}\cline{12-16}
                N4-1-1-1-4-1-3&  &  &  &  &  &  &  &  &  &  &
                $\bar{6}$  & $8$ & $3$ & $1$& $6$\\ \cline{1-1}\cline{12-16}
                N4-1-1-1-4-1-4&  &  &  &  &  &  &  &  &  &  &
                $1$ & $\bar{3}$ & $\bar{3}$ & $1$ & $6$\\ \cline{1-1}\cline{12-16}
                N4-1-1-1-4-1-5&  &  &  &  &  &  &  &  &  &  &
                $8$ & $\bar{3}$ & $\bar{3}$ & $1$ & $6$\\ \cline{1-1}\cline{12-16}
                N4-1-1-1-4-1-6&  &  &  &  &  &  &  &  &  &  &
                $3$ & $8$ & $\bar{6}$ & $1$ & $6$\\ \cline{1-1}\cline{12-16}
                N4-1-1-1-4-1-7&  &  &  &  &  &  &  &  &  &  &
                $8$ & $6$ & $6$ & $1$ & $6$\\ \cline{1-1}\cline{7-16}
                N4-1-1-1-4-2-1& &  &  &  &  &
                \multirow{7}{*}{$2$} & \multirow{7}{*}{$2$} & \multirow{7}{*}{$2$} & \multirow{7}{*}{$1$} & \multirow{7}{*}{$1$} &
                $\bar{3}$ & $\bar{6}$ & $1$ & $1$ & $6$\\ \cline{1-1}\cline{12-16}
                N4-1-1-1-4-2-2&  &  &  &  &  &  &  &  &  &  &
                $3$  & $8$ & $3$ & $1$& $6$\\ \cline{1-1}\cline{12-16}
                N4-1-1-1-4-2-3&  &  &  &  &  &  &  &  &  &  &
                $\bar{6}$  & $8$ & $3$ & $1$& $6$\\ \cline{1-1}\cline{12-16}
                N4-1-1-1-4-2-4&  &  &  &  &  &  &  &  &  &  &
                $1$ & $\bar{3}$ & $\bar{3}$ & $1$ & $6$\\ \cline{1-1}\cline{12-16}
                N4-1-1-1-4-2-5&  &  &  &  &  &  &  &  &  &  &
                $8$ & $\bar{3}$ & $\bar{3}$ & $1$ & $6$\\ \cline{1-1}\cline{12-16}
                N4-1-1-1-4-2-6&  &  &  &  &  &  &  &  &  &  &
                $3$ & $8$ & $\bar{6}$ & $1$ & $6$\\ \cline{1-1}\cline{12-16}
                N4-1-1-1-4-2-7&  &  &  &  &  &  &  &  &  &  &
                $8$ & $6$ & $6$ & $1$ & $6$\\ \cline{1-1}\cline{7-16}
                N4-1-1-1-4-3-1& &  &  &  &  &
                \multirow{7}{*}{$3$} & \multirow{7}{*}{$3$} & \multirow{7}{*}{$3$} & \multirow{7}{*}{$1$} & \multirow{7}{*}{$1$} &
                $\bar{3}$ & $\bar{6}$ & $1$ & $1$ & $6$\\ \cline{1-1}\cline{12-16}
                N4-1-1-1-4-3-2&  &  &  &  &  &  &  &  &  &  &
                $3$  & $8$ & $3$ & $1$& $6$\\ \cline{1-1}\cline{12-16}
                N4-1-1-1-4-3-3&  &  &  &  &  &  &  &  &  &  &
                $\bar{6}$  & $8$ & $3$ & $1$& $6$\\ \cline{1-1}\cline{12-16}
                N4-1-1-1-4-3-4&  &  &  &  &  &  &  &  &  &  &
                $1$ & $\bar{3}$ & $\bar{3}$ & $1$ & $6$\\ \cline{1-1}\cline{12-16}
                N4-1-1-1-4-3-5&  &  &  &  &  &  &  &  &  &  &
                $8$ & $\bar{3}$ & $\bar{3}$ & $1$ & $6$\\ \cline{1-1}\cline{12-16}
                N4-1-1-1-4-3-6&  &  &  &  &  &  &  &  &  &  &
                $3$ & $8$ & $\bar{6}$ & $1$ & $6$\\ \cline{1-1}\cline{12-16}
                N4-1-1-1-4-3-7&  &  &  &  &  &  &  &  &  &  &
                $8$ & $6$ & $6$ & $1$ & $6$\\ \hline \hline
        \end{tabular}
\caption{\label{tab:N41114} The one-loop models for the $0\nu\beta\beta$ decay operator N4 arising from the diagram N4-1-1-1-4 shown in figure~\ref{fig:external-N4}.}
\end{table}
\begin{table}[htbp]
        \renewcommand{\tabcolsep}{0.5mm}
        \renewcommand{\arraystretch}{1.3}
        \centering
        \begin{tabular}{|c|c|c|c|c|c||c|c|c|c|c||c|c|c|c|c|}
                \hline\hline
                \multirow{2}{*}{Model Name} &
                \multicolumn{5}{c||}{$U(1)_{Y}$}& \multicolumn{5}{c||}{$SU(2)_{L}$} & \multicolumn{5}{c|}{$SU(3)_{C}$}\\ \cline{2-16}
                & $F_{1}$ & $S_{1}$ & $S_{2}$ & $S_{3}$ & $S_{4}$ &
                $F_{1}$ & $S_{1}$ & $S_{2}$ & $S_{3}$ & $S_{4}$ &
                $F_{1}$ & $S_{1}$ & $S_{2}$ & $S_{3}$ & $S_{4}$\\ \hline
                N4-1-1-1-5-1-1 & \multirow{21}{*}{$-\frac{2}{3}+\alpha$} & \multirow{21}{*}{$\alpha$} & \multirow{21}{*}{$\frac{4}{3}-\alpha$} & \multirow{21}{*}{$-2$} & \multirow{21}{*}{$\frac{2}{3}$} &
                \multirow{7}{*}{$1$} & \multirow{7}{*}{$1$} & \multirow{7}{*}{$1$} & \multirow{7}{*}{$1$} & \multirow{7}{*}{$1$} &
                $3$ & $6$ & $1$ & $1$ & $\bar{6}$\\ \cline{1-1}\cline{12-16}
                N4-1-1-1-5-1-2&  &  &  &  &  &  &  &  &  &  &
                $1$  & $3$ & $3$ & $1$& $\bar{6}$\\ \cline{1-1}\cline{12-16}
                N4-1-1-1-5-1-3&  &  &  &  &  &  &  &  &  &  &
                $8$  & $3$ & $3$ & $1$& $\bar{6}$\\ \cline{1-1}\cline{12-16}
                N4-1-1-1-5-1-4&  &  &  &  &  &  &  &  &  &  &
                $\bar{3}$ & $8$ & $\bar{3}$ & $1$ & $\bar{6}$\\ \cline{1-1}\cline{12-16}
                N4-1-1-1-5-1-5&  &  &  &  &  &  &  &  &  &  &
                $6$ & $8$ & $\bar{3}$ & $1$ & $\bar{6}$\\ \cline{1-1}\cline{12-16}
                N4-1-1-1-5-1-6&  &  &  &  &  &  &  &  &  &  &
                $8$ & $\bar{6}$ & $\bar{6}$ & $1$ & $\bar{6}$\\ \cline{1-1}\cline{12-16}
                N4-1-1-1-5-1-7&  &  &  &  &  &  &  &  &  &  &
                $\bar{3}$ & $8$ & $6$ & $1$ & $\bar{6}$\\ \cline{1-1}\cline{7-16}
                N4-1-1-1-5-2-1& &  &  &  &  &
                \multirow{7}{*}{$2$} & \multirow{7}{*}{$2$} & \multirow{7}{*}{$2$} & \multirow{7}{*}{$1$} & \multirow{7}{*}{$1$} &
                $3$ & $6$ & $1$ & $1$ & $\bar{6}$\\ \cline{1-1}\cline{12-16}
                N4-1-1-1-5-2-2&  &  &  &  &  &  &  &  &  &  &
                $1$  & $3$ & $3$ & $1$& $\bar{6}$\\ \cline{1-1}\cline{12-16}
                N4-1-1-1-5-2-3&  &  &  &  &  &  &  &  &  &  &
                $8$  & $3$ & $3$ & $1$& $\bar{6}$\\ \cline{1-1}\cline{12-16}
                N4-1-1-1-5-2-4&  &  &  &  &  &  &  &  &  &  &
                $\bar{3}$ & $8$ & $\bar{3}$ & $1$ & $\bar{6}$\\ \cline{1-1}\cline{12-16}
                N4-1-1-1-5-2-5&  &  &  &  &  &  &  &  &  &  &
                $6$ & $8$ & $\bar{3}$ & $1$ & $\bar{6}$\\ \cline{1-1}\cline{12-16}
                N4-1-1-1-5-2-6&  &  &  &  &  &  &  &  &  &  &
                $8$ & $\bar{6}$ & $\bar{6}$ & $1$ & $\bar{6}$\\ \cline{1-1}\cline{12-16}
                N4-1-1-1-5-2-7&  &  &  &  &  &  &  &  &  &  &
                $\bar{3}$ & $8$ & $6$ & $1$ & $\bar{6}$\\ \cline{1-1}\cline{7-16}
                N4-1-1-1-5-3-1& &  &  &  &  &
                \multirow{7}{*}{$3$} & \multirow{7}{*}{$3$} & \multirow{7}{*}{$3$} & \multirow{7}{*}{$1$} & \multirow{7}{*}{$1$} &
                $3$ & $6$ & $1$ & $1$ & $\bar{6}$\\ \cline{1-1}\cline{12-16}
                N4-1-1-1-5-3-2&  &  &  &  &  &  &  &  &  &  &
                $1$  & $3$ & $3$ & $1$& $\bar{6}$\\ \cline{1-1}\cline{12-16}
                N4-1-1-1-5-3-3&  &  &  &  &  &  &  &  &  &  &
                $8$  & $3$ & $3$ & $1$& $\bar{6}$\\ \cline{1-1}\cline{12-16}
                N4-1-1-1-5-3-4&  &  &  &  &  &  &  &  &  &  &
                $\bar{3}$ & $8$ & $\bar{3}$ & $1$ & $\bar{6}$\\ \cline{1-1}\cline{12-16}
                N4-1-1-1-5-3-5&  &  &  &  &  &  &  &  &  &  &
                $6$ & $8$ & $\bar{3}$ & $1$ & $\bar{6}$\\ \cline{1-1}\cline{12-16}
                N4-1-1-1-5-3-6&  &  &  &  &  &  &  &  &  &  &
                $8$ & $\bar{6}$ & $\bar{6}$ & $1$ & $\bar{6}$\\ \cline{1-1}\cline{12-16}
                N4-1-1-1-5-3-7&  &  &  &  &  &  &  &  &  &  &
                $\bar{3}$ & $8$ & $6$ & $1$ & $\bar{6}$\\ \hline \hline
        \end{tabular}
\caption{\label{tab:N41115}The one-loop models for the $0\nu\beta\beta$ decay operator N4 arising from the diagram N4-1-1-1-5 shown in figure~\ref{fig:external-N4}.}
\end{table}

\end{appendix}

\clearpage
%\newpage

%\bibliographystyle{utphys}
%\bibliography{references.bib}

\providecommand{\href}[2]{#2}\begingroup\raggedright\endgroup

\end{document}